\documentclass[letterpaper,11pt]{article}
\usepackage[utf8]{inputenc}
\usepackage{amsmath}
\usepackage{amsthm}
\usepackage{mathrsfs}
\usepackage{fullpage}
\usepackage{color}
\usepackage{bbold}
\newtheorem{theorem}{Theorem}[section]
\newtheorem{corollary}{Corollary}[section]
\newtheorem{lemma}[theorem]{Lemma}

\theoremstyle{definition}

\newtheorem{definition}{Definition}
\usepackage{algorithm}
\usepackage{algorithmic}
\usepackage{url}
\definecolor{Blue}{rgb}{0.0,0.2,0.6}
\usepackage{hyperref}
\hypersetup{
    linktocpage=true,
    colorlinks=true,				
    linkcolor=Blue,				
    citecolor=Blue,				
    urlcolor=Blue,			
}
\usepackage{graphicx}
\usepackage{subcaption}
\newcommand{\argmax}{\mathop{\mathrm{arg\,max}}}
\newcommand{\argmin}{\mathop{\mathrm{arg\,min}}}
\newcommand{\dist}{\mathop{\mathrm{dist}}}

\title{Concentrated Geo-Privacy}

\author{
Yuting Liang\thanks{Computer Science and Engineering, Hong Kong University of Science and Technology. \href{mailto:yliangbs@cse.ust.hk}{yliangbs@cse.ust.hk}}
\and
Ke Yi\thanks{Computer Science and Engineering, Hong Kong University of Science and Technology. \href{mailto:yike@cse.ust.hk}{yike@cse.ust.hk}}
}
\date{}
\begin{document}
\maketitle

\begin{abstract}
This paper proposes \textit{concentrated geo-privacy (CGP)}, a privacy notion that can be considered as the counterpart of \textit{concentrated differential privacy (CDP)} for geometric data.  Compared with the previous notion of geo-privacy \cite{andres2013geo,chatzikokolakis2013broadening}, which is the counterpart of standard differential privacy, CGP offers many benefits including simplicity of the mechanism, lower noise scale in high dimensions, and better composability known as \textit{advanced composition}.  The last one is the most important, as it allows us to design complex mechanisms using smaller building blocks while achieving better utilities.  To complement this result, we show that the previous notion of geo-privacy inherently does not admit advanced composition even using its approximate version.  Next, we study three problems on private geometric data: the identity query, $k$ nearest neighbors, and  convex hulls.  While the first problem has been previously studied, we give the first mechanisms for the latter two under geo-privacy.  For all three problems, composability is essential in obtaining good utility guarantees on the privatized query answer. 
\end{abstract}

\section{Introduction}

\textit{Differential privacy (DP)} is the de facto privacy model for protecting personal information. Under DP, the output of a query is perturbed randomly such that its probability distribution does not change much on datasets that are ``similar''.  In standard DP, this similarity is defined in terms of \textit{neighboring inputs}:
\begin{definition}[Differential Privacy \cite{dwork2006calibrating}] 
\label{defn:dp}
Let $\varepsilon,\delta \geq 0$. A randomized mechanism ${M}:U\rightarrow V$ is $(\varepsilon,\delta)$-differentially private, or simply $(\varepsilon,\delta)$-DP, if for any neighboring inputs $x\sim x'\in U$ and any measurable $S\subseteq V$,
    \[\Pr[{M}(x)\in S]\leq e^{\varepsilon}\cdot \Pr[{M}(x')\in S]+\delta.\]
\end{definition}
Smaller values of $\varepsilon,\delta$ correspond to better privacy.  Typically, $\varepsilon$ is set to a constant, but may be set smaller for individual mechanisms during compositions (as discussed later). Roughly speaking\footnote{To be precise, the probabilistic interpretation is slightly stronger than the stated definition, where it implies the definition and is implied by it up to a small loss in the privacy parameters \cite{kasiviswanathan2014semantics, zhao2019reviewing}.}, $\delta$ is the probability with which $M$ breaches privacy, so it should be negligible. In particular, the mechanism is said to satisfy \textit{pure}-DP if $\delta=0$; the $\delta>0$ case is then called \textit{approximate}-DP. 

Depending on how the neighboring relationship $\sim$ is defined, we arrive at different DP variants.  In the \textit{central} model of DP, $x \sim x'$ iff they differ by a single record.  This models the situation where there is a trusted data curator who collects and publishes privatized query results.  Alternatively, if we set $x\sim x'$ between all pairs of distinct inputs, then the model becomes what is known as \textit{local} DP \cite{erlingsson2014rappor}. This corresponds to the scenario where each data owner privatizes his/her input before sending it to the curator, who is potentially malicious.

In local DP, any single message satisfying Definition \ref{defn:dp} has little utility, as the output distributions on any two inputs are required to be indistinguishable.  So all mechanisms under local DP has to aggregate messages from a large number of users for the result to be useful.  However, for many applications involving geometric data, e.g., the current or past locations of the user, we are interested in obtaining some privatized version of the data on a per user basis. The standard DP definition is thus not appropriate in such scenarios.

\subsection{From DP to GP}
In standard DP\footnote{For an account of differences between standard $(\varepsilon,\delta)$-DP and other DP-related definitions, the reader is referred to \cite{Desfontaines2020SoK}.}, two inputs $x, x'$ are either neighbors or not.  Such a binary relationship cannot capture the more quantitative relationships on geometric data.   
In the example above, requiring two faraway points $x,x'$ (e.g., in different cities) to also have indistinguishable output distributions is likely an overkill.  In seeing this, Andrés et al. \cite{andres2013geo} and Chatzikokolakis et al. \cite{chatzikokolakis2013broadening} extend the neighboring relationship to a metric space $(U, \dist)$, and stipulate that privacy should degrade gracefully as $\dist(x,x')$ increases. They have only considered the pure case $\delta=0$, but their definition extends to the approximate version, i.e.,  $\delta\ge 0$, in the natural way:

\begin{definition}[Geo-Privacy\footnote{This notion is called \textit{geo-indistinguishability} in \cite{andres2013geo} and  \textit{$d_{\chi}$-privacy} in \cite{chatzikokolakis2013broadening}, respectively.} \cite{andres2013geo,chatzikokolakis2013broadening}] 
\label{def:GP} Let $\varepsilon, \delta \geq 0$. A randomized mechanism ${M}:U\rightarrow V$ is $(\varepsilon,\delta)$-geo-private, or simply $(\varepsilon,\delta)$-GP, if for any $x,x'\in U$ and any measurable $S\subseteq V$,
    \[\Pr[{M}(x)\in S]\leq e^{\varepsilon\cdot \mathrm{dist}(x,x')}\cdot \Pr[{M}(x')\in S]+\delta.\]
\end{definition}

Note that $(\varepsilon,0)$-GP incorporates both central DP and local DP by choosing an appropriate distance function $\mathrm{dist}(\cdot,\cdot)$. To recover the former, we use the Hamming metric\footnote{This follows from the group privacy property of DP \cite{dwork2014algorithmic}. However, $(\varepsilon,\delta)$-GP under the Hamming metric is stronger than $(\varepsilon,\delta)$-DP, as the group size allowed in DP group privacy is limited to $O(\log(1/\delta)/\varepsilon)$ for $\delta>0$.}; for the latter, we use the discrete metric $\mathrm{dist}(x,x')=1$ for all $x\ne x'$.  However, the power of GP lies in the flexibility in choosing a distance function suitable for the input domain $U$.  For example, when $U=\mathbb{R}^d$, 
the Euclidean metric $\mathrm{dist}(x,x'):=\|x-x'\|_2$ is arguably a more appropriate distance function than the Hamming or the discrete metric.  Note that in terms of privacy, GP is at least as strong as DP for all $x,x'$ such that $\mathrm{dist}(x,x')\le 1$, while weaker as $\mathrm{dist}(x,x')$ becomes larger than $1$.  This is also why the $\varepsilon$ parameter in GP measures the privacy loss per unit distance, as explained in \cite{andres2013geo,chatzikokolakis2013broadening}.  However, it is exactly due to this weakening of privacy (for faraway inputs) that allows GP to enjoy higher utility.  In particular, due to the use of the discrete metric in local DP, the data collected from any single user has little utility by definition; only the aggregated result from many users may yield some useful information.  In contrast, GP mechanisms offer much better utility.  For example, the noise scale of the Laplace mechanism, the canonical GP mechanism, is\footnote{The $\tilde{O}(\cdot)$ notation suppresses dependency on $\varepsilon, \rho$, and polylogarithmic factors in $n, d, 1/\delta, 1/\beta$.}  $\tilde{O}(d)$ for the identity query (see Section \ref{sec:prelimGP} for details).

\subsection{Our Proposal: CGP}
We propose a new definition of geo-privacy using R\'{e}nyi divergences, which we call \textit{concentrated geo-privacy (CGP)}:
\begin{definition}[Concentrated Geo-Privacy] 
\label{def:CGP}
Let $\rho\geq 0$. A mechanism ${M}:U\rightarrow V$ satisfies $\rho$-concentrated-geo-privacy, or simply $\rho$-CGP, if for any $x, x'\in U$ and all $\alpha > 1$
    \[D_\alpha(\mathcal{M}(x)\|\mathcal{M}(x'))\leq \alpha \rho \cdot \mathrm{dist}(x,x')^2.\]
\end{definition}
In the definition above, $\mathcal{M}(x)$ and $\mathcal{M}(x')$ are the output distributions of the mechanism on $x$ and $x'$, respectively, and $D_\alpha(\cdot \| \cdot)$ denotes the R\'{e}nyi divergence of order $\alpha$ (see Section \ref{sec:prelimDP} for the detailed definition).  Our definition can be considered as the counterpart and generalization of \textit{concentrated differential privacy (CDP)} \cite{bun2016concentrated} in the geometric setting. Note that the definition of CGP stipulates that the R\'{e}nyi divergence grows quadratically as $\mathrm{dist}(x, x')$.  This is due to the relationship between the privacy parameter $\varepsilon$ used in GP and the $\rho$ used in CGP (see Lemma~\ref{lm:GPCGP} and \ref{lm:cgp_dDGP}). More precisely, we show that the privacy of $\rho$-CGP is at least as strong as that of $(\tilde{O}(\sqrt{\rho}), \delta)$-GP, except for faraway points (see Lemma \ref{lm:cgp_dDGP} for the formal statement).  Correspondingly, $\rho$ measures the privacy loss per unit distance squared.  

Compared with GP, CGP offers the following benefits:
\begin{enumerate}
    \item The canonical mechanism under CGP is the Gaussian mechanism, which is much easier to implement than the Laplace mechanism in $d\ge 2$ dimensions.
    \item The noise scale of the Gaussian mechanism is $\tilde{O}(\sqrt{d})$, while that of the Laplace mechanism is $\tilde{O}(d)$. 
    \item Gaussians have many nice properties, in particular, linearity.   Many geometric measurements are assumed to have some Gaussian noise already.  Then the linearity of Gaussians makes it easy to add more noise, if needed, to achieve any desired level of privacy.
    \item CGP admits \textit{advanced composition}, i.e., the noise scale is proportional to $\sqrt{k}$ as opposed to $k$ for GP, in any adaptive $k$-fold composition.  
\end{enumerate}

Point (4) above requires some elaboration.  Composition theorems are an important tool for privacy-preserving data analytics, as they allow us to ask multiple queries, possibly adaptively, on the same input under a given privacy budget.  Furthermore, it makes the modular design of complex mechanisms easy, as we can combine smaller building blocks while tracking the overall privacy loss.  

Indeed, one of the most important reasons why $(\varepsilon,\delta>0)$-DP is of interest is that it allows advanced composition \cite{dwork2014algorithmic}, while basic composition (i.e., noise scale grows linearly in $k$) is the best one can do under pure-DP. 

However, prior work on GP \cite{andres2013geo,chatzikokolakis2013broadening} has omitted the $\delta>0$ case in Definition \ref{def:GP}.  In fact, our initial goal was to derive an advanced composition theorem for GP.  In failing to do so, we actually proved that basic composition is the best possible for GP under the Euclidean metric, \textit{even for $\delta>0$} (see Theorem \ref{th:negative}).  This may have explained why  \cite{andres2013geo,chatzikokolakis2013broadening} did not consider the $\delta>0$ case.  

The intuitive reason why $(\varepsilon,\delta)$-DP admits advanced composition, but $(\varepsilon,\delta)$-GP (under the Euclidean metric) does not, is the following.  The advanced composition theorem for DP \cite{dwork2014algorithmic} actually consists of two terms\footnote{We use $\log$ to denote the natural logarithm with base $e$ throughout this paper.}: 
\begin{equation}
\label{eq:advcomp}
\varepsilon=\sqrt{2k\log(1/\delta)}\cdot \varepsilon'+k\varepsilon'(e^{\varepsilon'}-1),
\end{equation}
where $\varepsilon'$ is the privacy parameter for each of the $k$ mechanisms and $\varepsilon,\delta$ are the privacy parameters of the composed mechanism.  For the typical parameter regime $\varepsilon \le O(1)$, we need to set $\varepsilon' = O\left(\varepsilon/\sqrt{k \log(1/\delta)}\right)$, so that the second term is dominated by the first, yielding the square-root growth often quoted for advanced composition: $\varepsilon = \tilde{O}(\sqrt{k} \cdot \varepsilon')$.  However, in Definition \ref{def:GP} where $\varepsilon$ is replaced by $\varepsilon \cdot \dist(x,x')$, \eqref{eq:advcomp} should hold for all values of $\varepsilon \cdot \dist(x,x')$, which is unbounded.  Thus, the second term will eventually dominate as $x, x'$ get farther away from each other, breaking this square-root (actually, any sub-linear) relationship.  It is possible to fix this issue by adding a third privacy parameter $\Delta$ to limit $\dist(x,x')$ (see Definition \ref{def:deltaGP}), but our CGP definition using R\'{e}nyi divergences solves the problem in a more elegant way using a single parameter $\rho$. 

\subsection{Problems under GP and CGP}
In Section \ref{sec:applications}, we develop GP and CGP mechanisms for queries where the input is an $n$-tuple of points in $\mathbb{R}^d$.  
Such a tuple may represent a trajectory of a user, a collection of points visited, or the features of objects belonging to the user.  This setting corresponds to the local model, where each user privatizes his/her data before sending it to a malicious data curator. Prior works under GP \cite{andres2013geo,liu2021privacy,yang2022k} have only studied the identity query or applied post-processing to it.  In this paper, in addition to the identity query, we have also designed algorithms for the $k$ nearest neighbor ($k$NN) and the convex hull problem.  Both these problems can be solved using the identity query with post-processing, but our new algorithms offer much better utility.  Our algorithms can be made to satisfy either GP or CGP, but thanks to the benefits of CGP, there are some polynomial improvements in the utility as we switch from GP to CGP:

\begin{enumerate}
    \item For the identity query, the GP algorithm has an error of $\tilde{O}(dn)$, while the CGP algorithm has  error $\tilde{O}(\sqrt{dn})$.
    \item For the $k$NN problem, the GP algorithm has an error of $\tilde{O}(k)$, while the CGP algorithm has error $\tilde{O}(\sqrt{k})$.  Both error bounds hold for any $d$. 
    \item For the two-dimensional convex hull problem, the GP algorithm has an error of $\tilde{O}(\sqrt{\omega(x)} +1)$, while the CGP algorithm has an error of $\tilde{O}(\omega(x)^{1/3} +1)$, where $\omega(x)$ denotes the diameter (the farthest distance between any two points) in $x$.  The gap reduces in higher dimensions, though.
\end{enumerate}

Finally, we performed a set of experiments using trajectory data.  The results have confirmed the utility improvement of our $k$NN and convex hull algorithms over the baseline method, as well as the difference between the GP and the CGP versions. 

\subsection{Related Work}

While \cite{andres2013geo,chatzikokolakis2013broadening} are the most related to our work, privacy over geometric data has attracted much attention in the literature, with the following works also relevant. 
\cite{liu2021privacy} privatizes each location on the trajectory under GP, and applies post-processing steps on the privatized trajectory for travel time prediction. \cite{yang2022k} privatizes a collection of points in $\mathbb{R}^2$ and performs $k$-means clustering on the privatized data.
\cite{zhou2018achieving} proposes a variant of DP by defining the neighboring relationship between two points within a given distance of $r$.  However, their definition does not impose any relationship between the privacy and the distance, as GP and CGP do.

Another definition more closely related to standard DP is differential privacy on $\delta$-location set \cite{xiao2015protecting}, which requires a location set $X_{\delta}$ to be given in advance at each timestamp $t$. $X_{\delta}$ is the set of locations such that with probability $1-\delta$, the user is at one of the locations in $X_{\delta}$ at timestamp $t$. $X_{\delta}$ defines neighborhood in the sense that every pair of locations in $X_{\delta}$ needs to satisfy the standard DP requirement.

Many problems on geometric data have been studied under the central DP model.  However, for the privatized output to have any utility, the query must have low \textit{sensitivity}, i.e., the maximum amount of change in the query output under an (arbitrary) change in one record.  Examples of such queries include range counting\footnote{Given a collection of points and a range space, e.g., all halfspaces or axis-parallel rectangles, the problem is to release a privatized data structure from which the number of points inside each range can be approximately counted. } \cite{muthukrishnan2012optimal,nikolov2013geometry,huang2021approximate}, Tukey depth\footnote{Given a set of points, the Tukey depth of a given point $p$ is the minimum number of points on one side of a hyperplane that contains $p$. The vertices on the convex hull thus has Tukey depth $1$.} \cite{kaplan2020find,beimel2019private,gao2021differentially}, and the mean\footnote{For the mean problem to have low sensitivity, we need to further assume that the domain is bounded, e.g., the unit ball.} \cite{huang2021instance,levy2021learning}.  However, problems like $k$NN or convex hull have unbounded sensitivity: For the former, observe that a $k$NN query degenerates into the maximum problem in one dimension when the query point is at $+\infty$, which already has unbounded sensitivity.  For the latter, moving a point to $\infty$ will cause the convex hull to enlarge infinitely.  Note that \cite{kaplan2020find, beimel2019private} only return some point inside the convex hull while \cite{gao2021differentially} aims at producing an approximate convex hull in terms of the Tukey depth. Their algorithms do not return an approximate convex hull in terms of geometric distances, which is only possible under GP or CGP, which restrict (more precisely, lower the privacy requirement of) such large changes in the locations of the points.

There are several works for privatizing trajectory data in the central DP model \cite{chen2012differentially,he2015dpt,gursoy2018utility}. There, the input is a database of trajectories and two inputs are considered neighbors if they differ in a single trajectory. These works aim to generate a collection of synthetic trajectories resembling the original collection in distribution. To this end, they construct a privatized representation of the empirical distribution of chains of locations via private spatial decomposition techniques \cite{cormode2012differentially} which ultimately privatizes counts \cite{dwork2009complexity}. The problem has also been studied in the local DP model where the input consists of a single trajectory \cite{cunningham2021real}. However, their mechanisms do not actually satisfy the local DP requirement, since as mentioned, no mechanism can offer meaningful utility under local DP. By relaxing the privacy model to GP or CGP, our algorithms return privatized trajectories with meaningful utility guarantees.

\section{Preliminaries}
\subsection{Notation}

Any (nontrivial) mechanism $M:U \rightarrow V$ under differential privacy must be randomized.  More formally, such a randomized mechanism is a map $\mathcal{M}: U \rightarrow \mathbb{D}(V)$, where $\mathbb{D}(V)$ denotes the space of all probability distributions over $V$.  We shall use uppercase letters in scripts to denote probability distributions, e.g., $\mathcal{M}(x)$ denotes the output probability distribution of mechanism $\mathcal{M}$ on input $x$.   The same letter in uppercase (without scripts) represents a random variable drawn from this distribution, i.e., $M(x)$ is a random variable drawn from $\mathcal{M}(x)$.   The same letter in lowercase will be used to denote the corresponding probability density function (pdf), i.e., the pdf of $\mathcal{M}(x)$ is $m(x)(\cdot)$. 

\subsection{Differential Privacy}
\label{sec:prelimDP}
In addition to Definition \ref{defn:dp}, another popular variant for DP is based on R\'{e}nyi divergences:
\begin{definition}[R\'{e}nyi Divergences \cite{renyi1961measures, van2014renyi}]
    Let $\mathcal{P}, \mathcal{Q}$ be distributions on domain $R$ with pdf $p(\cdot)$ and $q(\cdot)$, respectively. The R\'{e}nyi divergence of order $\alpha\in (0,1)\cup (1,\infty)$ is defined as
    \[D_\alpha(\mathcal{P}\|\mathcal{Q})=\frac{1}{\alpha -1}\log \left(\int_Rp(y)^\alpha q(y)^{1-\alpha} dy \right).\]
    The max-divergence is
    \[
        D_{\infty}(\mathcal{P}\|\mathcal{Q})=\lim_{\alpha\rightarrow \infty}D_\alpha(\mathcal{P}\|\mathcal{Q}) = \sup_{y\in R}\log\left(\frac{p(y)}{q(y)}\right).
    \]
\end{definition}

The following properties of R\'{e}nyi divergences are well known:
\begin{lemma} [Properties of R\'{e}nyi divergences \cite{van2014renyi}] 
\label{lm:renyi_prop}
For $\alpha \geq 1$, R\'{e}nyi divergences satisfy the following properties:
    \begin{itemize}
        \item[(1)] $\mathrm{[Monotonicity]}$ $D_{\alpha}(\mathcal{P}\|\mathcal{Q})\leq D_{\alpha'}(\mathcal{P}\|\mathcal{Q})$ for $1\leq \alpha \leq \alpha'\leq \infty$.
        \item[(2)] $\mathrm{[Additivity]}$ $D_{\alpha}(\mathcal{P}_1\times \dotsb \times \mathcal{P}_k \| \mathcal{Q}_1\times \dotsb \times \mathcal{Q}_k) = \sum_{j=1}^k D_{\alpha}(\mathcal{P}_j\| \mathcal{Q}_j)$ for pairs of probability distributions $(\mathcal{P}_j,\mathcal{Q}_j)$ on $\sigma$-algebra $\mathscr{F}_j$, $j\in[k]$.
        \item[(3)] $\mathrm{[Data\; Processing\; Inequality]}$ $D_{\alpha}(P_{|\mathscr{G}}\|\mathcal{Q}_{|\mathscr{G}}) \leq D_{\alpha}(\mathcal{P}\|\mathcal{Q})$ for any sub-$\sigma$-algebra $\mathscr{G}\subseteq \mathscr{F}$.
        \item[(4)] $\mathrm{[Gaussian\; Distributions]}$ $D_\alpha\left(\mathcal{N}(\mu_1,\sigma^2 I_{d\times d})\|\mathcal{N}(\mu_2,\sigma^2 I_{d\times d})\right)=\frac{\alpha \|\mu_1-\mu_2\|_2^2}{2\sigma^2}$ for $\mu_1,\mu_2\in \mathbb{R}^d$. 
    \end{itemize}
\end{lemma}

\textit{Concentrated differential privacy} is defined in terms of  R\'{e}nyi divergences: 
\begin{definition} [Concentrated Differential Privacy\footnote{More commonly known as \textit{zero}-concentrated differential privacy. Note that there is a related two-parameter definition $(\mu,\tau)$-mCDP, also known by the name \textit{mean}-concentrated differential privacy \cite{dwork2016concentrated}, where $(\tau^2/2,\tau)$-mCDP implies $\tau^2/2$-CDP \cite{bun2016concentrated}.} \cite{bun2016concentrated}]
Let $\rho\geq 0$. A mechanism ${M}:U\rightarrow V$ satisfies concentrated differential privacy, denoted $\rho$-CDP, if for any neighboring inputs $x\sim x'\in U$, any measurable $S\subseteq V$, and all $\alpha > 1$,
    \[D_\alpha(\mathcal{M}(x)\|\mathcal{M}(x'))\leq \alpha \rho.\] 
\end{definition}

It is known that the privacy guarantee of CDP sits between that of pure and approximate DP \cite{bun2016concentrated}:
\begin{lemma}[Relationship between DP definitions \cite{bun2016concentrated}]
\label{lm:dp_relation}
Let $M$ be a randomized mechanism. Then
\begin{enumerate}
    \item if $M$ is $(\varepsilon,0)$-DP, then it is also $\rho$-CDP where $\rho=\frac{1}{2}\varepsilon^2$;
    \item if $M$ is $\rho$-CDP, then it is also $(\varepsilon,\delta)$-DP, where $\varepsilon=\rho+\sqrt{4\rho\log(1/\delta)}$ for all $\delta > 0$.
\end{enumerate}
\end{lemma}
The above relationships imply that $\varepsilon$ and $\rho$ are related quadratically in the typical parameter regime $\varepsilon\le O(1)$.

We can also characterize $(\varepsilon,\delta)$-DP in terms of the max-divergence.
\begin{lemma} [Characterization of Differential Privacy \cite{dwork2014algorithmic}]
    Let $\varepsilon, \delta \geq 0$. A randomized mechanism ${M}$ is
        $(\varepsilon,\delta)$-differentially private iff
        \[D_{\infty}^{\delta}(\mathcal{M}(x)\|\mathcal{M}(x')) \leq \varepsilon \mathrm{\;\;and\;\;} D_{\infty}^{\delta}(\mathcal{M}(x')\|\mathcal{M}(x)) \leq \varepsilon,\] where 
        \[D_{\infty}^{\delta}(\mathcal{P}\|\mathcal{Q}):=\sup_{S\subseteq R, \Pr[P\in S]> \delta}\log\left(\frac{\Pr[P\in S]-\delta}{\Pr[Q\in S]}\right),\]
for all $x\sim x'$.
\end{lemma}

For a query $g: U \rightarrow V = \mathbb{R}^d$, the simplest mechanism for achieving DP is to add noise drawn from a distribution with scale proportional to its \textit{sensitivity} $\Delta g$ to each dimension of the true query answer:
\begin{lemma} [Canonical DP mechanisms \cite{dwork2006calibrating, bun2016concentrated}]
    Let $g:U\rightarrow \mathbb{R}^d$ such that $\Delta g:=\sup_{x\sim x'} \|g(x)-g(x')\|_p < \infty$ exists. Then the mechanism $M(x):=g(x)+b\cdot[Z_1,\dotsb,Z_d]^T$ is 
    \begin{itemize}
        \item[1.] $\mathrm{[Laplace\; mechanism]}$ $\varepsilon$-DP for $p=1$, where $b=\frac{\Delta g}{\varepsilon}$ and $Z_j\sim \mathrm{Lap}(1)$ for $j\in [d]$;
        \item[2.] $\mathrm{[Gaussian\; mechanism]}$ $\rho$-CDP for $p=2$, where $b=\frac{\Delta g}{\sqrt{2\rho}}$ and $Z_j\sim \mathcal{N}(0,1)$ for $j\in [d]$.
    \end{itemize}
\end{lemma}
Note that the noise scales are proportional to $1/\varepsilon$ and $1/\sqrt{\rho}$, respectively, echoing the quadratic relationship between $\varepsilon$ and $\rho$. 

\begin{lemma} [Compositions of DP mechanisms \cite{dwork2014algorithmic,bun2016concentrated}]
\label{lm:standardDP_comp} Let $M_1$, $\dotsb$, $M_k$ be DP mechanisms. Let $M$ be a $k$-fold adaptive composition of the $M_j$'s. Then $M$ is
\begin{itemize}
    \item[(1)] $(\varepsilon,\delta)$-DP, where $\varepsilon:=\sum_{j\in[k]}\varepsilon_j$ and $\delta:=\sum_{j\in[k]}\delta_j$, if each $M_j$ is $(\varepsilon_j,\delta_j)$-DP for $j\in [k]$;
    \item[(2)] $(\varepsilon,k\delta'+\delta)$-DP for all $\delta> 0$, where $\varepsilon:=\sqrt{2k\log(1/\delta)}\varepsilon'+k\varepsilon'(e^{\varepsilon'}-1)$, if each $M_j$ is $(\varepsilon',\delta')$-DP for $j\in [k]$;
    \item[(3)] $\rho$-CDP, where $\rho:=\sum_{j\in[k]}\rho_j$, if each $M_j$ is $\rho_j$-CDP for $j\in [k]$.
\end{itemize}
\end{lemma}

Note that composition theorems are usually applied in the opposite direction, namely, for a given total privacy budget $\varepsilon$, how to allocate it to the $k$  mechanisms.  Thus, the factor-$k$ growth in $\varepsilon$ in Lemma~\ref{lm:standardDP_comp} (1) implies that the noise scale of each mechanism is $k$ times larger than that from a single invocation of the mechanism.  For this reason, it is often referred to as \textit{basic composition}.  In contrast, (2) is termed \textit{advanced composition}, as it yields a (quasi) square root growth. However, advanced composition leads to a $\delta>0$ for the composed mechanism, even if each individual $M_j$ satisfies pure DP.  In (3), the privacy loss grows linearly in terms of $\rho$.  But since $\rho=\tilde{\Theta}(\varepsilon^2)$, the growth in terms of $\varepsilon$, or equivalently the noise scale, is still square root.  In fact, by combining with Lemma \ref{lm:dp_relation}, when each $M_j$ is the Laplace (or any pure-DP) mechanism or the Gaussian mechanism, the composition result implied by (3) is no worse than that of (2).

\subsection{Geo-Privacy}
\label{sec:prelimGP}

Now suppose the output space is also a metric equipped with distance function $\mathrm{dist}_V(\cdot,\cdot)$.
A query $g: U \rightarrow V$ is \textit{$K$-Lipschitz} if $\mathrm{dist}_V(g(x), g(x')) \le K\cdot \mathrm{dist}(x, x')$ for all $x,x'\in U$.  A canonical mechanism for achieving $\varepsilon$-GP is the following, which can be considered as the instantiation of the \textit{exponential mechanism} \cite{mcsherry2007mechanism} with $\mathrm{dist}_V(g(x),y)$ as the utility score for $y\in V$:

\begin{lemma}[\cite{chatzikokolakis2013broadening}]
\label{lm:gp_basic_mech}
Let $g:U\rightarrow V$ be $K$-Lipschitz. The mechanism that, on input $x$, draws a $y\in V$ from a distribution with pdf $\propto e^{-\frac{\varepsilon}{K}\cdot\mathrm{dist}_V(g(x),y)}$, is $\varepsilon$-GP.
\end{lemma}

However, Lemma \ref{lm:gp_basic_mech} does not lend itself to an efficient implementation.  Andrés et al. \cite{andres2013geo} considered the special case where $U=V=\mathbb{R}^2$ (under the Euclidean metric) and $g(x)=x$ (which is $1$-Lipschitz), namely, one would like privatize a single point in $\mathbb{R}^2$, and showed that Lemma \ref{lm:gp_basic_mech} instantiates into the planar Laplace mechanism \cite{andres2013geo}.  In Appendix \ref{sec:Rd_gen}, we further show how this can be extended to $d$ dimensions. 

By arguments similar to those for standard DP, a basic composition theorem can be proved for GP.
\begin{lemma}[Basic composition for GP\footnote{\cite{andres2013geo} only proved the theorem for the pure case $\delta_j=0$, but extension to $\delta_j>0$ is straightforward.} \cite{andres2013geo}]
\label{lm:gp_basic_comp}
    Let $M_j$ be $(\varepsilon_j,\delta_j)$-GP for $j=1,\dots,k$, and let $M$ be a $k$-fold adaptive composition of the $M_j$'s. Then $M$ is
 $(\varepsilon,\delta)$-GP, where $\varepsilon:=\sum_{j\in[k]}\varepsilon_j$ and $\delta:=\sum_{j\in[k]}\delta_j$.
\end{lemma}

\section{Concentrated Geo-Privacy}

The definition of CGP is given in Definition \ref{def:CGP}.  In this section, we analyze its various properties.

\subsection{The Gaussian Mechanism}
\label{sec:Gauss}
The Gaussian mechanism is the canonical mechanism for CDP. We can extend it to CGP for queries $g:U\rightarrow \mathbb{R}^d$ equipped with the Euclidean metric after factoring in the Lipschitz constant:
\begin{theorem}
\label{lm:cgp_basic_mech}
Let $g:U\rightarrow \mathbb{R}^d$ be a $K$-Lipschitz function. Then the mechanism $G$ defined by $G(x):=g(x)+\frac{K}{\sqrt{2\rho}}Z$, where $Z\sim \mathcal{N}(0,I_{d\times d})$, is $\rho$-CGP.
\end{theorem}
\begin{proof} Let $x, x'\in U$,
let $\mathcal{G}_(x)$ and $\mathcal{G}(x')$ denote the distributions of $G(x)$ and $G(x')$, respectively. For $\alpha>1$,
    \begin{align*}
&{} D_\alpha\left(\mathcal{G}(x)\|\mathcal{G}(x')\right)\\
&= D_\alpha\left(\mathcal{N}\left(g(x), \frac{K^2}{2\rho}I_{d\times d}\right) \Bigg|\Bigg| \mathcal{N}\left(g(x'), \frac{K^2}{2\rho}I_{d\times d}\right)\right)\\
        &= \frac{\alpha\rho}{K^2} \|g(x)-g(x')\|_2^2 
        \qquad (\text{By Lemma \ref{lm:renyi_prop} (4)}) \\
        &\leq \alpha \rho \dist(x,x')^2.
    \end{align*}
\end{proof}

Compared with the $d$-dimensional planar Laplace distribution, 
$\mathcal{N}(0,I_{d\times d})$ is a distribution much easier to draw from: just generate each coordinate from the standard Gaussian.  Furthermore, the noise scale (i.e., $\|G(x) - g(x)\|_2$) of the Gaussian mechanism is $\tilde{O}(\sqrt{d})$ with high probability assuming $K=1$, while that of the Laplace mechanism is $\tilde{O}(d)$ (see Appendix~\ref{sec:Rd_gen}).

\subsection{Compositions}

We can show that our notion of CGP enjoys the same (adaptive) composition property as CDP.  We only state and prove the result for composing two mechanisms (proof follows similar arguments as those in \cite{mironov2017renyi} and is included in  Appendix~\ref{sec:cgp_comp_proof}); a simple induction leads to $k$-fold compositions.  Consequently, by Lemma \ref{lm:cgp_basic_mech}, the noise scale of each mechanism is $\sqrt{k}$ times larger than that on a single mechanism. 

\begin{theorem}
    \label{lm:cgp_comp}
    Let ${M}_1:U\rightarrow V_1$ be $\rho_1$-CGP, let ${M}_2:U\times V_1 \rightarrow V_2$ be $\rho_2$-CGP w.r.t. its first argument. Then the mechanism ${M}:U\rightarrow V_1\times V_2$ defined by ${M}(x)=({M}_1(x),{M}_2(x,{M}_1(x))$ is $(\rho_1+\rho_2)$-CGP.
\end{theorem}
Note that if $\rho_2=0$ in Lemma \ref{lm:cgp_comp}, which means that $M_2(x, y_1)$ does not depend on $x$, then the lemma degenerates into the post-processing property of CGP, which also follows from property (3) of Lemma~\ref{lm:renyi_prop}.

\paragraph{A negative result on the composability of $(\varepsilon,\delta)$-GP.}
It turns out that basic composition (Lemma \ref{lm:gp_basic_comp}) is the best we can do for GP, i.e., the privacy loss in a $k$-fold composition has to grow linearly in $k$, even if the composed mechanism is allowed to have $\delta>0$.  This stands in contrast with $(\varepsilon,\delta)$-DP, where advanced composition theorems are known to achieve an $\tilde{O}(\sqrt{k})$ growth\footnote{Thus, composition results tighter than that of Dwork and Roth \cite{dwork2014algorithmic} under standard $(\varepsilon,\delta)$-DP (e.g., \cite{kairouz2015composition, murtagh2015complexity}) also do not have counterparts under $(\varepsilon,\delta)$-GP.} for any negligible $\delta>0$. 

To formalize this result, we first note that $(\varepsilon,\delta)$-GP can also be characterized by R\'{e}nyi divergences, by replacing $\varepsilon$ with $\varepsilon\cdot\dist(x,x')$ in the characterization of standard DP:
\begin{lemma}
\label{lm:GP_renyi}
A mechanism $M:U\rightarrow V$ is $(\varepsilon,\delta)$-GP iff for all $x,x'\in U$,  $D_{\infty}^{\delta}(\mathcal{M}(x)\|\mathcal{M}(x')) \leq \varepsilon\cdot \mathrm{dist}(x,x')$ and $D_{\infty}^{\delta}(\mathcal{M}(x')\|\mathcal{M}(x)) \leq \varepsilon\cdot \mathrm{dist}(x,x')$. 
\end{lemma}

For a negative result, it suffices to consider the special case $U=V=\mathbb{R}$ with $g(x)=x$, where we use the canonical GP mechanism in Lemma \ref{lm:gp_basic_mech}.  In this case, on every input $x$, the mechanism returns an output $M(x)$ drawn from a distribution $\mathcal{M}(x)$ with pdf $m(x)(y)\propto e^{-\varepsilon\cdot |y-x|}$.  

\begin{lemma}
\label{lm:eps_GP_comp}
    Let each ${M}_j:\mathbb{R}\rightarrow \mathbb{R}$ be the $\varepsilon$-GP mechanism defined above. For any $k\ge 2$, define ${M}=(M_1(x),\dotsb,M_k(x))$. Then for any $\varepsilon>0$, $0\le \delta \le \frac{1}{4}$, $0<\lambda <1$,  $x\in\mathbb{R}$, there is $x'\in\mathbb{R}$ such that $D_{\infty}^{\delta}(\mathcal{M}(x)\|\mathcal{M}(x'))> (1-\lambda)k\varepsilon\cdot \|x-x'\|$.
\end{lemma}
\begin{proof}
    Let $x'=x-\Delta$ where $\Delta := \frac{2}{\varepsilon \lambda}\log(k)$. Let $s$ be a point between $x'$ and $x$ given by $s:=x-\frac{\lambda}{2}\Delta$, let $S^*:=\{(y_1,\dotsb,y_k)\in \mathbb{R}^k: y_j\geq s\; \forall j\in[k]\}$. Let $Y=(Y_1,\dotsb,Y_k), Y'=(Y'_1,\dotsb,Y'_k)$ be random vectors corresponding to ${M}(x)$ and ${M}(x')$, respectively.
     Then $\Pr[Y\in S^*] = \Pi_{j\in[k]}\Pr[Y_j\geq s] = (1-\frac{1}{2}e^{-\varepsilon\frac{\lambda}{2}\Delta})^k=(1-\frac{1}{2k})^k\geq \frac{1}{2}$ and $\Pr[Y'\in S^*]=\Pi_{j\in[k]}\Pr[Y'_j\geq s]=(\frac{1}{2}e^{-\varepsilon(1-\frac{\lambda}{2})\Delta})^k$. Thus,
    \begin{align*}
    D_{\infty}^{\delta}(\mathcal{M}(x)\|\mathcal{M}(x')) &= \max_{S\subseteq\mathbb{R}:\Pr[Y\in S]>\delta}\log\left(\frac{\Pr[Y\in S]-\delta}{\Pr[Y'\in S]}\right)\\
     &\geq \log\left(\frac{\Pr[Y\in S^*]-\delta}{\Pr[Y'\in S^*]}\right),\\ 
        e^{D_{\infty}^{\delta}(\mathcal{M}(x)\|\mathcal{M}(x'))} &\geq \frac{\Pr[Y\in S^*]-\delta}{\Pr[Y'\in S^*]} = \frac{(1-\frac{1}{2}e^{-\varepsilon\frac{\lambda}{2}\Delta})^k-\delta}{(\frac{1}{2}e^{-\varepsilon(1-\frac{\lambda}{2})\Delta})^k}\\
        &\geq \frac{1/2-\delta}{(1/2)^ke^{-k\varepsilon(1-\lambda/2)\Delta}} \\
        &\geq \frac{1}{4}2^ke^{k\varepsilon(1-\lambda/2)\Delta}
        > e^{k\varepsilon(1-\lambda)\|x-x'\|}.
    \end{align*}
\end{proof}

Combining Lemma \ref{lm:GP_renyi} and \ref{lm:eps_GP_comp}, we obtain:
\begin{theorem}
\label{th:negative}
    For any $\varepsilon>0, 0\le \delta \le {1\over 4}, 0< \lambda <1$, and any $k\ge 2$, there is an $(\varepsilon,0)$-GP mechanism $M$ such that its $k$-fold composition does not satisfy $((1-\lambda)k\varepsilon,\delta)$-GP. 
\end{theorem}

As a negative result, Theorem \ref{th:negative} does not preclude the possibility that a particular GP mechanism under in particular metric space may have better composability.  In fact, a general impossibility result is not true, since a GP mechanism under the discrete metric satisfies local DP, for which advanced composition holds.  Nevertheless, the negative result means that one cannot use GP mechanisms as black boxes to compose more complex GP mechanism with a privacy loss better than linear.  In particular, this is the case for the canonical GP mechanism in Lemma \ref{lm:gp_basic_mech}.

\subsection{Relationships between GP and CGP}

In terms of the relationship between GP and CGP, we have a partial analogue of Lemma~\ref{lm:dp_relation} in geo-privacy.

\begin{lemma} 
\label{lm:GPCGP}
Any $(\varepsilon,0)$-GP mechanism  is also $\frac{\varepsilon^2}{2}$-CGP.
\end{lemma}
\begin{proof}
    It has been shown in \cite{bun2016concentrated} that given $e^{-\varepsilon_0} \leq \frac{p(y)}{q(y)} \leq e^{\varepsilon_0}$ for all $y$, then $e^{(\alpha-1)D_{\alpha}(\mathcal{P}\|\mathcal{Q})}\leq e^{(\alpha-1)\alpha \frac{{\varepsilon_0}^2}{2}}$ for all $\alpha > 1$. Let $x, x'\in U$. 
    Let $\varepsilon_0:=\varepsilon\cdot \mathrm{dist}(x,x')$, then $e^{-\varepsilon_0}\leq \frac{m(x)(y)}{m(x')(y)} \leq e^{\varepsilon_0}$, which implies $e^{(\alpha-1)D_{\alpha}(\mathcal{M}(x)\|\mathcal{M}(x'))}\leq e^{(\alpha-1)\alpha \frac{{\varepsilon_0}^2}{2}}=e^{(\alpha-1)\alpha \frac{{\varepsilon}^2\cdot \mathrm{dist}(x,x')^2}{2}}$. I.e. $D_{\alpha}(\mathcal{M}(x)\|\mathcal{M}(x')) \leq \alpha\frac{{\varepsilon}^2}{2}\cdot \mathrm{dist}(x,x')^2$ for all $\alpha > 1$.
\end{proof}

However, $\rho$-CGP does not imply $\tilde{O}(\sqrt{\rho}, \delta)$-GP: If it did, we would be able to compose $k$ $\varepsilon$-GP mechanisms, which also satisfy $O(\varepsilon^2)$-CGP by Lemma \ref{lm:GPCGP}, into an $O(k\varepsilon^2)$-CGP mechanism by Theorem \ref{lm:cgp_comp}, hence an $(\tilde{O}(\sqrt{k}\varepsilon, \delta)$-GP mechanism, violating Theorem \ref{th:negative}.  Fundamentally, it is exactly due to this weakening in privacy that gives CGP the better composability. 

Nevertheless, below we show that this weakening in privacy only happens for faraway inputs.  Specifically, we show that CGP is at least as strong as the following natural relaxation of $(\varepsilon,\delta)$-GP:

\begin{definition}[$(\varepsilon,\delta,\Delta)$-GP]
\label{def:deltaGP}
    Let $\varepsilon,\delta \geq 0$ and $\Delta > 0$. A mechanism ${M}:U\rightarrow V$ is $(\varepsilon,\delta,\Delta)$-GP, if for any measurable $S\subseteq V$ and any $x,x'\in U$ satisfying $ \mathrm{dist}(x,x')\leq \Delta$,
    \[\Pr[\mathcal{M}(x)\in S]\leq e^{\varepsilon \cdot \mathrm{dist}(x,x')}\Pr[\mathcal{M}(x')\in S]+\delta.\]
\end{definition}
Note that the original $(\varepsilon,\delta)$-GP definition is the special case where $\Delta=\infty$.

\begin{lemma}
\label{lm:cgp_dDGP}
    Any mechanism ${M}$ that is $\rho$-CGP is also $(\varepsilon,\delta,\Delta)$-GP, for any $\varepsilon, \delta, \Delta$ such that $\varepsilon \ge \rho \Delta + 2\sqrt{\rho\log(1/\delta)}$. 
\end{lemma}
Plugging in typical values of $\varepsilon=1, \delta=10^{-10}$, Lemma \ref{lm:cgp_dDGP} implies that the privacy of CDP with $\rho\approx 0.01$ is at least as strong as that of $(\varepsilon, \delta, \Delta \approx 10)$-GP.  Note that for $\dist(x,x') \ge 10$, the privacy provided by GP between $M(x)$ and $M(x')$ is already negligible: The adversary can distinguish between $x$ and $x'$ with probability $1-e^{-10}$.  So weakening it further will not introduce any noticeable differences. The proof of Lemma~\ref{lm:cgp_dDGP} uses a similar derivation as that in \cite{bun2016concentrated} and can be found in Appendix~\ref{sec:cgp_dDgp_proof}.

\paragraph{An additional note on $(\varepsilon,\delta)$-GP vs. $(\varepsilon,\delta)$-DP.}
{One of the main advantages of standard $(\varepsilon,\delta>0)$-DP is that it allows advanced composition of $(\varepsilon,\delta\ge 0)$-DP mechanisms, which is not possible for $(\varepsilon,\delta)$-GP in general as we've shown in the negative result above. In addition, standard $(\varepsilon,\delta>0)$-DP admits natural mechanisms such as the Gaussian mechanism. Below we give a similar construction as in the proof of Lemma~\ref{lm:eps_GP_comp} which shows that adding constant Gaussian noise will not achieve $(\varepsilon,\delta)$-GP. Thus, currently we do not know of any mechanism that satisfies $(\varepsilon,\delta>0)$-GP but not $\varepsilon$-GP.

\begin{lemma}
    \label{lm:gauss_not_approx_gp}
    Fix $\varepsilon> 0$ and $\frac{1}{4}>\delta\geq 0$. Let ${M}:\mathbb{R}\rightarrow\mathbb{R}$ be the mechanism defined by $M(x) := x+\sigma Z$, where $Z\sim\mathcal{N}(0,1)$ and $\sigma=\sigma(\varepsilon,\delta)>0$ is any constant which may depend on $\varepsilon$ and $\delta$. Fix $x\in\mathbb{R}$. Then there is $x'\in \mathbb{R}$ such that $D_{\infty}^{\delta}(\mathcal{M}(x)\|\mathcal{M}(x'))> \varepsilon\|x-x'\|$.
\end{lemma}
\begin{proof}
    Let $x'=x-\Delta$ for some $\Delta >0$ to be decided, let $S^*:=\{y\in \mathbb{R}: y\geq x\}$. Then $\Pr[{M}(x)\in S^*]=\frac{1}{2}$. Let $Z\sim \mathcal{N}(0,\sigma^2)$, then $\Pr[Z\geq\Delta]\leq e^{-\frac{\Delta^2}{2\sigma^2}}$ by Lemma~\ref{ft:univariate_normal_ineq}. Thus, $\Pr[{M}(x')\in S^*]=\Pr[{M}(x')-x'\geq \Delta]=\Pr[Z\geq \Delta]\leq e^{-\frac{\Delta^2}{2\sigma^2}}$. Now choose $\Delta > \max(4\varepsilon\sigma^2, \frac{\ln 4}{\varepsilon})$, then
     \begin{align*}
        e^{D_{\infty}^{\delta}(\mathcal{M}(x)\|\mathcal{M}(x'))} &\geq \frac{\Pr[{M}(x)\in S^*]-\delta}{\Pr[{M}(x')\in S^*]} \\
        &\geq \frac{1/2-\delta}{e^{-\frac{\Delta^2}{2\sigma^2}}} \geq \frac{1}{4}e^{\frac{\Delta^2}{2\sigma^2}} > \frac{1}{4}e^{2\varepsilon \Delta} > e^{\varepsilon \|x-x'\|},
     \end{align*}
     where the second last inequality is due to $\Delta > 4\varepsilon\sigma^2$ and the last inequality is due to $\Delta > \frac{\ln 4}{\varepsilon}$.
\end{proof}
}

\section{Applications}
\label{sec:applications}
So far we've only described basic mechanisms for GP and CGP, where both require the function to be Lipschitz. In general, differentiable functions with bounded first-order derivatives are Lipschitz. Some examples are: point functions such as computing the mean, median and general linear maps; functions which deal with distance such as (min or max) distance or projection of a point to a line or collection of points (see also Lemma~\ref{lm:lipschitz_properties}). However, many functions of broad interest - such as identifying the nearest neighbor and computing the convex hull - cannot be easily posed as such. For these functions, we have to design multi-step algorithms which leverage composition and more advanced technical tools, as will be demonstrated in our applications.

In this section, we develop mechanisms for queries where the input is an $n$-tuple of points, i.e., the input domain is $U=(\mathbb{R}^d)^n$.  For simplicity, we focus on the $d=2$ case; extension to $d>2$ dimensions is briefly discussed at the end of each subsection.

Both GP and CGP require a definition of the distance function $\dist(\cdot, \cdot)$.  Let $x=(x_1,\dotsb,x_n)$ and $x'=(x'_1,\dotsb,x'_n)$ be two tuples of points, where each $x_i,x'_i\in \mathbb{R}^2$ for $i\in[n]$.  While the Euclidean distance is often the default distance function in $\mathbb{R}^2$, there are different ways to combine $n$ Euclidean distances into one.  The most natural ones are: $\mathrm{dist}_{\infty}(x,x') := \max_i\mathrm{dist}(x_i,x'_i)$, $\mathrm{dist}_1(x,x'):=\sum_i\mathrm{dist}(x_i,x'_i)$, and $\mathrm{dist}_2(x,x'):=\sqrt{\sum_i\|x_i-x'_i\|_2^2}$. 
Among these, $\mathrm{dist}_{\infty}$ provides the strongest privacy guarantee and is also the distance function adopted by \cite{andres2013geo}.  
In the remaining sections, we work mainly with $\dist_{\infty}$ and also briefly discuss $\dist_1$ and $\dist_2$ in Appendix~\ref{sec:metrics_central}.

\subsection{Technical Lemmas}

The utility analyses in the later subsections depend on some technical lemmas, whose proofs are deferred to Appendix~\ref{sec:technical_proofs}. The first lemma characterizes the magnitude of $\|\tilde{x}-x\|$, where $x\in \mathbb{R}^2$, and $\tilde{x}$ is the privatization of $x$ obtained from either the planar Laplace mechanism or Gaussian mechanism (see Appendix~\ref{sec:Rd_gen} for the case $d\ge 2$).

\begin{lemma}[Generalized Gamma Distribution~\cite{stacy1962generalization, stacy1965parameter}]
\label{lm:gengamma}
Let $G\sim \mathcal{G}(\lambda,k,p)$, which has pdf $g(r)=\frac{p/\lambda^k}{\Gamma(k/p)}r^{k-1}e^{-(r/\lambda)^p}$ for $r\in(0,\infty)$, where $\Gamma:\mathbb{R}_{\geq 0}\rightarrow \mathbb{R}$ is the gamma function defined by $\Gamma(z)=\int_0^{\infty}t^{z-1}e^{-t}dt$. Then $\mathbb{E}[G] = \lambda \frac{\Gamma(k/p + 1/p)}{\Gamma(k/p)}$ and $\Pr[G\leq r] = \frac{\gamma(k/p,(r/\lambda)^p)}{\Gamma(k/p)}$, where 
$\gamma:\mathbb{R}_{\geq 0}\times\mathbb{R}_{\geq 0}\rightarrow \mathbb{R}$ defined by $\gamma(z,v)=\int_0^v t^{z-1}e^{-t} dt$ is the lower incomplete gamma function. 
\end{lemma}

Recall the Lambert W function \cite{corless1996lambertw}, which computes the inverse of the function $f(w):=we^w$. We have the following inequality from \cite{chatzigeorgiou2013bounds}, {which will be used for bounding the growth of $\|\tilde{x}-x\|$ when $\tilde{x}$ is obtained from the planar Laplace mechanism.}
\begin{lemma}[\cite{chatzigeorgiou2013bounds}]
\label{lm:lambertw_bound}
Let $f^{-1}(\cdot)$ be the Lambert W function.  For $u>0$,  we have
 \[-1-\sqrt{2u}-u<f^{-1}(-e^{-u-1})<-1-\sqrt{2u}-2u/3.\]
\end{lemma}

{The analysis in Section~\ref{sec:kNN} involves the random variable $Y:=Z+W$, where $Z$, $W$ are i.i.d. $\mathrm{Lap}(b)$ random variables. We will use the following technical results:
\begin{lemma}
\label{lm:ZW_bound}
    Let $Z$, $W\sim \mathrm{Lap}(b)$, and $Y:=Z+W$. Then for $1>\beta>0$ with probability $1-\beta$, 
    $|Y| \leq b\left(\sqrt{2\log(1/\beta)}+\log(\frac{1}{\beta})\right)$.
\end{lemma}
}
\begin{lemma} \label{lm:V_leq_ZW_count}
Given $y\in \mathbb{R}$, suppose we draw a $V\sim \mathrm{Lap}(2b)$ until $V\leq y$. Let $r(y)$ be the number of draws given $y$ and let $R=r(Y)$, where $Y:=Z+W$ and $Z$, $W\sim \mathrm{Lap}(b)$. Then $\mathbb{E}[R]\leq 4$.
\end{lemma}

\subsection{The Identity Query}
\label{sec:identity}
The first problem we consider is the identity query $g(x):=x$, i.e., we wish to privatize the entire tuple of $n$ points.  

\paragraph{GP mechanism with $\dist_\infty$.}
With $\mathrm{dist}_{\infty}$, Lemma \ref{lm:gp_basic_mech} requires us to sample a $y$ from the pdf $m(x)(y) \propto e^{-\varepsilon \max_i\|y_i-x_i\|}$, which does not resemble any well-known distribution. Thus, implementing this would be difficult (a weakness typical of the exponential mechanism).  To get around this difficulty, one can use privacy composition to privatize each $g_i:(\mathbb{R}^2)^n\rightarrow \mathbb{R}^2$ defined by $g_i(x)=x_i$, which is $1$-Lipschitz with respect to $\mathrm{dist}_{\infty}$, using the 2-dimensional Laplace mechanism \cite{andres2013geo}. However, \cite{andres2013geo} did not give any utility analysis of this mechanism.  We provide one below.

\begin{lemma}
\label{lm:gp_r_bound}
Let $G:\mathbb{R}^2\rightarrow \mathbb{R}^2$ be the $\varepsilon$-GP mechanism which on input $x$ draws a $y\in \mathbb{R}^2$ from a distribution with pdf $\propto e^{-\varepsilon \|x-y\|}$. Then with probability at least $1-\beta$,
\[
\|G(x)-x\|\leq O\left({1\over \varepsilon}\log{1\over \beta}\right).
\]
\end{lemma}
\begin{proof}
The following identities of the lower incomplete gamma function can be easily derived: 
\begin{equation}
\nonumber
\gamma(1,v) = \int_0^v e^{-t} dt = 1-e^{-v},
\end{equation}
\begin{equation}
\nonumber
\gamma(z+1,v) = z\gamma(z,v)-v^{z}e^{-v},  z>0.
\end{equation}
Let $R$ denote the random variable $\|G(x)-x\|$. Now $R\sim \mathcal{G}(1/\varepsilon,2,1)$ and by Lemma~\ref{lm:gengamma},
\begin{align*}
\nonumber
\Pr[R > r] &= 1-\Pr[R \leq r]\\
\nonumber
&= 1-\frac{\gamma(2,r\varepsilon)}{\Gamma(2)} = 1- \frac{\gamma(1,r\varepsilon)-r\varepsilon e^{-r\varepsilon}}{1}\\
\label{eqn:rad_tail}
& = 1 - (1-e^{-r\varepsilon}-r\varepsilon e^{-r\varepsilon}) = (1+r\varepsilon)e^{-r\varepsilon}.
\end{align*}
We want to find $r$ so that the above probability is at most $\beta$. We will solve $(1+y)e^{-y}=\beta$ for $y \geq 0$ via the Lambert W function $f^{-1}$ where $f(w):=we^w$. We can write
\begin{align*}
-\beta &= (-1-y)e^{-y} = (-1-y)e^{-1-y}e^1\\
-\frac{\beta}{e} &= (-1-y)e^{-1-y}.
\end{align*}
Then we can set $w=-1-y\leq -1$, where $w=f^{-1}(\frac{-\beta}{e})$. Using the inequality  $-1-\sqrt{2u}-u<f^{-1}(-e^{-u-1})$ from Lemma~\ref{lm:lambertw_bound} for $u=\log(1/\beta)>0$, we have $y=-w-1 <\sqrt{2\log(1/\beta)}+\log(1/\beta)$. I.e. $r<\frac{1}{\varepsilon}(\sqrt{2\log(1/\beta)}+\log(1/\beta)).$
\end{proof}
\begin{theorem}
\label{th:basic_gp_maxerr}
Let $G:(\mathbb{R}^2)^n\rightarrow (\mathbb{R}^2)^n$ be defined by $G(x)=(G_1(x_1),\dotsb,G_n(x_n))$, where each $G_i(x_i)$  returns a $y_i\in \mathbb{R}^2$ from a distribution with pdf $\propto e^{-\frac{\varepsilon}{n}\|x_i-y_i\|}$, for $i\in[n]$. Then $G$ is $\varepsilon$-GP and with probability at least $1-\beta$,
\[
\textstyle{\dist_\infty(G(x), x)} \leq 
O\left({n\over \varepsilon} \log{n\over \beta}\right).
\]
\end{theorem}
\begin{proof}
    Follows from Lemmas \ref{lm:gp_basic_comp}, \ref{lm:gp_r_bound}, and a union bound.
\end{proof}

\paragraph{CGP mechanism with $\dist_\infty$.}
Under $\dist_\infty$, we simply apply use CGP composition and apply the Gaussian mechanism on each $g_i(x) = x_i$, i.e.,  adding a noise drawn from $\sqrt{n/2\rho} \cdot \mathcal{N}(0,1)$ to each coordinate of each point.  Below, we analyze its error in terms of $\dist_\infty$:

\begin{lemma}
\label{lm:cgp_r_bound}
Let $G:\mathbb{R}^2\rightarrow \mathbb{R}^2$ be the $\rho$-CGP mechanism defined by $G(x):=x+\frac{1}{\sqrt{2\rho}}Z$ where $Z\sim \mathcal{N}(0,I_{2\times 2})$. Then with probability at least $1-\beta$, we have
$
\|G(x)-x\| \leq \sqrt{\log(1/\beta)/\rho}.
$
\end{lemma}
\begin{proof}
Let $R$ denote the random variable $\|G(x)-x\|$. Then $R \sim \mathcal{G}(1/\sqrt{\rho},2,2)$ and by Lemma~\ref{lm:gengamma}
\begin{align*}
\Pr[R >r] &= 1-\Pr[P\leq r] \\
&= 1- \frac{\gamma(1,(\sqrt{\rho}r)^2)}{\Gamma(1)} = 1 - (1 - e^{-\rho r^2}) = e^{-\rho r^2}.
\end{align*}
Setting the above probability to be $\beta$ gives $r=\sqrt{\log(1/\beta)/\rho}$.
\end{proof}
\begin{theorem}
\label{thm:basic_cgp_maxerr}
Let $G:(\mathbb{R}^2)^n\rightarrow (\mathbb{R}^2)^n$ be defined by $G(x)=(G_1(x_1),\dotsb,G_n(x_n))$, where $G_i(x_i):=x_i+\frac{\sqrt{n}}{\sqrt{2\rho}}Z_i$ and $Z_i\sim \mathcal{N}(0,I_{2\times 2})$, for $i\in[n]$. Then $G$ is $\rho$-CGP and with probability at least $1-\beta$,
\[
\textstyle{\dist_\infty(G(x), x)} \le \sqrt{n \log({n}/{\beta}) \over \rho}.
\]
\end{theorem}
\begin{proof}
    Follows from Lemmas \ref{lm:cgp_comp}, \ref{lm:cgp_r_bound},  and a union bound.
\end{proof}

We see that the CGP mechanism is an $\tilde{O}(\sqrt{n})$-factor better than the GP mechanism. 

\paragraph{Extension to $d$ dimensions.}  All the mechanisms above extend to $d$ dimensions straightforwardly using the $d$-dimensional Laplace and Gaussian mechanism, respectively.  Correspondingly, the utility gap between GP and CGP enlarges to $\tilde{O}(\sqrt{dn})$.

\subsection{$k$ Nearest Neighbors}
\label{sec:kNN}
Given a query point $p\in \mathbb{R}^2$ and a tuple $x=(x_1,\dots,x_n)\in(\mathbb{R}^2)^n$, the $k$ nearest neighbors ($k$NN) query returns the $k$ indices of the points in $x$ that are the nearest to $p$.  This could be useful in applications where $x$ represents the locations visited by a user, say, in the past $l$ days, and the data collector wishes to know if and when s/he has gotten into the proximity of a point $p$ of interest (or danger).  Note that for modularity, we define the problem so that only the indices of the $k$ nearest neighbors are returned; if their locations are also needed (e.g., for computing the distance to $p$), then one can reserve part of the privacy budget to privatize these $k$ locations using the mechanism from Section \ref{sec:identity}.  

For the $k$NN query, the canonical GP or CGP mechanism cannot be applied, as it is not Lipschitz.  In fact, while we can still use $\dist_\infty$ for the input domain, it is not even clear what distance function to use for the output domain.    As a simple baseline algorithm, we can privatize the entire $x$ using the mechanisms from  Section \ref{sec:identity}.  Then finding the $k$ nearest neighbors of $p$ becomes just a post-processing step.  However, as shown in Section \ref{sec:identity}, this baseline method has an error of $\tilde{O}(n/\varepsilon)$ and $\tilde{O}(\sqrt{n/\rho})$ under GP and CGP, respectively. Below we develop an algorithm whose error depends (linearly or square-root) only on $k$, while the dependency on $n$ will be logarithmic.  

\paragraph{The sparse vector technique.}
As a technical tool, we first show that the \textit{sparse vector technique (SVT)} \cite{dwork2014algorithmic}, originally designed under standard DP, can be adapted to satisfy $\varepsilon$-GP, hence $\rho$-CGP for $\rho=\varepsilon^2/2$, which can be of independent interest. 

Under GP or CGP, we are given a (possibly infinite) sequence of $K$-Lipschitz queries $g_1,g_2,\dots$ with range in $\mathbb{R}$.  The goal of SVT is to find the (index of the) first query whose output exceeds a given threshold $T$.  The details are given in Algorithm \ref{alg:SVT}, in which we flip the goal, i.e., try to find the first query whose output is below $T$.  This does not introduce any real difference but makes our $k$NN algorithm more natural.
The proof that Algorithm \ref{alg:SVT} satisfies $\varepsilon$-GP is similar to the DP version \cite{lyu2016understanding, dwork2014algorithmic} and is included in Appendix~\ref{sec:svt_proof} for completeness.
\begin{algorithm}
\caption{Sparse Vector Technique}
    \label{alg:SVT}
    \begin{flushleft}
    \textbf{Input}: $x\in U$; $\varepsilon$; $T$; $K$; $g_1, g_2, \dotsb$ where each $g_j$ is $K$-Lipschitz\\
    \textbf{Output}: variable-length sequence $y_1, y_2, \dotsb$
    \end{flushleft}
    \begin{algorithmic}[1]
    \STATE $\varepsilon_1 \gets \varepsilon/2$, $\varepsilon_2 \gets \varepsilon/2$
    \STATE draw $W\sim\mathrm{Lap}(K/\varepsilon_1)$
    \FOR{$j = 1,\dotsb$}
    \STATE draw $V_j \sim\mathrm{Lap}(2K/\varepsilon_2)$
    \IF{$g_j(x)+V_j \leq T+W$}
    \STATE output $y_j = \top$ and \textbf{HALT}
    \ELSE
    \STATE output $y_j = \bot$
    \ENDIF
    \ENDFOR
    \end{algorithmic}
\end{algorithm}

\paragraph{Nearest neighbor.}
We first present an algorithm for finding the nearest neighbor, i.e., the case $k=1$. For $k>1$, we will iteratively invoke it $k$ times to find the $k$ nearest neighbors.  For generality, we describe our nearest neighbor algorithm as one that tries to find the nearest neighbor among a given subset of $m$ points in $x$.

The starting observation is that the function $g_i(x;p) := \|x_i - p\|$ is $1$-Lipschitz.  Below, we show that shortest distance $h(x;p) := \min_j g_j(x;p)$ is also $1$-Lipschitz. This allows us to first obtain a privatized $h(x;p)$. Then with this privatized shortest distance as $T$, we use SVT to cycle through the points in $x$, using their distances to $p$ as the queries. When the SVT terminates, a point whose distance close to $T$ will be identified with high probability.

    \begin{lemma}
    \label{lm:min_lipschitz}
            Let $g_j : U^n \rightarrow \mathbb{R}$ be $K$-Lipschitz (using $\dist_{\infty}$ on $U^n$) for each $j\in [m]$.  Then $h(x)=\min_{j\in [m]}g_j(x)$ is also $K$-Lipschitz.
    \end{lemma}
    \begin{proof}
        Let $x, x'\in (U)^n$. Let $j^*:=\argmin_{j\in [m]}g_j(x)$ and $l^*:=\argmin_{j\in [m]}g_j(x')$. Then
        \begin{align*}
            h(x)-h(x') &= g_{j^*}(x)-g_{l^*}(x') \leq g_{l^*}(x) - g_{l^*}(x')\\
            &\leq K\|x_{l^*}-x'_{l^*}\| \leq K\cdot\textstyle{\dist_{\infty}}(x,x'),
        \end{align*}
        and
        \begin{align*}
        h(x')-h(x) &= g_{l^*}(x')-g_{j^*}(x)
        \leq g_{j^*}(x')-g_{j^*}(x)\\
            &\leq K\|x'_{j^*}-x_{j^*}\| \leq K\cdot\textstyle{\dist_\infty}(x,x').
        \end{align*}
        Combining the inequalities above, $|h(x)-h(x')|\leq K\cdot\dist_{\infty}(x,x')$.
    \end{proof}

The private nearest neighbor (PNN) algorithm is described in Algorithm~\ref{alg:PNN2}.   In line 2, we assume the output of $\mathrm{SVT}$ is the index corresponding to the position of $\top$. The notation $\bar{l}_m$ denotes the $(l \bmod m)$-th index in $I$, so that it cycles through the $m$ indices in $I$. We use $x_I$ to denote the subset of points indexed by $I$. Since line 1 is $\varepsilon/3$-GP and line 2 is $2\varepsilon/3$-GP, Algorithm~\ref{alg:PNN2} is $\varepsilon$-GP.

\begin{algorithm}
    \caption{Private Nearest Neighbor ($\mathrm{PNN}$)}
        \label{alg:PNN2}
        \begin{flushleft}
        \textbf{Input}: $x\in(\mathbb{R}^2)^n$; $p\in\mathbb{R}^2$; $I\subseteq [n]$, $|I|=m$; $\varepsilon$; (optional) $\gamma_0 \ge 0$ \\
        \textbf{Output}: index of the privatized nearest neighbor in $I$
        \end{flushleft}
        \begin{algorithmic}[1]
        \STATE $T\gets h(x_I;p)+Z+\gamma_0$, $Z\sim \mathrm{Lap}(\frac{3}{\varepsilon})$
        \STATE $t\gets \mathrm{SVT}(x,2\varepsilon/3,T,K=1,g'_1,g'_2,\dots)$, $g'_l :=g_{\bar{l}_m}(\cdot\; ;p)$ \\
        \RETURN $\bar{t}_m$
        \end{algorithmic}
    \end{algorithm}

Now we show that the PNN algorithm has an error of $O({1\over \varepsilon}\log{m\over\beta})$ with probability $1-\beta$, i.e., the returned nearest neighbor is at most this much farther away than the true nearest neighbor. Note that this is optimal up to a logarithmic factor, since the PNN query degenerates into the identity query when $m=1$, which has an error of $O({1\over \varepsilon}\log{1\over\beta})$.

We first prove a looser but simpler result:
\begin{theorem}
\label{thm:pnn2}
    Fix any $0<\beta<1$, and let $\gamma:=\gamma_1+\gamma_2$, where $\gamma_1:=\frac{3}{\varepsilon}\left(\sqrt{2\log\left(\frac{m+1}{\beta}\right)}+\log\left(\frac{m+1}{\beta}\right)\right)$ and $\gamma_2:=\frac{6}{\varepsilon}\log\left(\frac{m+1}{\beta}\right)$. Set $\gamma_0=\gamma$ in Algorithm~\ref{alg:PNN2}, then it outputs a $\bar{t}_m$ such that $g_{\bar{t}_m}(x) \leq h(x)+2\gamma$ with probability at least $1-\beta$.  The algorithm terminates in $O(m)$ time with probability $1-\beta$.
\end{theorem}
\begin{proof}
    Let $W$, $V_1, V_2,\dotsb$ be the sequence of Laplace random variables in the $\mathrm{SVT}$ call. Let $t^*:=\argmin_{j\in[m]}g_j(x)$. When $\mathrm{SVT}$ visits $t^*$, it halts if $V_{t^*}\leq W+Z+\gamma_0$. We have $\Pr\left[|W+Z|\leq \gamma_1\right] \geq 1-\beta/(m+1)$ by Lemma~\ref{lm:ZW_bound}, and $\Pr\left[\max_{j\in [m]} |V_j| \leq \gamma_2\right] \geq 1-m\beta/(m+1)$ by the Laplace tail bound and a union bound, i.e., $V_{t^*}\leq \max_{j\in [m]} |V_j| \leq W+Z+\gamma_0$ with probability $1-\beta$. Consequently, we have simultaneously that (1) $\mathrm{SVT}$ halts the first time it visits $g_{t^*}$ or earlier; (2) $g_t(x) + V_t \leq T + W$, hence
    \begin{align*}
        g_{t}(x)&\leq T + W - V_t = h(x) + Z +\gamma_0 + W- V_t\\
        &\leq h(x) + |Z+W| + |V_t| +\gamma_0\\
        &\leq h(x) + \gamma_1+\gamma_2+\gamma_0 = h(x) + 2\gamma.
    \end{align*}
\end{proof}
 
The optional parameter $\gamma_0$ is used to control the trade-off between the error and running time of the algorithm.  In Theorem \ref{thm:pnn2}, we set $\gamma_0 =\gamma$ so that the SVT terminates within the first cycle of $x$ but the error is doubled.  Below, we give a more refined analysis, showing that setting $\gamma_0=0$ will not increase the asymptotic running time while reducing the error. 

 \begin{theorem}
 \label{thm:pnn_gamma0}
     Fix any $0<\beta<1$, and $0<\beta_1,\beta_2, \beta_3$ such that $\beta_1+\beta_2+\beta_3=\beta$. Let $\gamma:=\gamma_1+\gamma_2$, where $\gamma_1:=\frac{3}{\varepsilon}\left(\sqrt{2\log\left(\frac{1}{\beta_1}\right)}+\log\left(\frac{1}{\beta_1}\right)\right)$ and $\gamma_2:=\frac{6}{\varepsilon}\log\left(\frac{4m}{\beta_2\beta_3}\right)$. Set $\gamma_0=0$ in Algorithm~\ref{alg:PNN2}, then it outputs a $\bar{t}_m$ such that $g_{\bar{t}_m}(x)\leq h(x)+\gamma$ with probability at least $1-\beta$. The algorithm runs in expected $O(m)$ time.
 \end{theorem}

 \begin{proof}
     Let $t^*:=\argmin_{j\in[m]}g_j(x)$ and $t$ be the index (which is a random variable) returned by the $\mathrm{SVT}$ call. Let $k$ be positive integer. Then when $\mathrm{SVT}$ visits $kt^*$: it halts if $V_{kt^*} \leq W+Z$. By Lemma~\ref{lm:V_leq_ZW_count}, the expected number of times $t^*$ is visited is at most $4$, hence the expected running time is $O(m)$. Moreover, by Markov's inequality, with probability at least $1-\beta_2$ the number of times $t^*$ is visited is at most $4/\beta_2$. I.e. $t\leq 4m/\beta_2$. We also have $\Pr\left[\max_{j\leq 4m/\beta_2}|V_j|\leq \gamma_2\right] \leq 1-\beta_3$ and $\Pr[|Z+W|\leq \gamma_1]\geq 1-\beta_1$. Thus, by a union bound, we have with probability at least $1-(\beta_1+\beta_2+\beta_3)$:
     \begin{align*}
         g_{\bar{t}_m}(x) &\leq T+W-V_t = h(x)+Z+W-V_t\\
         &\leq h(x) + |Z+W| + \max_{j\leq 4m/\beta_2}|V_j|\\
         &\leq h(x) + \gamma_1+\gamma_2 = h(x) + \gamma.
     \end{align*}
 \end{proof}

\textbf{Remark}: We can in fact even set $\gamma_0<0$, which would further reduce the error at the cost of a longer running time.  Theoretically, the running time would increase exponentially in $|\gamma_0|$, in the pathological case where all points in $x$ but the nearest neighbor are very far away from $p$.  On most typical instances, however, the running time is still close to linear with a $\gamma_0<0$.

\paragraph{$k$ nearest neighbors.}
Now, to find the $k$ nearest neighbors of $p$, we simply invoke Algorithm \ref{alg:PNN2} $k$ times, as shown in Algorithm \ref{alg:kPNN}.
It follows from Theorem \ref{lm:cgp_comp} that Algorithm~\ref{alg:kPNN} is $\rho$-CGP. We can similarly obtain an $\varepsilon$-GP version of the algorithm by replacing line 3 of the algorithm with $t = \mathrm{PNN}(x,p,{I_{j-1}},\varepsilon/k)$.

    \begin{algorithm}
        \caption{$k$-Private Nearest Neighbors ($k$-$\mathrm{PNN}$)}
            \label{alg:kPNN}
            \begin{flushleft}
            \textbf{Input}: $p\in\mathbb{R}^2$; $x=(x_1,\dotsb,x_n)\in(\mathbb{R}^2)^n$; $k\in \mathbb{Z}_{> 0}$; $\rho$ \\
            \textbf{Output}: $J\subseteq [n]$
            \end{flushleft}
            \begin{algorithmic}[1]
            \STATE $I_0 \gets [n] $, $J_0 \gets \emptyset$
            \FOR{$j = 1,\dotsb, k$}
            \STATE $t \gets \mathrm{PNN}(x,p,I_{j-1},\sqrt{2\rho/k})$
            \STATE $J_j \gets J_{j-1} \cup \{t\}$
            \STATE $I_j \gets I_{j-1} \setminus \{t\}$
            \ENDFOR\\
            \RETURN $J=J_k  $
            \end{algorithmic}
        \end{algorithm}

The error of the $k$-PNN is $O\left(\sqrt{k/\rho} \log(n/\beta)\right)$, as shown in the following theorem.

    \begin{theorem}
    \label{thm:kpnn}
        Fix $1>\beta>0$, let $\gamma :=\frac{15\sqrt{k}}{\sqrt{2\rho}}\log({(4n+2)}/{\beta})+\frac{3\sqrt{k}}{\sqrt{\rho}}\sqrt{\log({(4n+2)}/{\beta})}$. Fix $1\leq j \leq k$, let $t_j\in J$ be the index corresponding to the $j$th nearest neighbor output by Algorithm~\ref{alg:kPNN}. Let $t^*_j$ be the true $j$th nearest neighbor. Then with probability at least $1-\beta$, 
        \[
            \|x_{t_j}-p\| \leq \|x_{t^*_j}-p\| + \gamma.
        \]
    \end{theorem}
    \begin{proof}
By Theorem~\ref{thm:pnn_gamma0}, with $\varepsilon_j=\sqrt{2\rho/k}$, $\beta_1=\beta_3=\beta/(4n+2)$ and $\beta_2=4n\beta/(4n+2)$, we have with probability at least $1-\beta$, $g_{t_j}=\|x_{t_j}-p\|\leq \min_{i\in[n] \setminus J_{j-1}}\|x_i-p\|+\gamma$. It remains to show $\min_{i\in[n] \setminus J_{j-1}}\|x_i-p\|\leq\|x_{t^*_j}-p\|$. Let $J^*_{j-1}=\{t^*_1,\dotsb,t^*_{j-1}\}$ be the set of indices containing the true $j-1$ nearest neighbors. If $J_{j-1}=J^*_{j-1}$ then $\min_{i\in[n] \setminus J_{j-1}}\|x_i-p\|=\|x_{t^*_j}-p\|$; otherwise there is $t^*_l$ where $1\leq l \leq j-1$ which is not in $J_{j-1}$, then $\min_{i\in[n] \setminus J_{j-1}}\|x_i-p\|\leq \|x_{t^*_l}-p\|\leq \|x_{t^*_j}-p\|$.
    \end{proof}

On the other hand, the error of the $\varepsilon$-GP version of the algorithm has an error of $O(k/\varepsilon\cdot \log(n/\beta))$.  The proof is similar and omitted.

\paragraph{Extension to $d$ dimensions.} Both the GP and CGP algorithms for the $k$NN problem extend to $d$ dimensions verbatim, as the function $g_i(x) = \|x_i -p \|_2$ is $1$-Lipschitz in any dimensions, so the noise required does not depend on $d$.  Consequently, the error guarantee remains the same for any $d$. 

\subsection{Convex Hull}
\label{sec:pch}

In the last application, we are interested in releasing a privatized convex hull of a collection of points. Given a tuple $x=(x_1,\dotsb,x_n)\in (\mathbb{R}^2)^n$, let $\mathrm{CONV}(x)$ denote the convex hull of the points in $x$.  As in the $k$NN algorithm, we first privately find a set of $k$ indices $A\subseteq [n]$.  Let $x_A$ be the subset of points indexed by $A$.  Then we release a privatized $x_A$, denoted $\tilde{x}_A$, and compute $\mathrm{CONV}(\tilde{x}_A)$ as a post-processing step.  However, unlike the $k$NN problem where $k$ is given, here $k$ is an internal parameter that controls the balance of two sources of errors: A large $k$ will make $\mathrm{CONV}(x_A)$ close to $\mathrm{CONV}(x)$ but enlarge the gap between $\mathrm{CONV}(x_A)$ and $\mathrm{CONV}(\tilde{x}_A)$ as the noise injected in $\tilde{x}_A$ is $\tilde{O}(\sqrt{k})$ by Theorem \ref{thm:basic_cgp_maxerr}.  These two sources of errors must be quantified in order to find an optimal $k$.  

\paragraph{Finding $A$ privately.}
We first consider the problem of, for a given $k$, how to privately find a subset of $k$ points, indexed by $A$, so that $\mathrm{CONV}(x_A)$ is as close to $\mathrm{CONV}(x)$ as possible.  The intuition is that we want to include in $A$ those points that are near the boundary of $\mathrm{CONV}(x)$.  Meanwhile, the points selected to $A$ should not be too close to each other, so as to maximize their coverage. Thus, the idea is to find ``equally'' spaced points near the boundary of $\mathrm{CONV}(x)$.  We do so in three steps: 
\begin{enumerate}
    \item Let $c(x)$ be the center of $x$, whose coordinates are the midpoints between the smallest and largest coordinates in each dimension.  In Appendix~\ref{sec:center_lipschitz_proof} we show that $c(\cdot)$ is $\sqrt{2}$-Lipschitz, so we can invoke the Gaussian mechanism to obtain a privatized $\tilde{c}$.
    \item Next, we identify a privatized and large enough radius $\tilde{R}$ so that a circle of radius $\tilde{R}$ around $\tilde{c}$ encloses $x$.  This is done by applying the Gaussian mechanism on the function $g(x) := \max_{i\in [n]} \|x_i-\tilde{c}\|=-\min_{i\in [n]} (-\|x_i-\tilde{c}\|)$, which is $1$-Lipschitz by Lemmas~\ref{lm:lipschitz_properties} and \ref{lm:min_lipschitz}.  Then we enlarge it by $\tilde{O}(1)$ so that it encloses all points in $x$ with high probability.
    \item Finally, we place $k$ points equally spaced on the boundary of the circle, and find the nearest neighbor in $x$ to each of the $k$ points, using the $\mathrm{PNN}$ algorithm from the previous section.
\end{enumerate}

The detailed private convex hull (PCH) algorithm is given in Algorithm \ref{alg:PCH}.  We use a privacy budget of ${2\over 3}\rho_0$ for line 2, and  ${1\over 3}\rho_0$ for line 3.  The $k$ calls of $\mathrm{PNN}$ at line 8 consume a total of $(\rho-\rho_0)$ privacy budget.  So the whole algorithm satisfies $\rho$-CGP.

\begin{algorithm}
    \caption{Private Convex Hull (PCH)}
        \label{alg:PCH}
        \begin{flushleft}
        \textbf{Input}: $x=(x_1,\dotsb,x_n)\in(\mathbb{R}^2)^n$; $k$; $\rho$; $\beta$ \\
        \textbf{Output}: $A\subseteq [n]$
        \end{flushleft}
        \begin{algorithmic}[1]
        \STATE $\rho_0 \gets \rho/20, \rho_1 \gets (\rho-\rho_0)/k$
        \STATE $\tilde{c} \gets c(x) + Z_c$, $Z_c\sim \mathcal{N}(0,\frac{3}{2\rho_0}I_{2\times 2})$ 
        \STATE $\tilde{R} \gets \max_i(\|x_i-\tilde{c}\|) + \sqrt{3\log(2/\beta)/\rho_0} + Z_{R}$, $Z_R\sim \mathcal{N}(0,\frac{3}{2\rho_0})$
        \STATE $A \gets \emptyset$;
        \FOR{$j = 1,\dotsb, {k}$}
            \STATE $\theta_j \gets \frac{2\pi(j-1)}{k}$
            \STATE $P_j \gets \tilde{c}+[\tilde{R}\cos(\theta_j), \tilde{R}\sin(\theta_j)]^T$
            \STATE $a_j \gets \mathrm{PNN}(x, P_j, \{1,\dotsb,n\}, \sqrt{2\rho_1})$
            \STATE $A \gets A \cup \{a_{j}\}$
        \ENDFOR\\
        \RETURN $A$
        \end{algorithmic}
    \end{algorithm}

To bound the difference between $\mathrm{CONV}(x_A)$ and $\mathrm{CONV}(x)$, we will show that a small expansion of $\mathrm{CONV}(x_A)$ will enclose $\mathrm{CONV}(x)$.  This is formalized using the Minkowski sum. Recall for two sets of vectors $V, V'\subseteq \mathbb{R}^d$, their Minkowski sum is: $V+V':=\{v+v': v\in V, v'\in V'\}$, while $V-V'$ is defined such that $(V-V')+V'=V$. Let $\mathcal{B}_r$ denote the ball of radius $r$ centered at the origin, and let $\omega(x):=\max_{i,j}\|x_i-x_j\|$ be the diameter of $x$.
    \begin{lemma}
        \label{thm:PCH_error}
            Let $A=\mathrm{PCH}(x, k, \rho, \beta)$ be the output of Algorithm~\ref{alg:PCH} and $x_A:=\{x_{a_j}\}_{a_j\in A}$. Then with probability at least $1-\beta$, \[\mathrm{CONV}(x_A)\subseteq \mathrm{CONV}(x)\subseteq \mathrm{CONV}(x_A)+\mathcal{B}_{\gamma},\] where $\gamma={O}\left(\frac{\sqrt{k}\log(n/\beta)}{\sqrt{\rho}}+\frac{\omega(x)}{k}\right)$.
        \end{lemma}

{We first show that, given a large enough circle which encloses all points in $x$, the set of nearest neighbors in $x$ of $k$ equally spaced points on the boundary of the circle provides a good approximation to $\mathrm{CONV}(x)$. Let $V^*$ denote the indices of the set of vertices of $\mathrm{CONV}(x)$, and let $p:[0,2\pi]\rightarrow \mathbb{R}^2$ be a function defined by $p(\theta;r,o):=o+[r\cos(\theta),r\sin(\theta)]^T$.

\begin{figure*}[h]
      \centering
            \includegraphics[width=0.6\textwidth]{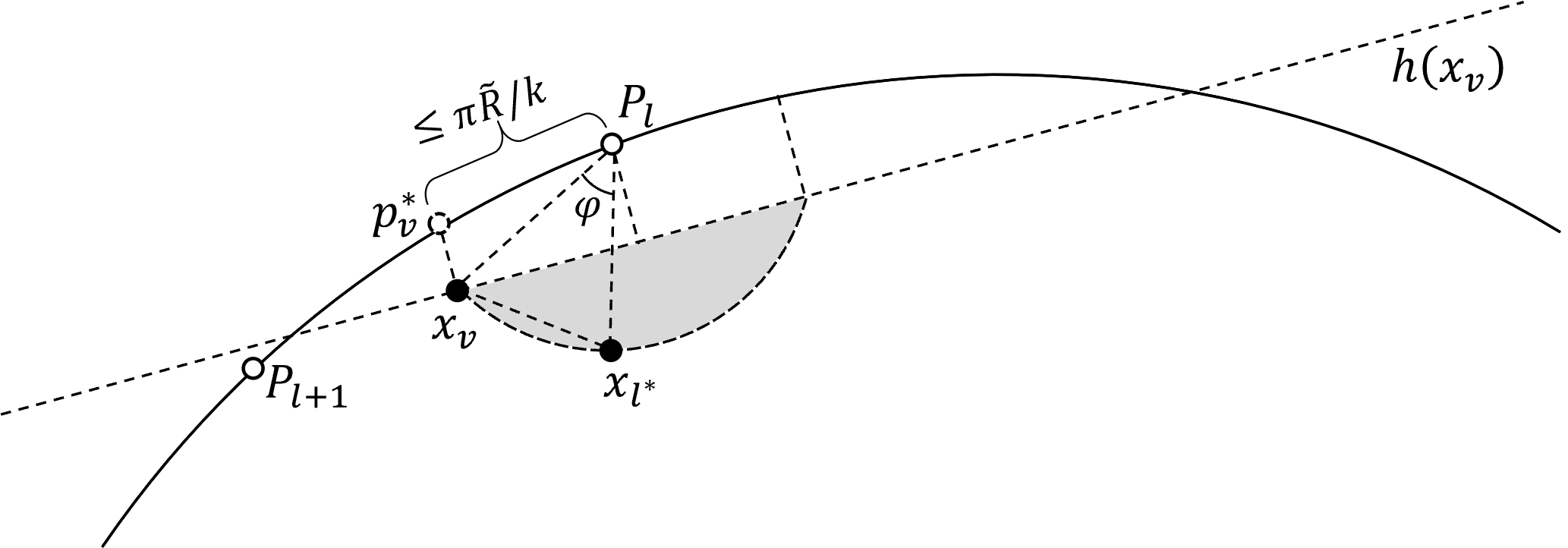}
         \caption{Illustration of the position of $x_{l^*}$ in Lemma~\ref{lm:pch_nonpriv_error}.}
         \label{fig:approx_dist}
    \end{figure*}
    
\begin{lemma}
\label{lm:pch_nonpriv_error}
Let $C=(o,R)$ be a circle of radius $R$ centered at $o$ which encloses all points in $x$. Let $k$ be a positive integer, and let $P_1,\dotsb,P_j$ be $k$ equally spaced points on the boundary of $C$, where $P_j:=p(2\pi(j-1)/k;R,o)$ for $j\in[k]$. Let $V_k:=\{\argmin_{i\in [n]}\|x_i-P_j\|: j\in [k]\}$ be the set of nearest neighbors to the $k$ points. Then for any $v\in V^*\setminus V_k$, there is $u\in V_k$ such that $x_u$ is at a distance of at most $\gamma_2:={2\pi \over k} R$ from $x_v$.
\end{lemma}
}
\begin{proof}
Let $v\in V^*\setminus V_k$. Recall a point $x_i$ in $x$ is a vertex of $\mathrm{CONV}(x)$ if there is a hyperplane (a line in $\mathbb{R}^2$) that separates $x_j$ from the rest of $x$; or equivalently, there is a hyperplane that just touches $x_i$ such that all of $x$ are on the same side of the hyperplane. Let $h(x_v)$ denote such a hyperplane for $x_v$. By construction, there are no points between $h(x_v)$ and the boundary of ${C}$; if there were such a point, it would be crossed by $h(x_v)$ before $x_v$, contradicting the definition of $h(x_v)$. Consider a line perpendicular to $h(x_v)$ extended from $x_v$ until it meets the boundary of ${C}$, let $p^*_v=p(\theta^*_v;{R},{c})$ denote this point of intersection, for some $\theta^*_v\in [0,2\pi]$. Then $p^*_v$ is on the boundary between the points $P_l$ and $P_{l+1}$ for some $l\in [k]$. Without loss of generality, assume $p^*_v$ lies closer to $P_l$ on the boundary. Then $\|P_l-p^*_v\|\leq \frac{1}{2}\cdot\frac{2\pi\tilde{R}}{k}$. Let ${l^*}:=\argmin_{i\in [n]}\|P_l-x_i\|$. Next, we will show that $x_{l^*}$ is at most a distance of $\gamma_2$ from $x_v$, and since $x_{l^*}\in V_k$, it is the $x_u$ desired in the lemma.

To this end, consider the triangle with vertices $(x_v,P_l,x_{l^*})$. Let $a:=\|P_l-x_v\|$, $b:=\|P_l-x_{l^*}\|$ and $c:=\|x_v-x_{l^*}\|$. Let $\varphi\in [0,\pi]$ be the angle between the edges $\overline{x_vP_l}$ and $\overline{x_{l^*}P_l}$. By the law of cosines, $c^2=a^2+b^2-2ab\cos(\varphi)$, which is maximized when $\cos(\varphi)$ is minimized and $b$ is maximized. Since $\cos(\varphi)$ is strictly decreasing for $\varphi\in [0,\pi]$, $c$ is maximized when $\varphi$ is as large as possible without crossing $h(x_v)$. Also, $b=\|P_l-x_{l^*}\|=\min_{i \in [n]}\|P_l-x_i\|\leq \|P_l-x_v\|=a$. Thus, $c$ is at most equal to the distance between $x_v$ and the other point where a circle of radius $a$ centered at $P_l$ intersects $h(x_v)$, which can be obtained by mirroring $x_v$ about a line that is perpendicular to $h(x_v)$ and crosses $P_l$ (see Fig.~\ref{fig:approx_dist} for an illustration). Thus $c$ is at most $2\cdot\|P_l-p^*_v\|\leq\frac{2\pi}{k}{R}=\gamma_2$.
\end{proof}
{
Now, to prove Lemma~\ref{thm:PCH_error}, it remains to show that the set of indices $A$ returned by Algorithm~\ref{alg:PCH} is a good approximation to $V_k$.}
\begin{proof} [Proof of Lemma~\ref{thm:PCH_error}]
By the Gaussian tail bound, $\max_i{\|x_i-\tilde{c}\|} \le \tilde{R} \le \omega(x) + O\left(\sqrt{\log(1/\beta)/\rho}\right)$ with probability at least $1-\beta/2$. Then, a circle $\tilde{C}$ centered at $\tilde{c}$ with radius $\tilde{R}$ encloses all of the points in $x$. The analysis below is conditioned upon this happening. By Lemma~\ref{lm:pch_nonpriv_error}, with $P_j:=p(2\pi(j-1)/k;\tilde{R},\tilde{c})$ for $j\in[k]$ and $V_k:=\{\argmin_{i\in [n]}\|x_i-P_j\|: j\in [k]\}$, then for each $v\in V_k\setminus V^*$ we can find $u\in V_k$ such that $x_u$ is at a distance of at most $\gamma_2={2\pi \over k} \tilde{R}$ from $x_v$.  Let $\gamma_1:=\frac{15\log(2(4n+2)k/\beta)+3\sqrt{2\log(2(4n+2)k/\beta)}}{\sqrt{2 \rho_1}}={O}(\frac{\sqrt{k}\log(n/\beta)}{\sqrt{\rho}})$. Note that $\max_{j\in [k]}\|x_{a_j}-x_{j^*}\|\leq \gamma_1$ with probability at least $1-\beta/2$, which follows from Theorem~\ref{thm:kpnn} and a union bound, where $j^*:=\argmin_{i\in [n]}\|P_j-x_i\|$ corresponds to the true nearest neighbor of $P_j$. Thus, for each $v\in V^*$, there is $a_u\in A$ such that $\|x_{a_u}-x_v\|\leq \|x_{a_u}-x_{u}\|+\|x_{u}-x_v\|\leq \gamma_1+\gamma_2={O}\left(\frac{\sqrt{k}\log(n/\beta)}{\sqrt{\rho}}+\frac{\omega(x)}{k}\right)$. 
\end{proof}

\paragraph{Finding the convex hull privately.}
We first run Algorithm~\ref{alg:PCH} with, say, half of the privacy budget.  Then, we use the other half of budget to privatize the points $x_A$ using the algorithm from Section \ref{sec:identity}, denoted $\tilde{x}_A$.  More precisely, for each $a\in A$, we release $\tilde{x}_a:=x_a+\frac{1}{\sqrt{\rho/k}} \cdot \mathcal{N}(0,I_{2\times 2})$.
Finally, we return $\mathrm{CONV}(\tilde{x}_A)$.

We can also bound the error of $\mathrm{CONV}(\tilde{x}_A)$ in terms of the Minkowski sum:

\begin{lemma}
    \label{corr:PCH_error2}
With probability at least $1-\beta$, 
    \[\mathrm{CONV}(\tilde{x}_A)-\mathcal{B}_{\gamma_3} \subseteq \mathrm{CONV}(x)\subseteq \mathrm{CONV}(\tilde{x}_A)+\mathcal{B}_{\gamma},\] where $\gamma={O}\left(\frac{\sqrt{k}\log(n/\beta)}{\sqrt{\rho}}+\frac{\omega(x)}{k}\right)$ and $\gamma_3={O}\left(\frac{\sqrt{k\log(k/\beta)}}{\sqrt{\rho}}\right)$.
\end{lemma}
\begin{proof}
Let $\gamma_1:=\frac{15\log(4(4n+2)k/\beta)+3\sqrt{2\log(4(4n+2)k/\beta)}}{\sqrt{\rho_1}}$ and $\gamma_2:=2\pi \tilde{R}/k$. Let $x_A\subseteq x$ be the collection of "anchor" points corresponding to the indices in $A$. Following the arguments in the proof of Theorem~\ref{thm:PCH_error}, with probability $1-\beta/2$, $\mathrm{CONV}(x_A)\subseteq \mathrm{CONV}(x)\subseteq \mathrm{CONV}(x_A)+\mathcal{B}_{(\gamma_1+\gamma_2)}$. Let $\gamma_3:=\frac{\sqrt{2k\log(2k/\beta)}}{\sqrt{\rho}}$, then $\max_{j\in[k]}\|\tilde{x}_{a_j}-x_{a_j}\|\leq \gamma_3$ with probability at least $1-\beta/2$. Thus, \[\mathrm{CONV}(\tilde{x}_A)-\mathcal{B}_{\gamma_3}\subseteq \mathrm{CONV}(x_A) \subseteq \mathrm{CONV}(\tilde{x}_A)+\mathcal{B}_{\gamma_3}.\] By a union bound, together we have with probability $1-\beta$,
\[\mathrm{CONV}(\tilde{x}_A)-\mathcal{B}_{\gamma_3} \subseteq \mathrm{CONV}(x_A) \subseteq \mathrm{CONV}(x)\subseteq \mathrm{CONV}(\tilde{x}_A)+\mathcal{B}_{\gamma},\]
where $\gamma = \gamma_1+\gamma_2+\gamma_3$.
\end{proof}

\paragraph{Setting $k$.}
To minimize the error, we can choose $k$ so as to balance the two terms $\frac{\sqrt{k}\log(n/\beta)}{\sqrt{\rho}}$ and $\frac{\omega(x)}{k}$, yielding $k=\left( {\omega(x) \sqrt{\rho} \over \log(n/\beta)}\right)^{2/3}$.  One caveat is that we cannot use $\omega(x)$ directly as it is private information.  So we replace it with $\tilde{R}$, which is  larger than $\omega(x)$ by an additive $O\left(\sqrt{\log(1/\beta)/\rho}\right)$ term.  Plugging this value of $k$ into Lemma \ref{corr:PCH_error2}, we obtain:

\begin{theorem}
The private convex hull algorithm above satisfies $\rho$-CGP. It returns an $\tilde{x}_A$ such that with probability at least $1-\beta$,
     \[\mathrm{CONV}(\tilde{x}_A)-\mathcal{B}_{\gamma} \subseteq \mathrm{CONV}(x)\subseteq \mathrm{CONV}(\tilde{x}_A)+\mathcal{B}_{\gamma},\]
     where $\gamma = O\left(\left(\omega(x) + \sqrt{\log(1/\beta)/\rho}\right)^{1/3} \log^{2/3}(n/\beta)\over \rho^{1/3} \right)  = 
  \tilde{O}\left(\omega(x)^{1/3} + 1\right)$.
\end{theorem}

Note that the extra $\tilde{O}(1)$ term, resulting from the technical replacement of $\omega(x)$ with $\tilde{R}$, is actually unavoidable: In the extreme case where $x$ consists of just one point, we have $\omega(x)=0$ while the convex hull problem degenerates into the identity query, which must have $\tilde{O}(1)$ error.

The algorithm can also be made to satisfy $\varepsilon$-GP.  Similar to the $k$NN problem, the $\varepsilon$-GP version of the algorithm can only use basic composition to allocate the privacy to the $k$ PNN queries in Algorithm \ref{alg:PCH}, as well as for privatizing the $k$ points in $x_A$.  This leads to an overall error of  $O\left(\frac{k\log(n/\beta)}{\varepsilon}+\frac{\omega(x)}{k}\right)$, which is 
$\tilde{O}\left(\sqrt{\omega(x)} + 1\right)$ after balancing these two terms similarly as above.

{\bf Remark 1:}  If $\omega(x) > \tilde{\Omega}(n^{3/2})$, i.e., the point set is very sparse (relative to the unit distance), then the optimal value of $k$ would be greater than $n$.  In this case, the algorithm essentially degenerates into the naive algorithm that privatizes all $n$ points and then computes the convex hull.  Then the $O(\omega(x)/k)$ terms goes away, and the error just becomes $\tilde{O}(\sqrt{n})$, same as that for the identity query.  For the GP version, the error is then $\tilde{O}(n)$.

{\bf Remark 2:} In the analysis above, we simply balanced the two error terms for finding a good $k$.  This gets the asymptotic result right.  In our implementation, we derive the precise constant coefficients in the error expression of Lemma \ref{corr:PCH_error2}, and minimize it by setting its derivative to $0$ to find an optimal $k$.  

\paragraph{Extending to $d$ dimensions.}  
Algorithm~\ref{alg:PCH} can be generalized to work in $d$ dimensions as follows: (1) find a privatized center $\tilde{c}$, by computing the midpoint of each dimension, which is $\sqrt{d}$-Lipschitz (alternatively, the $d$-dimensional mean can be used, and computing the mean is $1$-Lipshitz); (2) find a privatized and large enough radius $\tilde{R}$, by computing $\max_{i\in [n]}\|x_i-\tilde{c}\|$, which is $1$-Lipschitz; (3) place $k$ ``evenly spaced'' points on the hypersphere of radius $\tilde{R}$ centered at $\tilde{c}$ and find their private nearest neighbors; and (4) release the $k$ nearest neighbors by adding Gaussian noise of scale $\tilde{O}(\sqrt{dk})$ and compute their convex hull. 
Omitting the details, we can show that the total error becomes $\tilde{O}\left(\sqrt{dk} + \omega(x)/k^{1/(d-1)}\right)$, which minimizes to $\tilde{O}\left(\omega(x)^{d-1 \over d+1} d^{1 \over d+1} + 1\right)$.  The error of the GP version is $\tilde{O}\left(\omega(x)^{d-1 \over d} d^{1 \over d} + 1\right)$.

\section{Experiments}

We perform experiments on a dataset containing mobility traces of taxi cabs 
\cite{epfl-mobility-20090224}. The dataset contains approximately $500$ trajectories which were collected over 30 days in the Bay Area of San Francisco. Most ($> 95\%$) of these trajectories contain $5000-30000$ points with an average of $20000$ points. Each trajectory corresponds to one collection of points in $(\mathbb{R}^2)^n$.
We convert the GPS coordinates into $\mathbb{R}^2$ coordinates via the Mercator projection, with meters ($\mathrm{m}$) being the units of distance.  As argued in \cite{andres2013geo,chatzikokolakis2013broadening}, the choice of the distance unit is unimportant as all it matters is $\varepsilon \cdot \dist(x,x')$ ($\rho \cdot \dist(x,x')^2$ for CGP). If for example kilometers are used, 
we can just scale up $\varepsilon$ by a factor of $10^3$ (resp. scale up $\rho$ by a factor of $10^6$). {Also, mechanisms employing different distance units can be composed after converting to the same distance unit.} The values of $\rho$ used in these experiments are in the range $[5\cdot 10^{-8},5\cdot 10^{-4}]$, corresponding to $\varepsilon$ values in the range of approximately $0.002$ (i.e. $2$ per $1000$m) to $0.2$ (i.e. $2$ per $10$m). We also include the same set of experiments for larger privacy budgets ($\rho$ ranging from $10^{-4}$ to $1$) in Appendix~\ref{sec:experiments_lp}. The code for all the experiments can be found at:  \url{https://github.com/hkustDB/ConcentratedGeoPrivacy}.
 
\begin{figure}[htbp]
     \centering
         \begin{subfigure}[t]{0.23\textwidth}
            \centering
            \includegraphics[width=\textwidth]{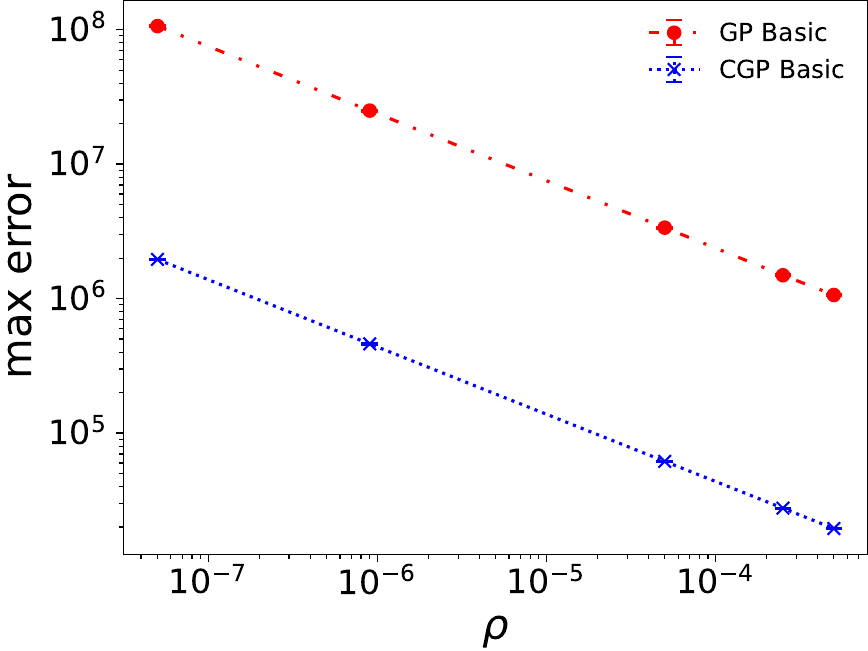}
             \vskip -.08in
            \subcaption{$n=20000$.}
            \;
         \end{subfigure}
        \hfill
         \begin{subfigure}[t]{0.23\textwidth}
            \centering
            \includegraphics[width=\textwidth]{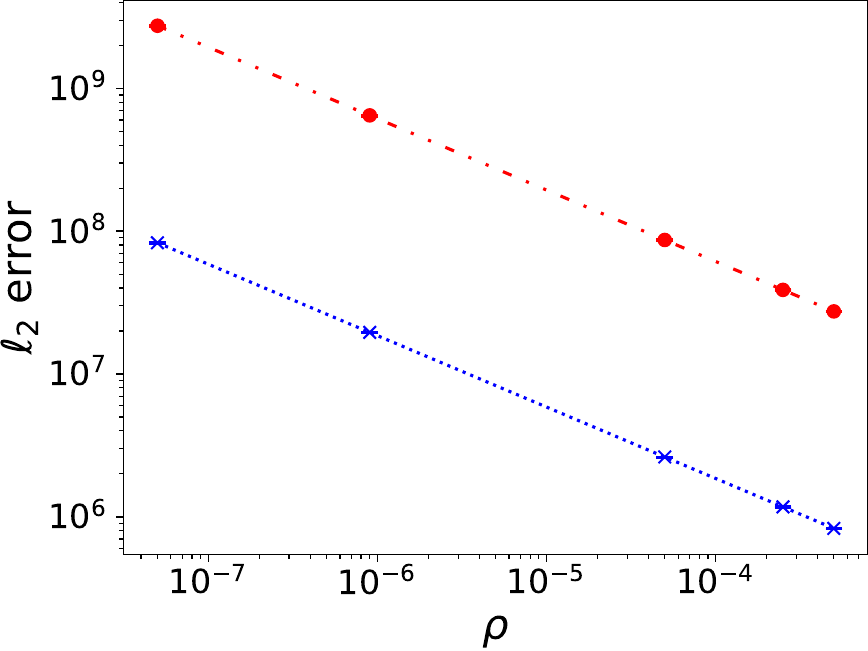}
            \vskip -.08in
            \subcaption{$n=20000$.}
            \;
         \end{subfigure}
         \hfill
         \begin{subfigure}[t]{0.23\textwidth}
            \centering
            \includegraphics[width=\textwidth]{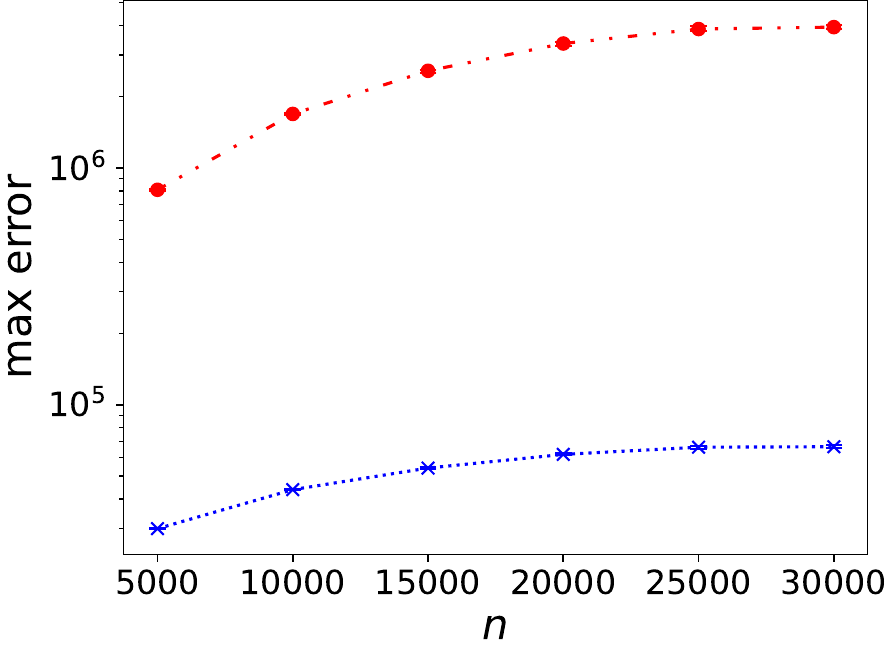}
            \vskip -.08in
            \subcaption{$\rho = 0.00005$.}
         \end{subfigure}
         \hfill
         \begin{subfigure}[t]{0.23\textwidth}
            \centering
            \includegraphics[width=\textwidth]{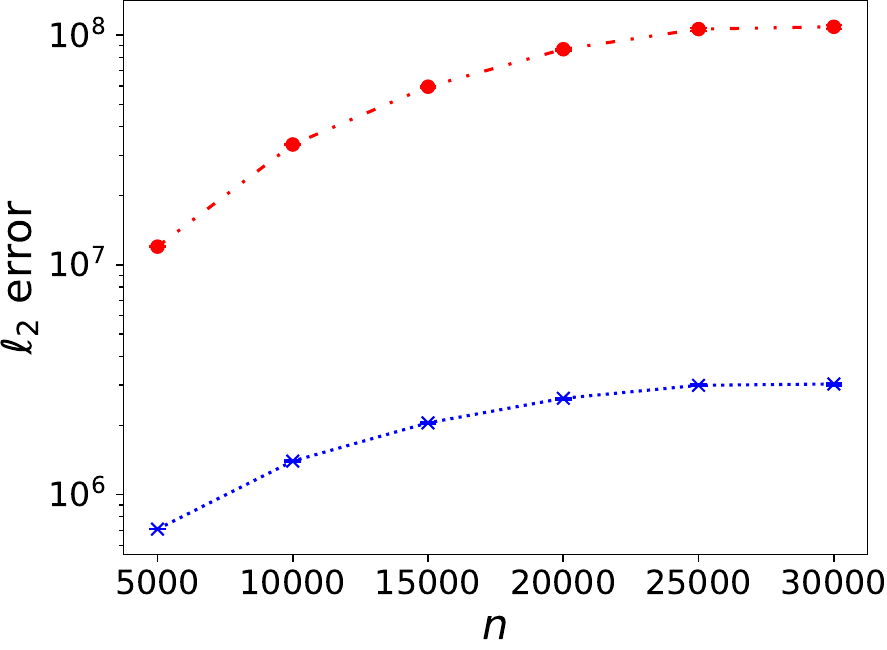}
            \vskip -.08in
            \subcaption{$\rho = 0.00005$.}
         \end{subfigure}
          \vskip -.2in
         \caption{Releasing a collection of points : error with respect to privacy level (top) and to tuple size (bottom).}
         \label{fig:trajerr}
    \end{figure}

We examine the error with respect to various parameter settings. The parameters of interest are the number of input points ($n$), privacy level ($\rho$) and the number of neighbors ($k$). When we compare $\rho$-CGP algorithms with $\varepsilon$-GP algorithms, we advantage the latter with a larger privacy budget using the relationship described in Lemma~\ref{lm:cgp_dDGP} with $\delta=10^{-10}$ and $\varepsilon \Delta\ge 10$. Following the discussion after Lemma~\ref{lm:cgp_dDGP}, the differences in the privacy guarantees between GP and CGP are thus negligible. 
In each experiment, we randomly select $50$ collections of points and we sample $n$ points from each collection to be used as inputs to the algorithms. If a collection has less than $n$ points, all of its points will be used. Each experiment is repeated $25$ times and we report the mean, the $25$th and $75$th percentiles. We group the experiments by application as follows:

\paragraph{The identity query (Fig.~\ref{fig:trajerr}).} We evaluate the algorithms which privatize each location with $(\varepsilon/n)$-GP ($\mathrm{GP\;Basic}$) and $(\rho/n)$-CGP ($\mathrm{CGP\;Basic}$), respectively. We report both the max error on a single point and the $\ell_2$ error across the entire tuple.
    
Both algorithms have smaller error as the privacy budget $\rho$ increases and have larger error as the number of points $n$ increases.  When $n$ is held fixed, the error of $\mathrm{GP\;Basic}$ is close to $\sqrt{n}$ times that of $\mathrm{CGP\;Basic}$, as can be seen in Fig. \ref{fig:trajerr}.  This is consistent with our utility analysis in Section \ref{sec:identity}.

\begin{figure}[htbp]
     \centering
         \begin{subfigure}[t]{0.23\textwidth}
            \centering
            \includegraphics[width=\textwidth]{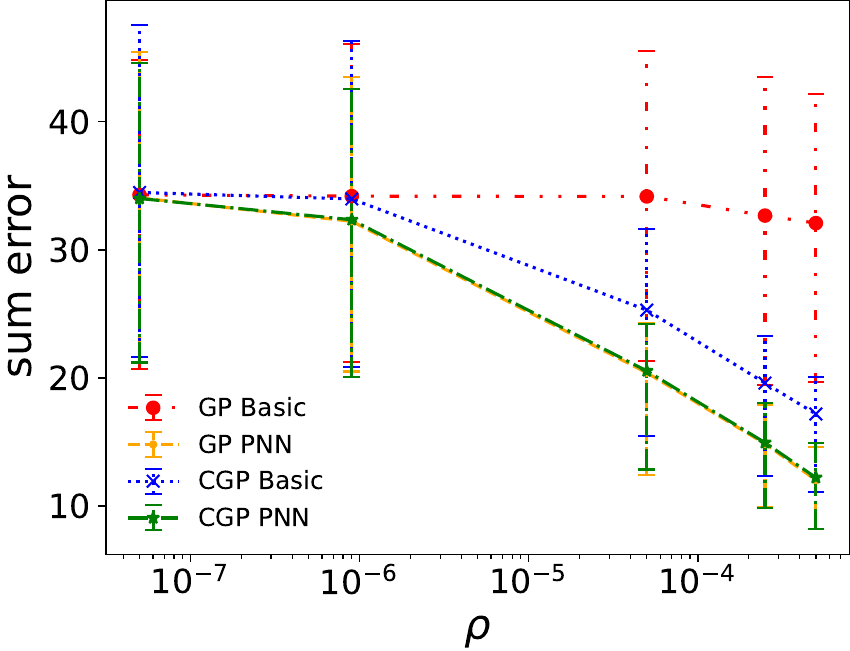}
             \vskip -.08in
            \subcaption{$k=50$.}
            \;
         \end{subfigure}
        \hfill
         \begin{subfigure}[t]{0.23\textwidth}
            \centering
            \includegraphics[width=\textwidth]{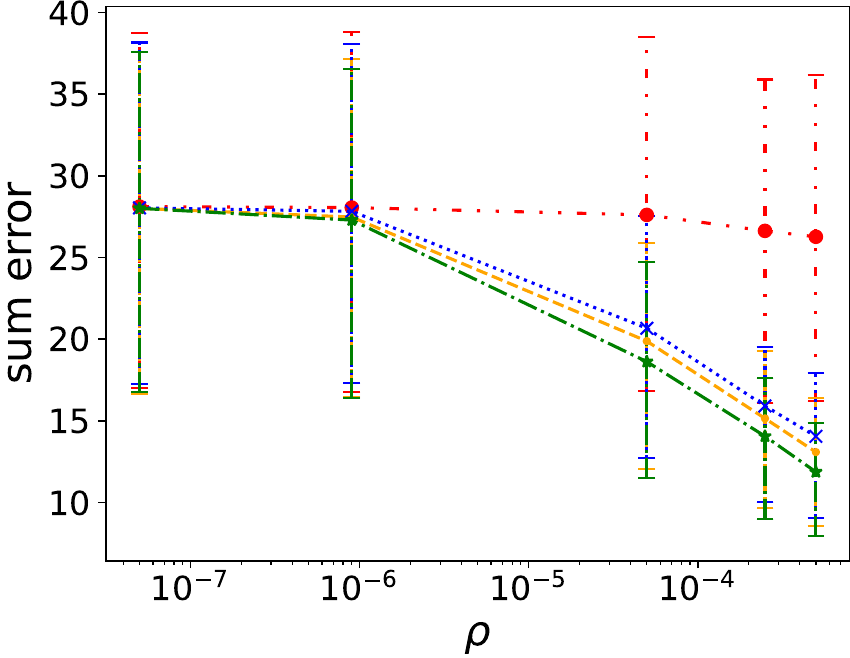}
             \vskip -.08in
            \subcaption{$k=75$.}
            \;
         \end{subfigure}
         \hfill
         \begin{subfigure}[t]{0.23\textwidth}
            \centering
            \includegraphics[width=\textwidth]{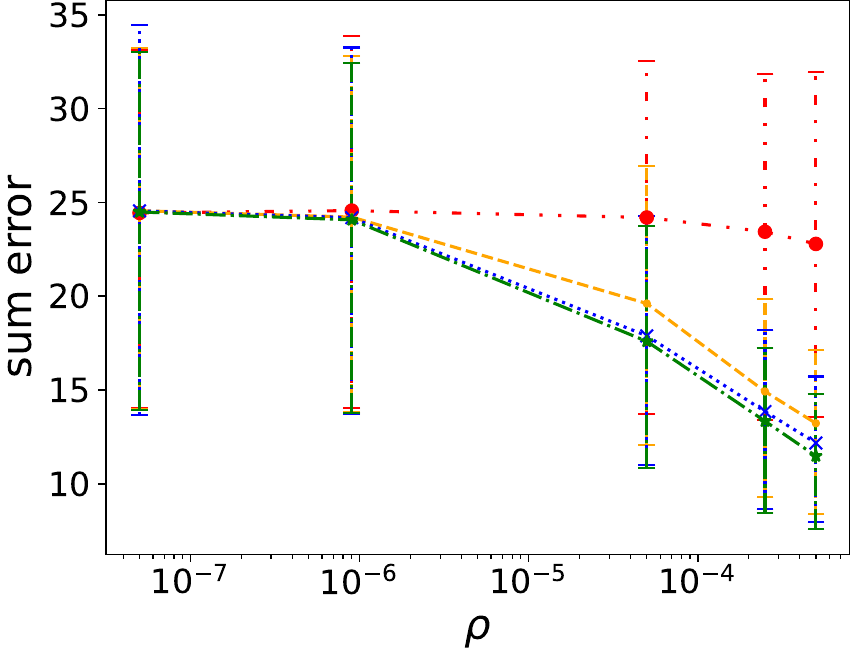}
             \vskip -.08in
            \subcaption{$k=100$.}
         \end{subfigure}
         \hfill
         \begin{subfigure}[t]{0.23\textwidth}
            \centering
            \includegraphics[width=\textwidth]{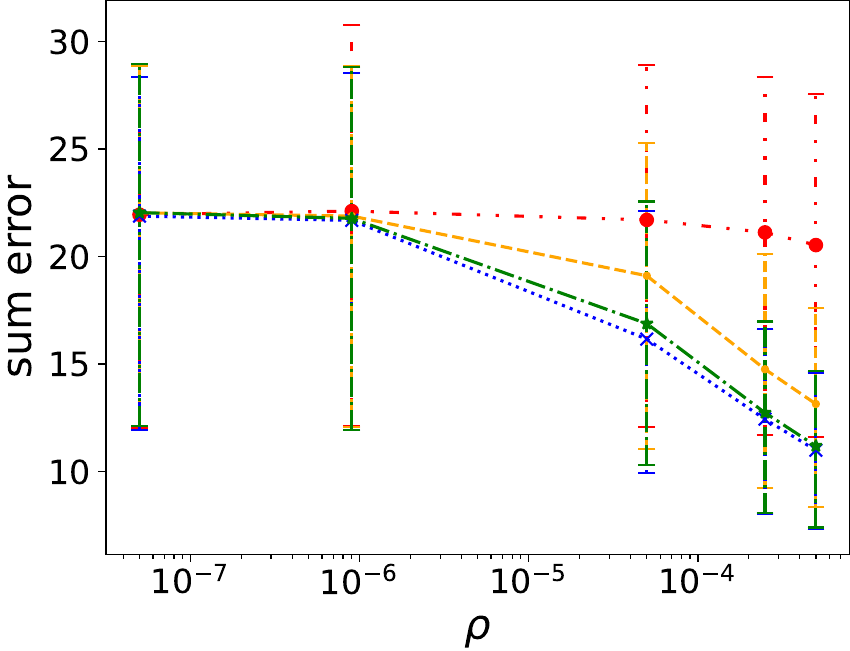}
            \vskip -.08in
            \subcaption{$k=125$.}
         \end{subfigure}
          \vskip -.2in
         \caption{Nearest neighbor: error with respect to privacy level, fixing $n=20000$.}
         \label{fig:sumerr_rho}
    \end{figure}

\begin{figure}[htbp]
     \centering
         \begin{subfigure}[t]{0.23\textwidth}
            \centering
            \includegraphics[width=\textwidth]{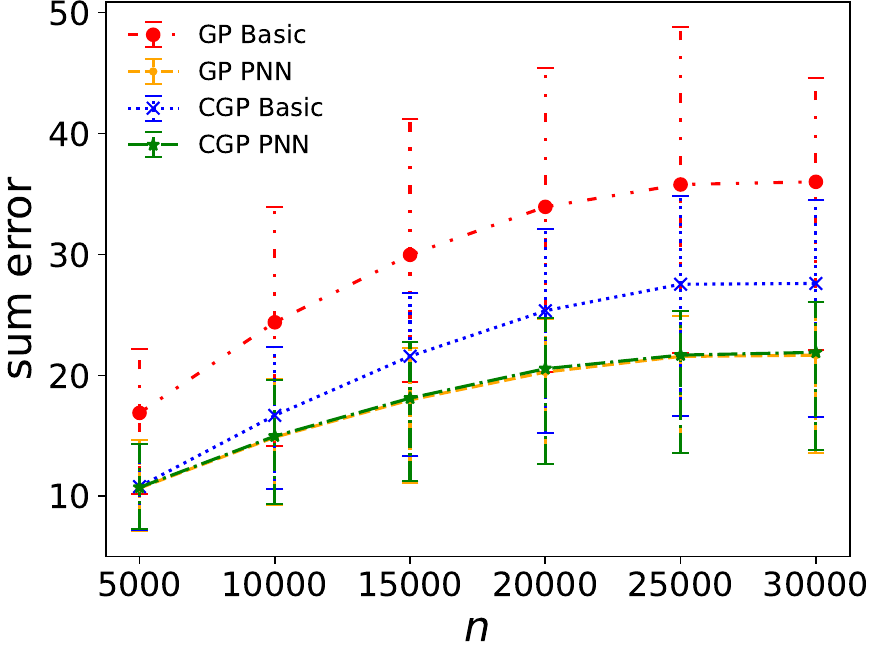}
             \vskip -.08in
            \subcaption{$k=50$.}
            \;
         \end{subfigure}
        \hfill
         \begin{subfigure}[t]{0.23\textwidth}
            \centering
            \includegraphics[width=\textwidth]{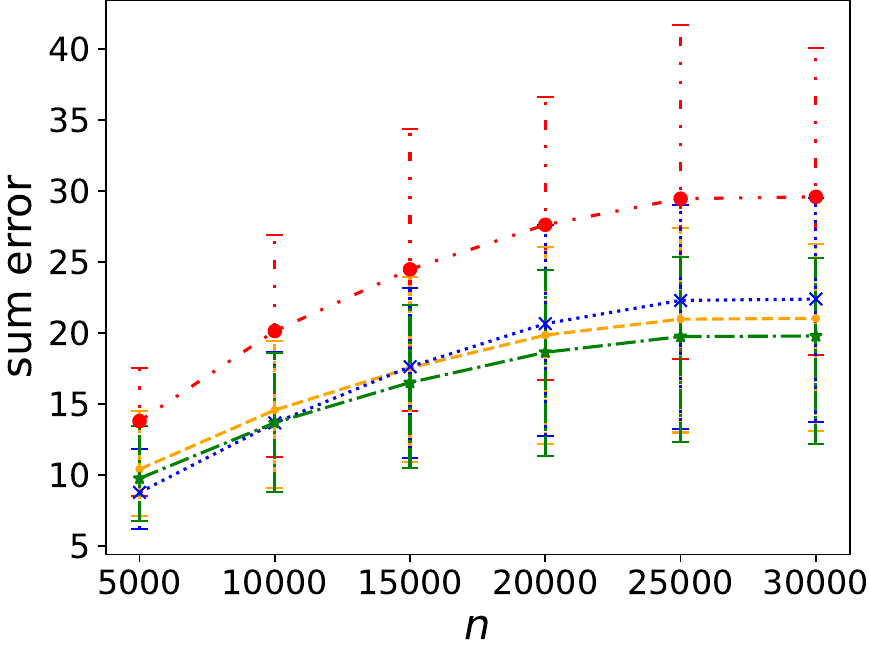}
            \vskip -.08in
            \subcaption{$k=75$.}
            \;
         \end{subfigure}
         \hfill
         \begin{subfigure}[t]{0.23\textwidth}
            \centering
            \includegraphics[width=\textwidth]{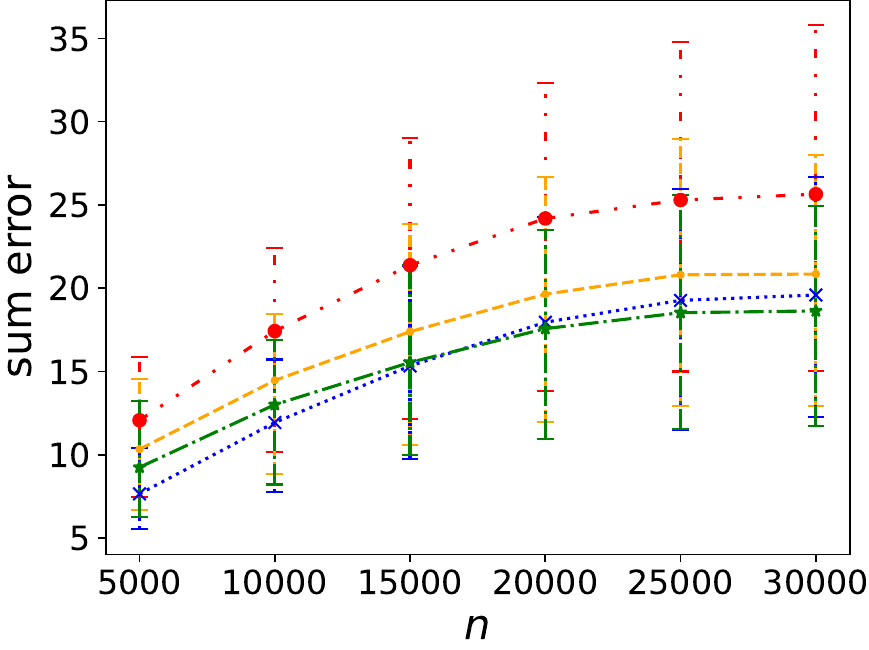}
            \vskip -.08in
            \subcaption{$k=100$.}
         \end{subfigure}
         \hfill
         \begin{subfigure}[t]{0.23\textwidth}
            \centering
            \includegraphics[width=\textwidth]{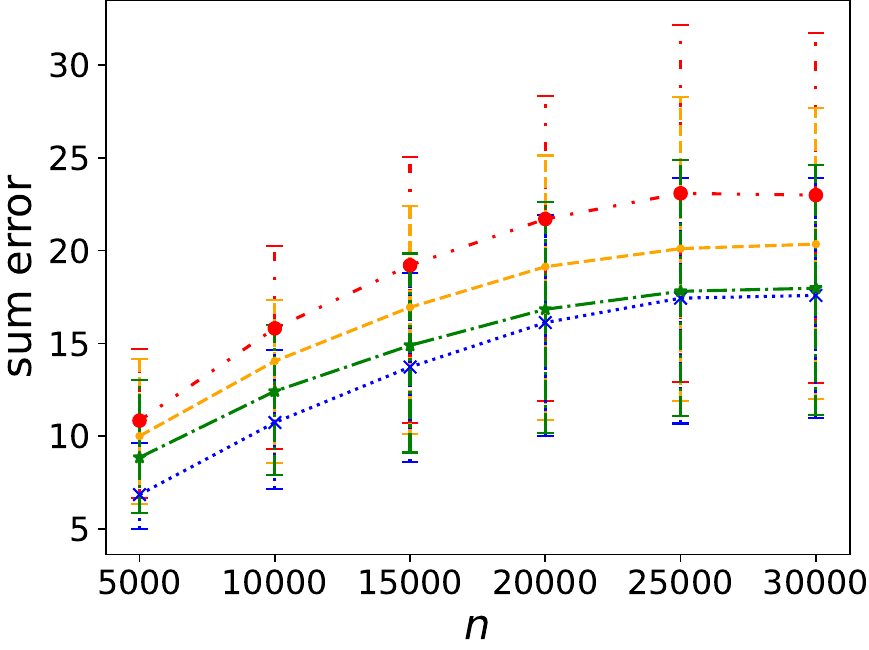}
            \vskip -.08in
            \subcaption{$k=125$.}
         \end{subfigure}
          \vskip -.2in
         \caption{Nearest neighbor: error with respect to tuple size, fixing $\rho=0.00005$.}
         \label{fig:sumerr_m}
    \end{figure}

 \paragraph{$k$ nearest neighbors (Fig.~\ref{fig:sumerr_rho}, \ref{fig:sumerr_m}).} We compare Algorithm~\ref{alg:kPNN} ($\mathrm{CGP\;PNN}$) and its $\varepsilon$-GP variant ($\mathrm{GP\; PNN}$) against the baseline algorithms $\mathrm{GP\;Basic}$ and $\mathrm{CGP\;Basic}$. In each experiment, a query point $p$ is generated uniformly at random from a list of coordinates composed of the centers of all $1\mathrm{m}\times 1\mathrm{m}$ squares that are crossed by some trajectory. The error is computed as the sum of distances of the output neighbors (to the query point), normalized by that of the true (non-private) nearest neighbors.

Across all of the plots in Fig.~\ref{fig:sumerr_rho} and \ref{fig:sumerr_m}, we see that for small $k$, the $\mathrm{PNN}$ algorithms perform better, consistent with our analysis. As $k$ increases, $\mathrm{CGP\;Basic}$ starts to outperform $\mathrm{GP\; PNN}$ and eventually $\mathrm{CGP\; PNN}$, since their errors scale with $k$ and $\sqrt{k}$, respectively, and logarithmically in $n$.  On the other hand, $\mathrm{CGP\;Basic}$ scales with $\sqrt{n}$ and is independent of $k$. $\mathrm{GP\;Basic}$ has the worst error, since it scales with $n$.
    
\begin{figure}[htbp]
     \centering
         \begin{subfigure}[t]{0.23\textwidth}
            \centering
            \includegraphics[width=\textwidth]{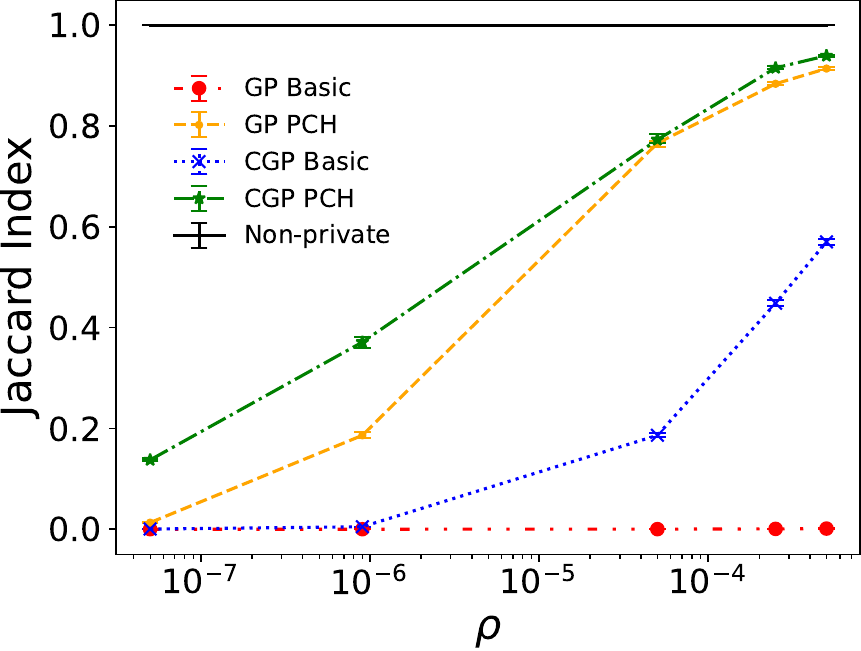}
             \vskip -.08in
            \subcaption{$n=20000$.}
         \end{subfigure}
         \;\;\;
         \begin{subfigure}[t]{0.24\textwidth}
            \centering
            \includegraphics[width=\textwidth]{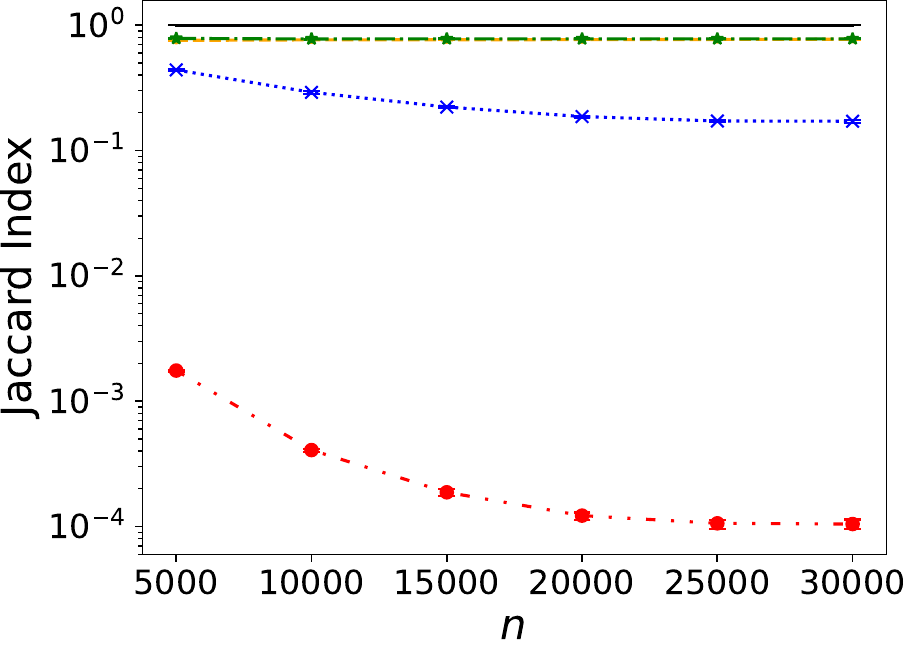}
             \vskip -.08in
            \subcaption{$\rho = 0.00005$.}
         \end{subfigure}
         \caption{Convex hull : error with respect to privacy level (left) and to tuple size (right).}
         \label{fig:convherr}
    \end{figure}

 \paragraph{Convex hull (Fig.~\ref{fig:convherr}).} We compare Algorithm~\ref{alg:PCH} ($\mathrm{CGP\;PCH}$) and its $\varepsilon$-GP variant ($\mathrm{GP\;PCH}$) against the baseline algorithms $\mathrm{GP\;Basic}$ and $\mathrm{CGP\;Basic}$. We compute the convex hulls from the privatized points returned by these algorithms. The $\mathrm{PCH}$ algorithms return $k$ points, while the baseline algorithms return $n$ points. The parameter $k$ is set as described in Section~\ref{sec:pch}, which has a different value in $\mathrm{CGP\;PCH}$ and $\mathrm{GP\;PCH}$. We also round it to the range $[16, 128]$ for efficiency reasons. 
 The utility is measured by the Jaccard similarity index between the privatized convex hull and the true convex hull (higher is better):
 \begin{align*}
 \mathrm{Jacc}\left(\mathrm{CONVH}(M(x),\mathrm{CONVH}(x)\right) = \frac{\mu\left(\mathrm{CONVH}(M(x))\cap \mathrm{CONVH}(x)\right)}{\mu\left(\mathrm{CONVH}(M(x))\cup \mathrm{CONVH}(x)\right)},
 \end{align*}
 where $\mu(C)$ is the area enclosed by the edges of $C$.

All four algorithms have better performance as $\rho$ increases; in the high privacy regime (lower privacy budget), the baseline algorithms have little utility, and $\mathrm{GP\;Basic}$ has essentially no utility even for the highest privacy budget considered. The baseline algorithms, especially GP Basic, lose utility as the number of points $n$ increases, since the area enclosed by the noisy convex hull becomes larger due to a smaller privacy budget being allotted to each point. The $\mathrm{PCH}$ algorithms do not suffer from this issue and maintain consistently high utility as $n$ increases.  While the CGP version of the PCH algorithm has some advantage over the GP version, the gap is not large.  Indeed, while the difference is a polynomial $\tilde{O}(\omega(x)^{1/6})$ theoretically, it is not as significant on real datasets. 


\section{Concluding Remarks}
We have shown that CGP offers many benefits over GP, including simplicity of the mechanism, lower noise scale in high dimensions, and better composability.  
This potentially opens a door for more problems on private geometric data to be studied.  One particularly interesting direction is when the input $x$ is an (unordered) \textit{set} of $n$ points as opposed to an (ordered) $n$-tuple. Between point sets, popular metrics include the earth mover distance or the Hausdorff distance.  Other metric spaces, such as curves using the Fr\'echet distance, or even non-geometric data such as strings using the edit distance, can also be studied under the CGP framework, and are of potential interest.

\section*{Acknowledgements}
This work has been supported by HKRGC under grants 16201819, 16205420, and 16205422. We thank the anonymous reviewers of CCS '23 for their valuable comments.

\bibliographystyle{alpha}
\bibliography{cgp_ccs_full}


\appendix 
\section{The $d$-dimensional Laplace Mechanism}
\label{sec:Rd_gen}

For a point $x\in\mathbb{R}^2$, the $\varepsilon$-GP mechanism in \cite{andres2013geo} for privately releasing $x$ produces a random variable $Y$ which follows a planar Laplacian distribution with pdf proportional to $e^{-\varepsilon\|y-x\|}$. 
Given $y=(y_1, y_2)\in\mathbb{R}^2$, it can be represented in polar coordinates as $(r,\theta)$ via the standard transformation $y_1=r\cos\theta$ and $y_2=r\sin\theta$ where $r\geq 0$ is the distance of $y$ from $x$, and $\theta\in [0,2\pi)$ is the angle between the horizontal axis and the line through $x$ and $y$. Thus, the pdf of the planar Laplacian distribution can be re-parameterized in polar coordinates such that
\[
\int_0^{2\pi} \int_0^{\infty} C_{\varepsilon} e^{-\varepsilon r} r dr d\theta = 1.
\]
It can be verified that $C_{\varepsilon}=\frac{\varepsilon^2}{2\pi}$ \cite{andres2013geo}. Moreover, the pdf has independent marginals for $R$ and $\Theta$:
\[
g_{\varepsilon}(r)=\varepsilon^2 r e^{-\varepsilon r}
\]
\[
h_{\varepsilon}(\theta)=\frac{1}{2\pi}.
\]

Let $y\in\mathbb{R}^d$ where $d\geq 2$; $y$ can be represented by $(r,\theta_1,\theta_2,\dotsb,\theta_{d-1})$, where $\theta_j\in [0,\pi)$ for $1\leq j\leq d-2$ and $\theta_{d-1}\in [0,2\pi)$. The $\varepsilon$-GP mechanism above can be generalized to $\mathbb{R}^d$ via a $d$-dimensional planar Laplacian distribution, which has pdf in spherical coordinates such that
\[
\int_0^{2\pi} \int_0^{\pi}\dotsb \int_0^{\pi} \int_0^{\infty} C_{d,\varepsilon} e^{-\varepsilon r} r^{d-1} \sin^{d-2}(\theta_1)\sin^{d-3}(\theta_2)\dotsb\sin(\theta_{d-2}) dr d\theta_1 d\theta_2 \dotsb d\theta_{d-1} = 1,
\]
for some constant $C_{d,\varepsilon} > 0$. Now, due to independence of the marginal distributions, the marginal pdf of $R$ is of the form $B_{d,\varepsilon} r^{d-1} e^{-\varepsilon r}$ for some constant $B_{d,\varepsilon} > 0$ such that
\[
\int_0^{\infty} B_{d,\varepsilon} r^{d-1} e^{-\varepsilon r} dr = 1.
\]
I.e., $R\sim \mathcal{G}(\frac{1}{\varepsilon}, d, 1)$. Note that $R$ can be efficiently simulated by rejection sampling with high efficiency \cite{marsaglia2000simple}, where the envelope distribution is Gaussian.

We can also re-parameterize the pdf of the Gaussian distribution for $\rho$-CGP in spherical coordinates, where
\[
\int_0^{2\pi} \int_0^{\pi}\dotsb \int_0^{\pi} \int_0^{\infty} C_{d,\rho} e^{-\rho r^2} r^{d-1} \sin^{d-2}(\theta_1)\sin^{d-3}(\theta_2)\dotsb\sin(\theta_{d-2}) dr d\theta_1 d\theta_2 \dotsb d\theta_{d-1} = 1.
\]
In this case, the marginal pdf of $R$ becomes $B_{d,\rho} r^{d-1} e^{-\rho r^2}$ for some constant $B_{d,\rho} > 0$. I.e., $R\sim \mathcal{G}(\frac{1}{\sqrt{\rho}}, d, 2)$.\\
From Lemma~\ref{lm:gengamma}, the following corollaries are immediate.

\begin{corollary}
Let ${M}_{\varepsilon}$ denote the planar Laplacian mechanism (for $\varepsilon$-GP) which, on input $x\in\mathbb{R}^d$, returns a $y\in\mathbb{R}^d$ with pdf $\propto e^{-\varepsilon\|y-x\|}$. Then, 
\[\mathbb{E}\left[\|\mathcal{M}_{\varepsilon}(x)-x\|\right] = \frac{1}{\varepsilon}\frac{\Gamma(d+1)}{\Gamma(d)}=\frac{d}{\varepsilon}.\]
\end{corollary}
\begin{corollary}
Let ${M}_{\rho}$ denote the Gaussian mechanism (for $\rho$-CGP) given by ${M}_{\rho}(x):=x+Z$ for $x\in\mathbb{R}^d$ where $Z\sim \mathcal{N}\left(0,\frac{1}{2\rho}I_{d\times d}\right)$. Then, 
\[\mathbb{E}\left[\|\mathcal{M}_{\rho}(x)-x\|\right] = \frac{1}{\sqrt{\rho}}\frac{\Gamma(d/2+1/2)}{\Gamma(d/2)}=O\left(\frac{\sqrt{d}}{\sqrt{\rho}}\right).\]
\end{corollary}

\section{Proof for CGP Composition}
\label{sec:cgp_comp_proof}

\begin{lemma} [Theorem ~\ref{lm:cgp_comp}]
    Let ${M}_1:U\rightarrow V_1$ be $\rho_1$-CGP, let ${M}_2:U\times V_1 \rightarrow V_2$ be $\rho_2$-CGP w.r.t. its first argument. Then the mechanism ${M}:U\rightarrow V_1\times V_2$ defined by ${M}(x)=({M}_1(x),{M}_2(x,{M}_1(x))$ is $(\rho_1+\rho_2)$-CGP.
\end{lemma}

\begin{proof}
 Let $m_1(x)(\cdot)$ denote the pdf of ${M}_1(x)$, let $m(x)(\cdot,\cdot)$ denote the joint pdf of ${M}(x)=({M}_1(x), {M}_2(x, {M}_1(x)))$. For $(y_1, y_2)\in V_1 \times V_2$, its probability density can be written as $m(x)(y_1, y_2) = m_1(x)(y_1)\cdot m_2(x, y_1)(y_2)$ where $m_2(x,y_1)(\cdot)$ is the pdf of ${M}_2(x, y_1)$. We also have $m(x')(y_1, y_2) = m_1(x')(y_1)\cdot m_2(x', y_1)(y_2)$ corresponding to input $x'$. Let $\mathcal{M}(x), \mathcal{M}({x'})$ be the distributions of $m(x)(\cdot,\cdot)$ and $m(x')(\cdot,\cdot)$, respectively.
    \begin{align*}
        e^{(\alpha-1)D_{\alpha}(\mathcal{M}(x)\|\mathcal{M}(x'))} &= \int_{V_1 \times V_2} m(x)(y_1, y_2)^{\alpha} m(x')(y_1, y_2)^{1-\alpha} dy_1 dy_2\\
        &= \int_{V_1 \times V_2} m_1(x)(y_1)^{\alpha}m_2(x, y_1)(y_2)^{\alpha} m_1(x')(y_1)^{1-\alpha}m_2(x', y_1)(y_2)^{1-\alpha} dy_1 dy_2\\
        &= \int_{V_1}m_1(x)(y_1)^{\alpha}m_1(x')(y_1)^{1-\alpha} \underbrace{\left(\int_{V_2}m_2(x, y_1)(y_2)^{\alpha}m_2(x', y_1)(y_2)^{1-\alpha} dy_2\right)}_{\leq e^{(\alpha-1)\alpha \rho_2 \cdot \mathrm{dist}(x,x')^2}} dy_1\\
        &\leq e^{(\alpha-1)\alpha \rho_2 \cdot \mathrm{dist}(x,x')^2}\int_{V_1}m_1(x)(y_1)^{\alpha}m_1(x')(y_1)^{1-\alpha} dy_1\\
        &\leq e^{(\alpha-1)\alpha (\rho_1+\rho_2) \cdot \mathrm{dist}(x,x')^2}
    \end{align*}
where the first inequality is due to ${M}_2$ being $\rho_2$-CGP (w.r.t. its first argument) and the last inequality is due to ${M}_1$ being $\rho_1$-CGP.
\end{proof}

\section{Proof for relationship between CGP and $(\varepsilon,\delta,\Delta)$-GP}
\label{sec:cgp_dDgp_proof}

\begin{lemma} [Lemma~\ref{lm:cgp_dDGP}]
    A mechanism ${M}$ that is $\rho$-CGP is also $(\varepsilon,\delta,\Delta)$-GP, for any $\varepsilon, \delta, \Delta$ such that $\varepsilon \ge \rho \Delta + 2\sqrt{\rho\log(1/\delta)}$.
\end{lemma}

\begin{proof}
    Let $x, x'\in U$ where $x\neq x'$. Let $\mathcal{M}(x), \mathcal{M}(x')$ denote the distributions of ${M}(x)$ and ${M}(x')$, with pdf $m(x)(\cdot)$ and $m(x')(\cdot)$, respectively. Let $Y, Y'$ be random variables drawn from $\mathcal{M}(x)$ and $\mathcal{M}(x')$, respectively. Given $D_{\alpha}(\mathcal{M}(x)\|\mathcal{M}(x')) \leq \alpha \rho \cdot \dist(x,x')$ for all $\alpha > 1$, we will show that, for a suitable choice of $\alpha$, for any measurable subset $S\subseteq V$,
    \[\Pr[Y\in S] \leq e^{\varepsilon \cdot \dist(x,x')}\Pr[Y'\in S] + \delta,\]
    where $\delta \leq \frac{2}{5}e^{-\frac{\varepsilon^2}{8\rho}}$ for $\mathrm{dist}(x,x')\leq \Delta$.
    
   Let $\mathtt{l}:V\rightarrow \mathbb{R}$ be defined by $\mathtt{l}(y)=\log\left(\frac{m(x)(y)}{m(x')(y)}\right)$. First, note that $\mathbb{E}[e^{(\alpha-1)\mathtt{l}(Y)}]=e^{(\alpha-1)D_{\alpha}(\mathcal{M}(x)\|\mathcal{M}(x'))}$, since
\[D_{\alpha}(\mathcal{M}(x)\|\mathcal{M}(x'))=\frac{1}{\alpha-1}\log\left(\int_V \left(\frac{m(x)(y)}{m(x')(y)}\right)^{\alpha-1}m(x)(y)dy \right)=\frac{1}{\alpha-1}\log(\mathbb{E}[(e^{\mathtt{l}(Y)})^{\alpha-1}]).\]
In particular, for $\alpha=2$, we have by Jensen's inequality
\[e^{\mathbb{E}[\mathtt{l}(Y)]}\leq \mathbb{E}[e^{\mathtt{l}(Y)}]=e^{D_{2}(\mathcal{M}(x)\|\mathcal{M}(x'))}\leq e^{2\rho\dist(x,x')^2}.\]
I.e. the random variable $\mathtt{l}(Y)$ has finite expectation (fixing $x$ and $x'$) and $|\mathtt{l}(Y)|$ is bounded almost surely. However, in order to find a suitable bound for $\varepsilon$, we allow $\delta > 0$ and restrict $\dist(x,x')$.
   We use a similar derivation as that in \cite{bun2016concentrated} to bound $\delta$ in terms of $\varepsilon$ and $\rho$. Let $S\subseteq V$ be any measurable subset. Let $S_{\varepsilon}:=\{y\in S: \mathtt{l}(y)\leq \varepsilon \cdot \mathrm{dist}(x,x')\}$. Then
   \begin{align*}
    \Pr[Y\in S] &= \int_{y\in S_{\varepsilon}} m(x)(y) dy + \int_{y\in S\setminus S_{\varepsilon}} m(x)(y) dy \\
    &\leq e^{\varepsilon \cdot \mathrm{dist}(x,x')} \int_{y\in S_{\varepsilon}} m(x')(y) dy + \int_{y\in S\setminus S_{\varepsilon}} m(x)(y) dy\\
    &= e^{\varepsilon \cdot \mathrm{dist}(x,x')} \left(\int_{y\in S} m(x')(y) dy - \int_{y\in S\setminus S_{\varepsilon}} m(x')(y) dy \right) + \int_{y\in S\setminus S_{\varepsilon}} m(x)(y) dy\\
    &= e^{\varepsilon \cdot \mathrm{dist}(x,x')}\Pr[Y'\in S] + \underbrace{\int_{y\in S\setminus S_{\varepsilon}}\left(m(x)(y) - e^{\varepsilon \cdot \mathrm{dist}(x,x')}m(x')(y)\right) dy}_{\delta}.
   \end{align*}
Let $l(\cdot)$ denote the pdf of $\mathtt{l}(Y)$.
\begin{align*}
    \delta &=\int_{y\in S\setminus S_{\varepsilon}}\left(m(x)(y) - e^{\varepsilon \cdot \mathrm{dist}(x,x')}m(x')(y)\right) dy
    = \int_{y\in S\setminus S_{\varepsilon}}\left(e^{\mathtt{l}(y)} - e^{\varepsilon \cdot \mathrm{dist}(x,x')}\right)m(x')(y) dy \\
    &= \int_{y\in S\setminus S_{\varepsilon}}\left(e^{\mathtt{l}(y)} - e^{\varepsilon \cdot \mathrm{dist}(x,x')}\right)\frac{m(x')(y)}{m(x)(y)}m(x)(y) dy
    = \int_{y\in S\setminus S_{\varepsilon}}\left(1 - e^{\varepsilon\cdot \mathrm{dist}(x,x')-\mathtt{l}(y)}\right)m(x)(y) dy\\
    &= \mathbb{E}\left[\left(1-e^{\varepsilon\cdot \mathrm{dist}(x,x')-\mathtt{l}(Y)}\right)\mathbb{1}_{Y\in S}\mathbb{1}_{\mathtt{l}(Y)> {\varepsilon \cdot \mathrm{dist}(x,x')}}\right]
    \leq \int_{\varepsilon \cdot \mathrm{dist}(x,x')}^{\infty} (1-e^{\varepsilon\cdot \mathrm{dist}(x,x')-z})l(z) dz.
\end{align*}
Let $u(z)=e^{-z}$, $v(z)=\int_{-\infty}^{z} l(w) dw = \Pr[\mathtt{l}(Y)\leq z]$. Integration by parts gives
\begin{align*}
\delta &\leq \Pr[\mathtt{l}(Y)>\varepsilon \cdot \mathrm{dist}(x,x')] - e^{\varepsilon \cdot \mathrm{dist}(x,x')}\left(\left[e^{-z}\Pr[\mathtt{l}(Y)\leq z]\right]_{z=\varepsilon \cdot \mathrm{dist}(x,x')}^{\infty} - \int_{\varepsilon \cdot \mathrm{dist}(x,x')}^{\infty} (-e^{-z})\Pr[\mathtt{l}(Y)\leq z]dz\right)\\
&= \Pr[\mathtt{l}(Y)>\varepsilon \cdot \mathrm{dist}(x,x')]+\Pr[\mathtt{l}(Y)\leq\varepsilon \cdot \mathrm{dist}(x,x')]-e^{\varepsilon \cdot \mathrm{dist}(x,x')}\int_{\varepsilon \cdot \mathrm{dist}(x,x')}^{\infty} e^{-z}(1-\Pr[\mathtt{l}(Y)> z]) dz\\
&=1 + e^{\varepsilon \cdot \mathrm{dist}(x,x')}\left[e^{-z}\right]_{z=\varepsilon \cdot \mathrm{dist}(x,x')}^{\infty}+e^{\varepsilon \cdot \mathrm{dist}(x,x')}\int_{\varepsilon \cdot \mathrm{dist}(x,x')}^{\infty} e^{-z}\Pr[\mathtt{l}(Y)> z] dz\\
&= e^{\varepsilon \cdot \mathrm{dist}(x,x')}\int_{\varepsilon \cdot \mathrm{dist}(x,x')}^{\infty} e^{-z}\Pr[\mathtt{l}(Y)> z] dz.
\end{align*}
For $z\geq \varepsilon \cdot \mathrm{dist}(x,x') > 0$ and $\alpha > 1$,
\[\Pr[\mathtt{l}(Y)>z]=\Pr[e^{(\alpha-1)\mathtt{l}(Y)}>e^{(\alpha-1)z}] \leq \frac{\mathbb{E}[e^{(\alpha-1)\mathtt{l}(Y)}]}{e^{(\alpha-1)z}} = \frac{e^{(\alpha-1)D_\alpha(\mathcal{M}(x)\|\mathcal{M}(x'))}}{e^{(\alpha-1)z}} = e^{(\alpha-1)(\alpha \rho \cdot \mathrm{dist}(x,x')^2-z)}\]
where the first inequality is due to Markov's inequality and the last equality follows from ${M}$ being $\rho$-CGP. Thus,
\begin{align*}
\delta &\leq
e^{\varepsilon \cdot \mathrm{dist}(x,x')}\int_{\varepsilon \cdot \mathrm{dist}(x,x')}^{\infty} e^{-z}\Pr[\mathtt{l}(Y)> z] dz \leq e^{\varepsilon \cdot \mathrm{dist}(x,x')}\int_{\varepsilon \cdot \mathrm{dist}(x,x')}^{\infty} e^{-z}e^{(\alpha-1)(\alpha \rho \cdot \mathrm{dist}(x,x')^2-z)} dz\\
&= e^{\varepsilon \cdot \mathrm{dist}(x,x')}e^{(\alpha-1)\alpha \rho \cdot \mathrm{dist}(x,x')^2}\int_{\varepsilon \cdot \mathrm{dist}(x,x')}^{\infty} e^{-\alpha z} dz = e^{\varepsilon \cdot \mathrm{dist}(x,x')}e^{(\alpha-1)\alpha \rho \cdot \mathrm{dist}(x,x')^2}\left(\frac{1}{\alpha} e^{-\alpha \varepsilon \cdot \mathrm{dist}(x,x')}\right)\\
&= \frac{1}{\alpha}e^{(\alpha\rho\cdot \mathrm{dist}(x,x')^2-\varepsilon\cdot \mathrm{dist}(x,x'))(\alpha-1)}.
\end{align*}
For $\mathrm{dist}(x,x')< \frac{\varepsilon}{s\rho}$, where $s>1$, choose $\alpha = \frac{\varepsilon\cdot \mathrm{dist}(x,x')+\rho\cdot \mathrm{dist}(x,x')^2}{2\rho\cdot \mathrm{dist}(x,x')^2}$. Then $\alpha >1$ and 
\[(\alpha\rho\cdot \mathrm{dist}(x,x')^2-\varepsilon\cdot \mathrm{dist}(x,x'))(\alpha-1)=-\frac{(\varepsilon\cdot \mathrm{dist}(x,x')-\rho\cdot \mathrm{dist}(x,x')^2)^2}{4\rho\cdot \mathrm{dist}(x,x')^2}=-\frac{(\varepsilon-\rho\cdot \mathrm{dist}(x,x'))^2}{4\rho}.\]
We also have $\frac{(s-1)\varepsilon}{s}<\varepsilon-\rho\cdot \mathrm{dist}(x,x')$ and $\frac{1}{\alpha} \leq \frac{2}{s+1}$. Thus,
\begin{align*}
\delta &\leq \frac{1}{\alpha}e^{(\alpha\rho\cdot \mathrm{dist}(x,x')^2-\varepsilon\cdot \mathrm{dist}(x,x'))(\alpha-1)}=\frac{1}{\alpha}e^{-\frac{(\varepsilon-\rho\cdot \mathrm{dist}(x,x'))^2}{4\rho}} \leq \frac{2}{s+1}e^{-\frac{((s-1)\varepsilon/s)^2}{4\rho}}=:\delta(s,\varepsilon;\rho).
\end{align*}
I.e., $M$ satisfies $\left(\varepsilon, \delta(s,\varepsilon;\rho), \Delta(s,\varepsilon;\rho)\right)$-GP, where $\delta(s,\varepsilon;\rho)$ is defined above, and $\Delta(s,\varepsilon;\rho) := \frac{\varepsilon}{s\rho}$, for all $s>1$. Equivalently, given $\delta, \Delta>0$, $M$ satisfies $\left(\varepsilon(s,\delta,\Delta;\rho), \delta, \Delta\right)$-GP, for all $s>1$, where $\varepsilon(s,\delta,\Delta;\rho):=\max\left\{\frac{s}{s-1}2\sqrt{\rho\log(2/(s+1)/\delta)},s\rho\Delta\right\}$. In particular, choosing $s=1+\frac{2\sqrt{\rho\log(1/\delta)}}{\rho\Delta}$, we have
\[
\frac{s}{s-1}2\sqrt{\rho\log(2/(s+1)/\delta)} = \left(1+\frac{\rho\Delta}{2\sqrt{\rho\log(1/\delta)}}\right)\cdot 2\sqrt{\rho\log(2/(s+1)/\delta)} \leq  2\sqrt{\rho\log(1/\delta)}+\rho\Delta,
\]
\[
s\rho\Delta = \left(1+\frac{2\sqrt{\rho\log(1/\delta)}}{\rho\Delta}\right)\cdot \rho\Delta \leq \rho\Delta + {2\sqrt{\rho\log(1/\delta)}}.
\]
I.e., $M$ satisfies $\left(\rho\Delta + {2\sqrt{\rho\log(1/\delta)}}, \delta, \Delta\right)$-GP.

\end{proof}
\begin{figure}[H]
     \centering
         \begin{subfigure}[t]{0.25\textwidth}
            \centering
            \includegraphics[width=\textwidth]{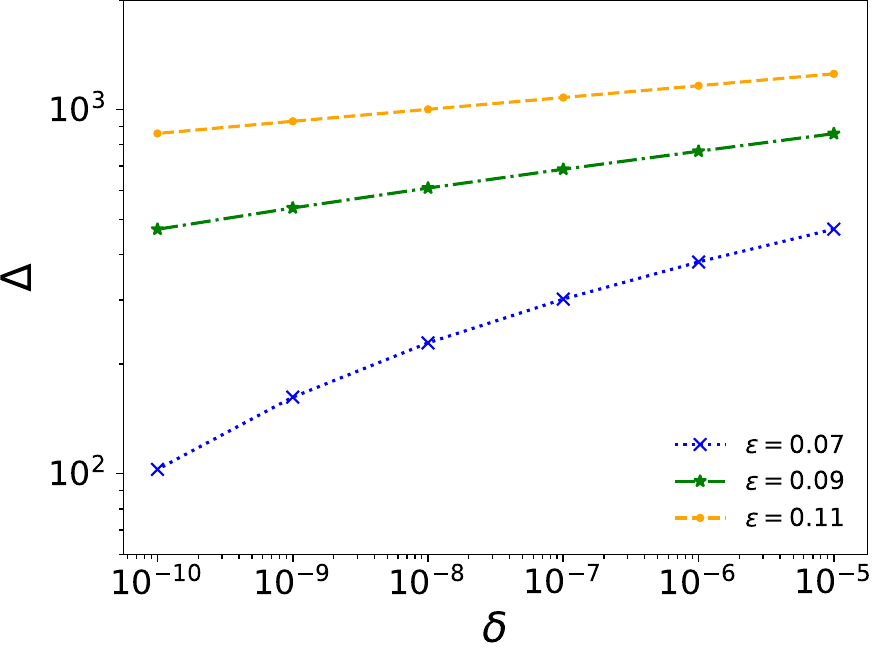}
             \vskip -.08in
            \subcaption{Values of $\Delta$ at various levels of $\delta$.}
         \end{subfigure}
        \;\;\;\;
         \begin{subfigure}[t]{0.25\textwidth}
            \centering
            \includegraphics[width=\textwidth]{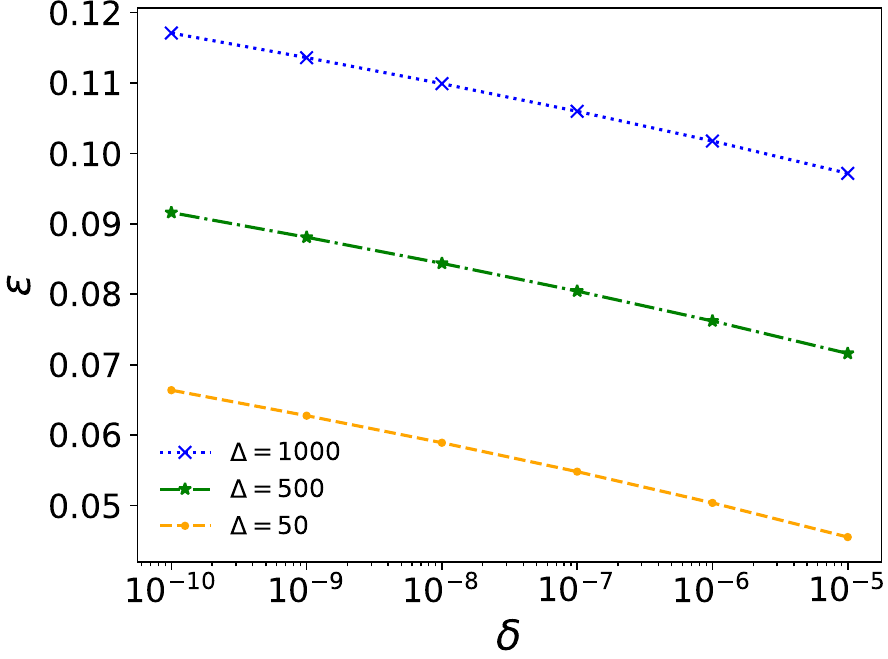}
             \vskip -.08in
            \subcaption{Values of $\varepsilon$ at various levels of $\delta$.}
         \end{subfigure}
          \vskip -.1in
        \caption{{$(\varepsilon, \delta, \Delta)$-GP implied by $\rho$-CGP, fixing $\rho=0.00005$.}}
         \label{fig:eps_delta_Delta_relationship}
    \end{figure}
    
\section{Proof for $\varepsilon$-GP variant of Sparse Vector Technique}
\label{sec:svt_proof}

\begin{theorem} 
Algorithm~\ref{alg:SVT} is $\varepsilon$-GP.
\end{theorem}

\begin{proof}Fix $x, x'\in U$. Let $M(x), M(x')$ denote the outputs of Algorithm~\ref{alg:SVT} corresponding to inputs $x$ and $x'$, respectively. Given $W$, for positive integer $j$, let $q_j(x,W), p_j(x,W)$ denote the probabilities of the events $g_j(x)+V_j>T+W$ and $g_j(x)+V_j\leq T+W$, respectively. There are two cases to consider: 1) if the algorithm executes line $6$ and halts, then its output is of the form $\{\bot\}^{t-1}\times \{\top\}$ for some positive integer $t$; 2) if the algorithm never executes line $6$, then we must show it's $\varepsilon$-GP with respect to output $\{\bot\}^{t}$ \textit{for all} positive integers $t$. Fix any such integer $t$, and consider a positive perturbation $\tau:= \max_{1\leq j \leq t} |g_j(x)-g_j(x')|\leq K\|x-x'\|$. Then for $1\leq j\leq t$,
\begin{align}
\nonumber
q_j(x,w) = \Pr[M(x)_{j}=\bot|W=w] &= \Pr[g_j(x)+V_j > T + w] \\
\nonumber
&= \Pr[g_j(x') + g_j(x)+V_j > g_j(x') + T + w]\\
\nonumber
&=\Pr[g_j(x') + V_j > g_j(x')-g_j(x) + T + w]\\
\nonumber
&\leq \Pr[g_j(x') + V_j > -|g_j(x')-g_j(x)| + T + w]\\
&\leq \Pr[g_j(x') + V_j > -\tau+T+w]=q_j(x',w-\tau).
\label{ineq:qj}
\end{align}
\begin{align}
\nonumber
p_j(x,w) = \Pr[M(x)_{j}=\top|W=w] &= \Pr[g_j(x)+V_j \leq T + w] \\
\nonumber
&= \Pr[g_j(x)+g_j(x')+V_j \leq  T + w+g_j(x')]\\
\nonumber
&\leq\Pr\left[g_j(x') + V_j \leq T + w+g_j(x')-g_j(x)\right]\\
\nonumber
&\leq\Pr\left[g_j(x') + V_j \leq T + w+|g_j(x')-g_j(x)|\right]\\
&=\Pr\left[g_j(x')+V_j \leq T+w+\tau\right]=p_j(x',w+\tau).
\label{ineq:pj}
\end{align}
Also, 
\begin{align}
\nonumber
p_j(x',w+\tau) &= \Pr[M(x')_j=\top | W=w+\tau]= \Pr[g_j(x')+V_j \leq T+w+\tau]\\
\nonumber
&= \Pr[g_j(x')-2\tau+V_j \leq T+w-\tau]\\
\nonumber
&= \int_{-\infty}^{T+w-\tau} l\left(g_j(x')-2\tau, {2K}/{\varepsilon_2}\right)(y) dy\\
\nonumber
&\leq \int_{-\infty}^{T+w-\tau} e^{2\tau {\varepsilon_2}/(2K)} l\left(g_j(x'), {2K}/{\varepsilon_2}\right)(y) dy\\
&= e^{\varepsilon_2 \tau/K}\Pr[g_j(x')+V_j \leq T+w-\tau] = e^{\varepsilon_2 \tau/K}p_j(x',w-\tau)
\label{ineq:pj2}
\end{align}
where the inequality follows from Lemma~\ref{lm:lap_shift_pdf} with shift $2\tau$.
Now, the $q_j's$ and $p_j's$ are independent given $W$. Let $u:=w-\tau$. Then in the first case we have
\begin{align*}
\Pr[M(x)=(\bot^{t-1})\top] &= \int_{-\infty}^{\infty} \Pi_{j=1}^{t-1} q_j(x,w)\cdot p_t(x,w)\cdot l(0,K/\varepsilon_1)(w) dw\\
&\leq \int_{-\infty}^{\infty} \Pi_{j=1}^{t-1} q_j(x',w-\tau)\cdot p_t(x',w+\tau)\cdot l(0,K/\varepsilon_1)(w) dw\\
&\leq e^{\varepsilon_2 \tau/K}\int_{-\infty}^{\infty} \Pi_{j=1}^{t-1} q_j(x',w-\tau)\cdot p_t(x',w-\tau)\cdot l(0,K/\varepsilon_1)(w) dw\\
&\leq e^{\varepsilon_2 \tau/K}\int_{-\infty}^{\infty} \Pi_{j=1}^{t-1} q_j(x',u)\cdot p_t(x', u) \cdot l(-\tau,K/\varepsilon_1)(u) du\\
&\leq e^{(\varepsilon_1+\varepsilon_2)\tau/K}\int_{-\infty}^{\infty} \Pi_{j=1}^{t-1} q_j(x',u)\cdot p_t(x', u) \cdot l(0,K/\varepsilon_1)(u) du\\
&\leq e^{\varepsilon\|x-x'\|}\Pr[M(x')=(\bot^{t-1})\top],
\end{align*}
where the first inequality follows from (\ref{ineq:qj}) and (\ref{ineq:pj}), the second inequality follows from (\ref{ineq:pj2}) and the third inequality is due to the change of variables $u=w-\tau$.
In the second case, we have
\begin{align*}
\Pr[M(x)=\bot^{t}] &= \int_{-\infty}^{\infty} \Pi_{j=1}^{t} q_j(x,w)\cdot l(0,K/\varepsilon_1)(w) dw\\
&\leq \int_{-\infty}^{\infty} \Pi_{j=1}^{t} q_j(x',w-\tau)\cdot l(0,K/\varepsilon_1)(w) dw\\
&\int_{-\infty}^{\infty} \Pi_{j=1}^t q_j(x',u)\cdot l(-\tau,K/\varepsilon_1)(u) du\\
&\leq e^{\varepsilon_1\tau/K}\int_{-\infty}^{\infty} \Pi_{j=1}^t q_j(x',u) \cdot l(0,K/\varepsilon_1)(u) du\\
&\leq e^{\varepsilon_1\|x-x'\|}\Pr[M(x')=\bot^{t}] .
\end{align*}
\end{proof}

\section{Technical Lemmas and Missing Proofs}
\label{sec:technical_proofs}

\begin{lemma}
\label{lm:lipschitz_properties}
Let $(U_0,\mathrm{dist}_{U_0})$, $(U_1,\mathrm{dist}_{U_1})$ and $(U_2,\mathrm{dist}_{U_2})$ be metric spaces. Let $g_1:{U_0}\rightarrow U_1$ and $g_2:U_1\rightarrow U_2$ be Lipschitz with constants $K_1$ and $K_2$ respectively. 
    \begin{enumerate}
        \item $\mathrm{[Composition]}$ $h:{U_0}\rightarrow U_2$ defined by $h(x):= g_2(g_1(x))$ is $(K_1K_2)$-Lipschitz;
        \item $\mathrm{[Linear\;maps]}$ $h:{U_0}\rightarrow U_1$ defined by $h(x)=a\cdot g_1(x)+b$ is $(K_1\cdot |a|)$-Lipschitz if $U_1$ is a real vector space, for any $a\in \mathbb{R}$ and $b\in U_1$.
    \end{enumerate}
\end{lemma}

\begin{lemma}[Univariate Gaussian tail bound]
    \label{ft:univariate_normal_ineq}
     Let $Z\sim \mathcal{N}(0,\sigma^2)$. Then 
     $\Pr[Z>t]\leq e^{-\frac{t^2}{2\sigma^2}}$ for all $t\geq 0$.
\end{lemma}

\begin{lemma} 
\label{lm:lap_shift_pdf}
Let $l(\mu,b)(\cdot)$ denote the pdf of a Laplace random variable with mean $\mu$ and scale $b$. For any shift $s\in \mathbb{R}$ in the mean and for any $y\in \mathbb{R}$: $
l(\mu,b)(y) \leq e^{|s|/b}\cdot l(\mu+s,b)(y).$
\end{lemma}
\begin{proof}
Since $-|y-\mu|\leq -|y-(\mu+s)|+|s|$, 
\[
l(\mu,b)(y) = \frac{1}{2b}e^{-|y-\mu|/b} \leq \frac{1}{2b}e^{-|y-(\mu+s)|/b+|s|/b} = e^{|s|/b}\cdot l(\mu+s,b)(y).
\]
\end{proof}

\begin{lemma} 
\label{lm:pdf_ZW}
Let $Z$, $W$ be i.i.d. $\mathrm{Lap}(b)$ random variables. Then the pdf of $Y:=Z+W$ is given by $\frac{1}{4b^2}(be^{-|y|/b}+|y|e^{-|y|/b})$ for $y\in \mathbb{R}$.
\end{lemma}

\begin{proof}
    Let $f_Z$ and $f_W$ denote the pdf of $Z$ and $W$, respectively. The pdf $f_Y$ of $Y$ can be computed by convolution:
    \begin{align*}
    f_Y(y) &= \int_{-\infty}^{\infty} f_Z(z)f_W(y-z) dz\\
    &= \int_{-\infty}^{\infty} \frac{1}{2b}e^{-|z|/b}\cdot \frac{1}{2b} e^{-|y-z|/b} dz\\
    &= \frac{1}{4b^2}\int_{-\infty}^{\infty} e^{-|z|/b-|y-z|/b} dz.
    \end{align*}
    For $y \leq 0$:
    \begin{align*}
        \int_{-\infty}^{\infty} e^{-|z|/b-|y-z|/b} dz &= \int_{-\infty}^{y} e^{z/b-(y-z)/b} dz + \int_{y}^{0} e^{z/b-(z-y)/b} dz + \int_{0}^{\infty} e^{-z/b-(z-y)/b} dz\\
        &= \left[\frac{b}{2}e^{2z/b-y/b}\right]_{z=-\infty}^{y} + \left[ze^{y/b}\right]_{z=y}^{0}+\left[-\frac{b}{2}e^{-2z/b+y/b}\right]_{z=0}^{\infty}\\
        &=\frac{b}{2}e^{y/b}-ye^{y/b}+\frac{b}{2}e^{y/b} = be^{y/b}-ye^{y/b} = be^{-|y|/b}+|y|e^{-|y|/b}.
    \end{align*}
    For $y\geq 0$:
    \begin{align*}
        \int_{-\infty}^{\infty} e^{-|z|/b-|y-z|/b} dz &= \int_{-\infty}^{0} e^{z/b-(y-z)/b} dz + \int_{0}^{y} e^{-z/b-(y-z)/b} dz + \int_{y}^{\infty} e^{-z/b-(z-y)/b} dz\\
        &= \left[\frac{b}{2}e^{2z/b-y/b}\right]_{z=-\infty}^{0} + \left[ze^{-y/b}\right]_{z=0}^{y} + \left[-\frac{b}{2}e^{-2z/b+y/b}\right]_{z=y}^{\infty}\\
        &= \frac{b}{2}e^{-y/b} + ye^{-y/b} + \frac{b}{2}e^{-y/b} = be^{-y/b}+ye^{-y/b} = be^{-|y|/b}+|y|e^{-|y|/b}.
    \end{align*}
\end{proof}

\begin{lemma} 
\label{lm:ZW_bound2}
    Let $Y$ be a random variable with pdf $\frac{1}{4b^2}(be^{-|y|/b}+|y|e^{-|y|/b})$ for $y\in \mathbb{R}$, where $b>0$. Then $|Y| \leq b\left(\sqrt{2\log(1/\beta)}+\log(\frac{1}{\beta})\right)$ with probability $1-\beta$.
\end{lemma}
\begin{proof}
    We compute $r\geq 0$ which satisfies $\Pr[-r\leq Y \leq r] \geq 1-\beta$.
    
\begin{align*}
    &{}\Pr[-r\leq Y \leq r] \\
    &= \frac{1}{4b^2}\int_{-r}^r \left(be^{-|y|/b}+|y|e^{-|y|/b}\right) dy \\
    &= \frac{1}{4b^2} \left(\int_{-r}^0 \left(be^{y/b}-ye^{y/b}\right) dy + \int_{0}^r \left(be^{-y/b}+ye^{-y/b}\right) dy\right)\\
    &= \frac{1}{4b^2} \left(\left[b^2e^{y/b}\right]_{y=-r}^0 -\left[ybe^{y/b}\right]_{y=-r}^0+\int_{-r}^0 be^{y/b}dy+\left[-b^2e^{-y/b}\right]_{y=0}^r+\left[-ybe^{-y/b}\right]_{y=0}^r+\int_0^r be^{-y/b}dy\right)\\
    &= \frac{1}{4b^2} \left(b^2-b^2e^{-r/b}-rbe^{-r/b}+\left[b^2 e^{y/b}\right]_{y=-r}^0 -b^2e^{-r/b}+b^2-rbe^{-r/b}+\left[-b^2e^{-y/b}\right]_{y=0}^r \right)\\
    &= \frac{1}{4b^2} \left(2b^2-2b^2e^{-r/b}-2rbe^{-r/b}+b^2-b^2e^{-r/b} -b^2e^{-r/b}+b^2 \right)\\
    &= \frac{1}{4b^2} \left(4b^2-4b^2e^{-r/b}-2rbe^{-r/b} \right) \geq 1-e^{-r/b}-(r/b)e^{-r/b}=1+e^{-r/b}(-1-r/b).
\end{align*}
Setting the last expression to be $1-\beta$, we have $1+e^{+1}\cdot e^{-1-r/b}(-1-r/b)=1-\beta$, or $we^{w}=-\beta/e$, where $w:=-1-r/b\leq -1$ can be solved via the Lambert W function as $w=f^{-1}(-\beta/e)$ where $f(w):=we^{w}$. Then $r/b=-w-1<\sqrt{2\log(1/\beta)}+\log(1/\beta)$ and the proof is complete.
\end{proof}

\begin{lemma} [Lemma~\ref{lm:ZW_bound}]
    Let $Z$, $W\sim \mathrm{Lap}(b)$, and $Y:=Z+W$. Then for $1>\beta>0$, with probability $1-\beta$, 
    $|Y| \leq b\left(\sqrt{2\log(1/\beta)}+\log(\frac{1}{\beta})\right)$.
\end{lemma}
\begin{proof}
Follows from Lemma~\ref{lm:pdf_ZW} and Lemma~\ref{lm:ZW_bound2}.
\end{proof}

\begin{lemma} [Lemma~\ref{lm:V_leq_ZW_count}]
Given $y\in \mathbb{R}$, suppose we draw a $V\sim \mathrm{Lap}(2b)$ until $V\leq y$. Let $r(y)$ be the number of draws given $y$ and let $R=r(Y)$, where $Y:=Z+W$ and $Z$, $W\sim \mathrm{Lap}(b)$. Then $\mathbb{E}[R]\leq 4$.
\end{lemma}
\begin{proof}
Given $y$, $r(y)$ follows a Geometric distribution with success probability $p(y)=\Pr[V\leq y]$. Thus, $\mathbb{E}[r(Y)|Y=y]=\frac{1}{\Pr[V\leq y]}$. Then,
    \begin{align*}
        \mathbb{E}[R]&=\mathbb{E}\left[\mathbb{E}\left[r(Y)|Y=y\right]\right]\\
        &= \int_{-\infty}^{\infty} \frac{1}{\Pr[V\leq y]}\cdot \frac{1}{4b^2}(be^{-|y|/b}+|y|e^{-|y|/b}) dy\\
        &= \frac{1}{4b^2} \int_{-\infty}^{0} \frac{1}{(1/2)e^{y/(2b)}} (be^{y/b}-ye^{y/b}) dy + \frac{1}{4b^2} \int_{0}^{\infty} \frac{1}{1-(1/2)e^{-y/(2b)}} (be^{-y/b}+ye^{-y/b}) dy\\
        &\leq \frac{1}{4b^2} \int_{-\infty}^{0} 2e^{-y/(2b)} (be^{y/b}-ye^{y/b}) dy + \frac{1}{4b^2} \int_{0}^{\infty} 2 (be^{-y/b}+ye^{-y/b}) dy\\
        &= \frac{1}{4b^2} \int_{-\infty}^{0} 2be^{y/(2b)}-2ye^{y/(2b)}) dy + \frac{1}{4b^2} \int_{0}^{\infty} 2 (be^{-y/b}+ye^{-y/b}) dy\\
        &= \frac{1}{4b^2} \left[12b^2\right] + \frac{1}{4b^2} \left[4b^2\right] = 4.
    \end{align*}
\end{proof}

\section{Proof for Lipschitzness of Computing the Center}
\label{sec:center_lipschitz_proof}

\begin{lemma} 
\label{lm:center_lipschitz}
    Let $h_l:(\mathbb{R}^2)^n\rightarrow \mathbb{R}$ be defined by $h_l(x)=\max_{i\in[n]}x_{i,l}$. Let $g_l:(\mathbb{R}^2)^n\rightarrow \mathbb{R}$ be defined by $g_l(x)=\max_{i\in[n]}(-x_{i,l})$, where $x_{i,l}$ denotes the $l$th coordinate of $x_i$ for $l=1, 2$. Let $c:(\mathbb{R}^2)^n\rightarrow \mathbb{R}^2$ be defined by $c(x) = \frac{1}{2}(h_1(x)-g_1(x), h_2(x)-g_2(x))$. Then $c$ is $\sqrt{2}$-Lipschitz.
\end{lemma}
\begin{proof} Fix $x, x'\in (\mathbb{R}^2)^n$. Let $i^*_l:=\argmax_i x_{i,l}$, $j^*_l:=\argmax_i (-x_{i_l})$. I.e., $h_l(x)=x_{i^*_l}$, $g_l(x)=-x_{j^*_l}$ for $l=1, 2$. 
     Then $h_l(x)-h_l(x')=x_{i^*_l}-h_l(x') \leq x_{i^*_l}-x'_{i^*_l}\leq \dist_{\infty}(x,x')$ and $g_l(x)-g_l(x') = -x_{j^*_l}-g_l(x') \leq -x_{j^*_l} - (-x'_{j^*_l})=x'_{j^*_l}-x_{j^*_l}\leq \dist_{\infty}(x,x')$. We can similarly show that $h_l(x')-h_l(x) \leq \dist_{\infty}(x,x')$ and $g_l(x')-g_l(x)\leq \dist_{\infty}(x,x')$. Thus, $|h_l(x)-h_l(x')| \leq \dist_{\infty}(x,x')$ and $|g_l(x)-g_l(x')| \leq \dist_{\infty}(x,x')$.
    \begin{align*}
        \|c(x)-c(x')\|^2 &= \frac{1}{4}\left\|(h_1(x)-g_1(x)-(h_1(x')-g_1(x')), h_2(x)-g_2(x)-(h_2(x')-g_2(x')))\right\|^2 \\
        &= \frac{1}{4} \bigl[(h_1(x)-h_1(x')-(g_1(x)-g_1(x')))^2+(h_2(x)-h_2(x')-(g_2(x)-g_2(x')))\bigr]^2\\
        &= \frac{1}{4} \bigl[(h_1(x)-h_1(x'))^2+(g_1(x)-g_1(x'))^2-2(h_1(x)-h_1(x'))(g_1(x)-g_1(x'))\\
        &\qquad +(h_2(x)-h_2(x'))^2+(g_2(x)-g_2(x'))^2-2(h_2(x)-h_2(x'))(g_2(x)-g_2(x'))\bigr] \\
        &\leq \frac{1}{4} \cdot 8\cdot {\dist}_{\infty}(x,x')^2 = 2\cdot {\dist}_{\infty}(x,x')^2.
    \end{align*}
\end{proof}

\section{Experiments in the low privacy regime}
\label{sec:experiments_lp}
\begin{figure}[H]
     \centering
         \begin{subfigure}[t]{0.23\textwidth}
            \centering
            \includegraphics[width=\textwidth]{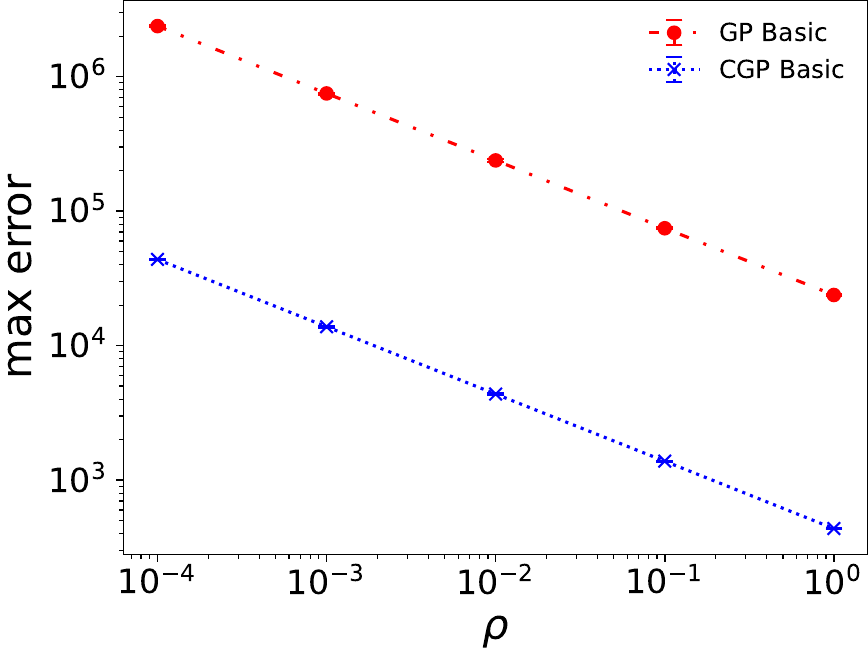}
             \vskip -.08in
            \subcaption{$n=20000$.}
            \;
         \end{subfigure}
        \hfill
         \begin{subfigure}[t]{0.23\textwidth}
            \centering
            \includegraphics[width=\textwidth]{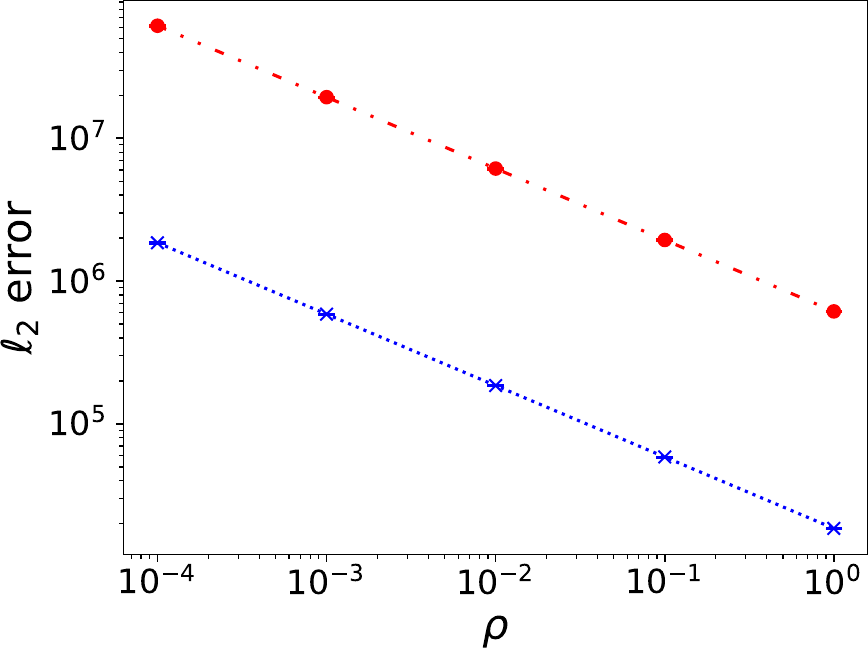}
            \vskip -.08in
            \subcaption{$n=20000$.}
            \;
         \end{subfigure}
         \hfill
         \begin{subfigure}[t]{0.23\textwidth}
            \centering
            \includegraphics[width=\textwidth]{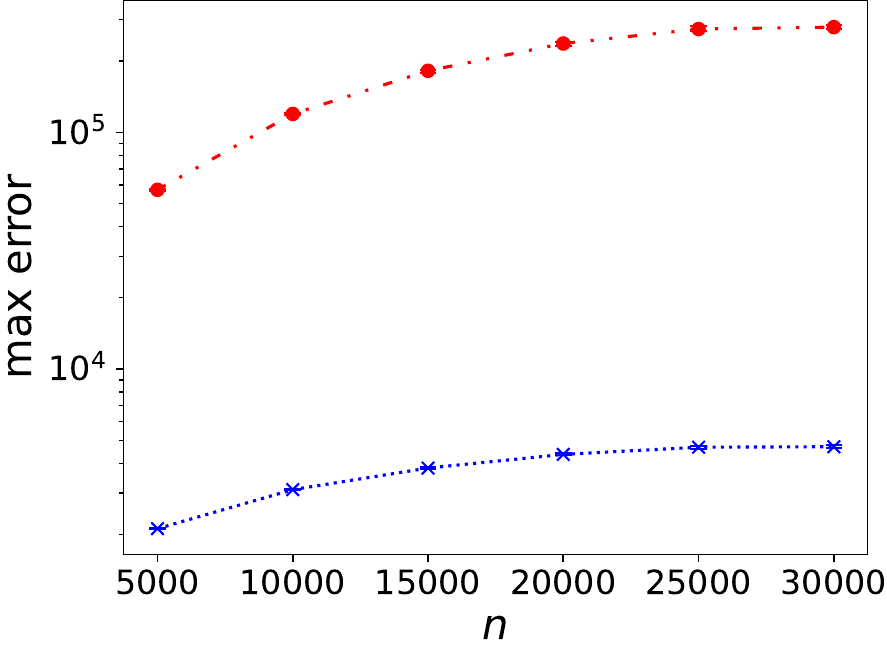}
            \vskip -.08in
            \subcaption{$\rho = 0.01$.}
         \end{subfigure}
         \hfill
         \begin{subfigure}[t]{0.23\textwidth}
            \centering
            \includegraphics[width=\textwidth]{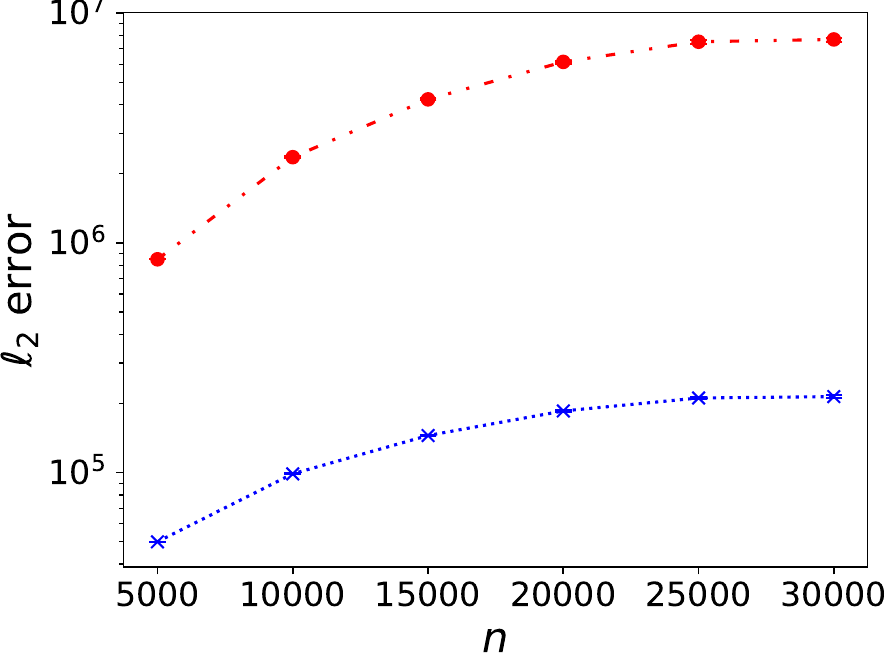}
            \vskip -.08in
            \subcaption{$\rho = 0.01$.}
         \end{subfigure}
         \vskip -.2in
         \caption{Releasing a collection of points : error with respect to privacy level (left) and to tuple size (right).}
         \label{fig:trajerr_lp}
    \end{figure}
    \begin{figure}[H]
     \centering
         \begin{subfigure}[t]{0.23\textwidth}
            \centering
            \includegraphics[width=\textwidth]{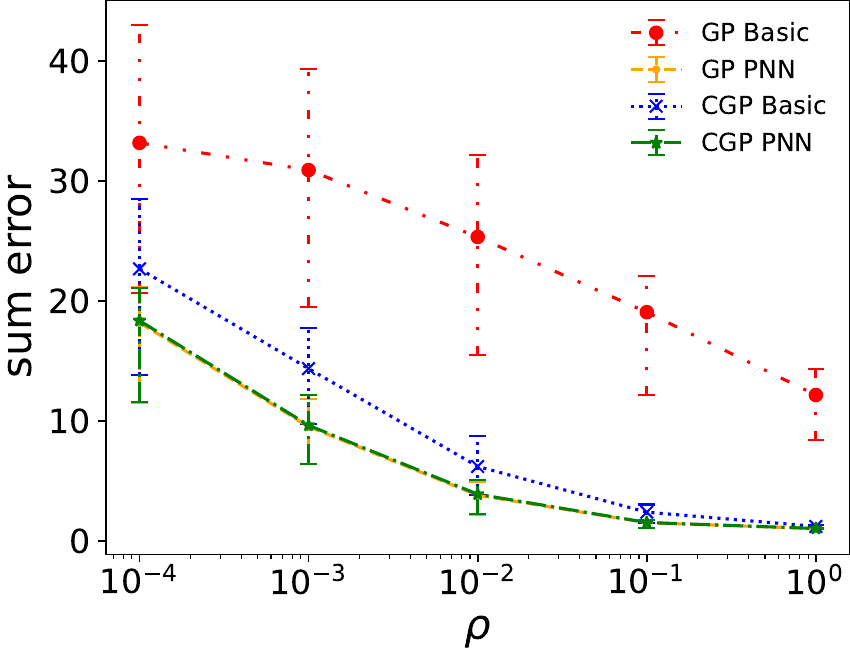}
             \vskip -.08in
            \subcaption{$k=50$.}
            \;
         \end{subfigure}
        \hfill
         \begin{subfigure}[t]{0.23\textwidth}
            \centering
            \includegraphics[width=\textwidth]{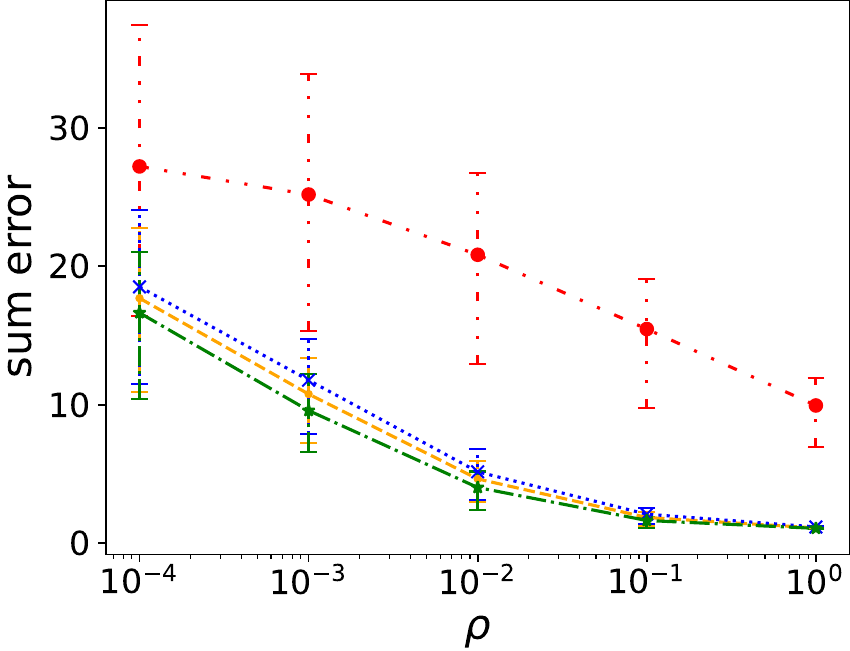}
             \vskip -.08in
            \subcaption{$k=75$.}
            \;
         \end{subfigure}
         \hfill
         \begin{subfigure}[t]{0.23\textwidth}
            \centering
            \includegraphics[width=\textwidth]{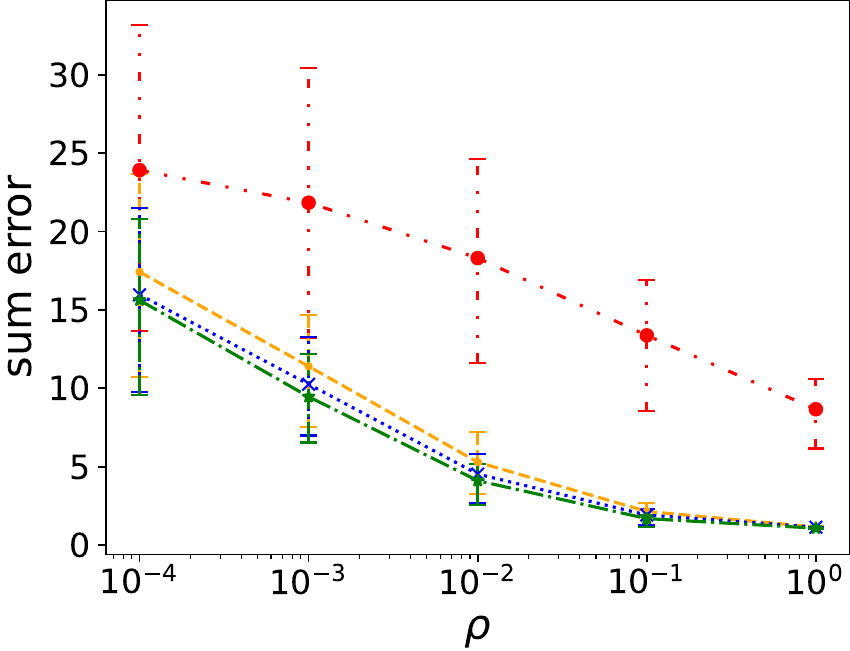}
             \vskip -.08in
            \subcaption{$k=100$.}
         \end{subfigure}
         \hfill
         \begin{subfigure}[t]{0.23\textwidth}
            \centering
            \includegraphics[width=\textwidth]{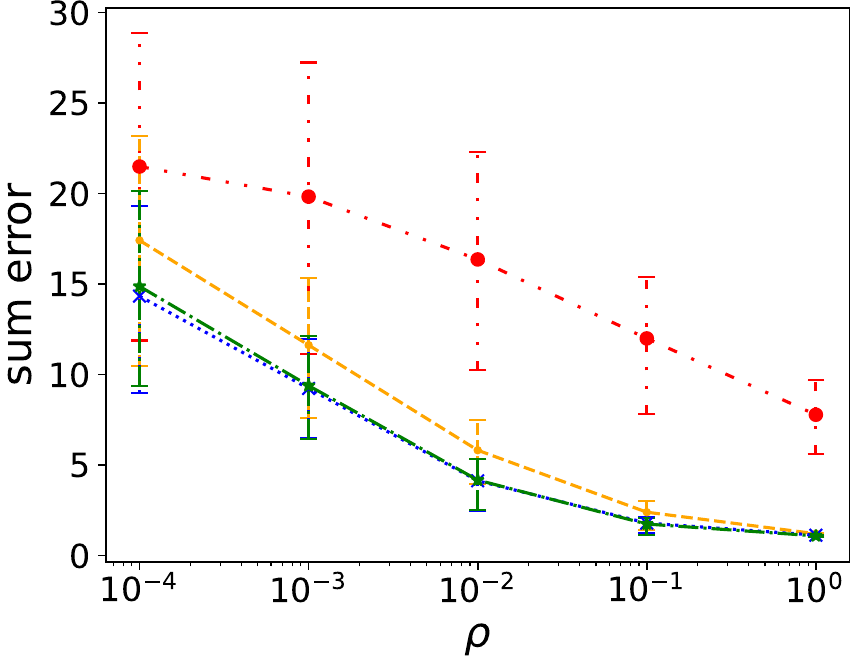}
            \vskip -.08in
            \subcaption{$k=125$.}
         \end{subfigure}
          \vskip -.2in
         \caption{Nearest neighbor: error with respect to privacy level, fixing $n=20000$.}
         \label{fig:sumerr_rho_lp}
    \end{figure}
\begin{figure}[H]
     \centering
         \begin{subfigure}[t]{0.23\textwidth}
            \centering
            \includegraphics[width=\textwidth]{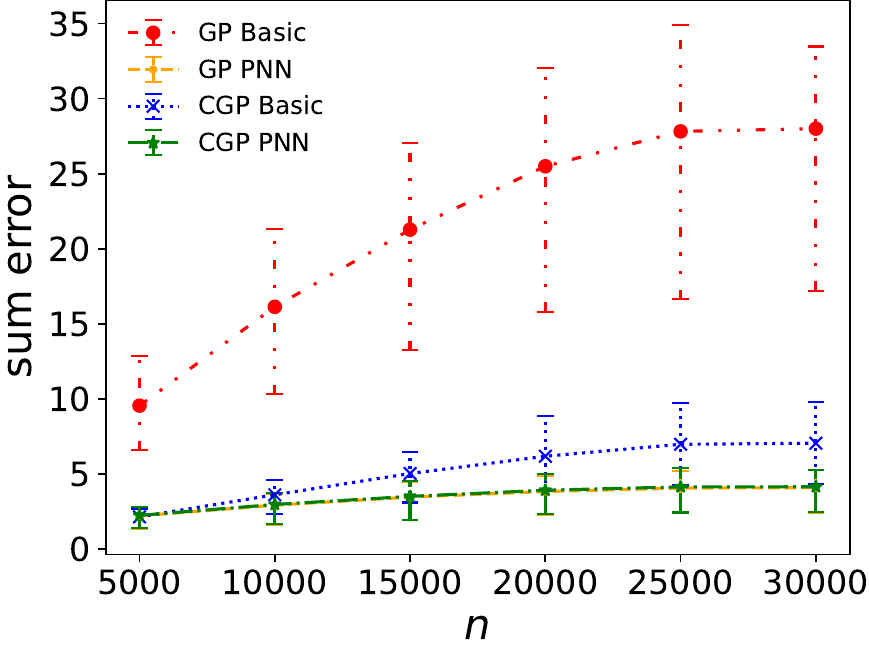}
             \vskip -.08in
            \subcaption{$k=50$.}
            \;
         \end{subfigure}
        \hfill
         \begin{subfigure}[t]{0.23\textwidth}
            \centering
            \includegraphics[width=\textwidth]{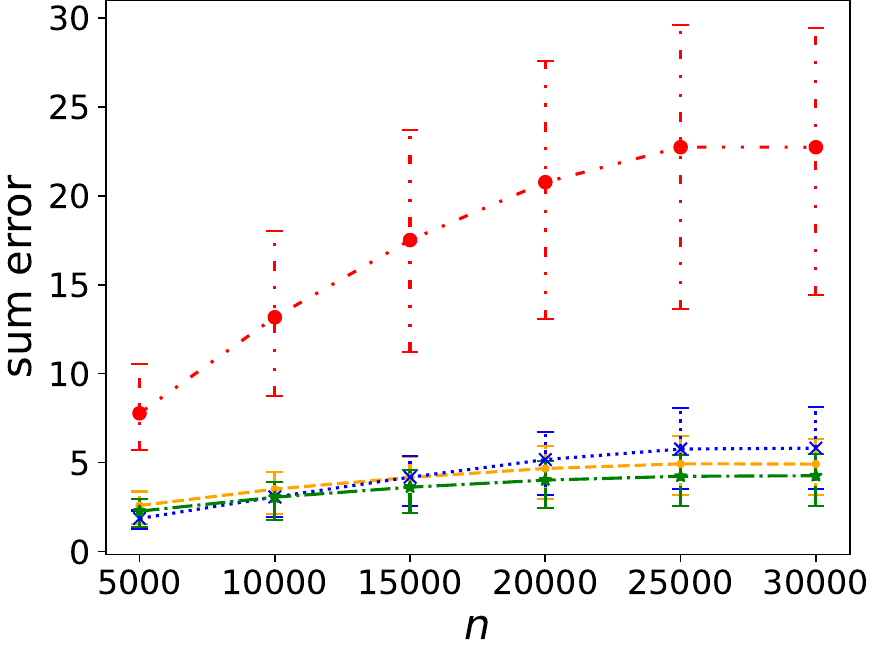}
            \vskip -.08in
            \subcaption{$k=75$.}
            \;
         \end{subfigure}
         \hfill
         \begin{subfigure}[t]{0.23\textwidth}
            \centering
            \includegraphics[width=\textwidth]{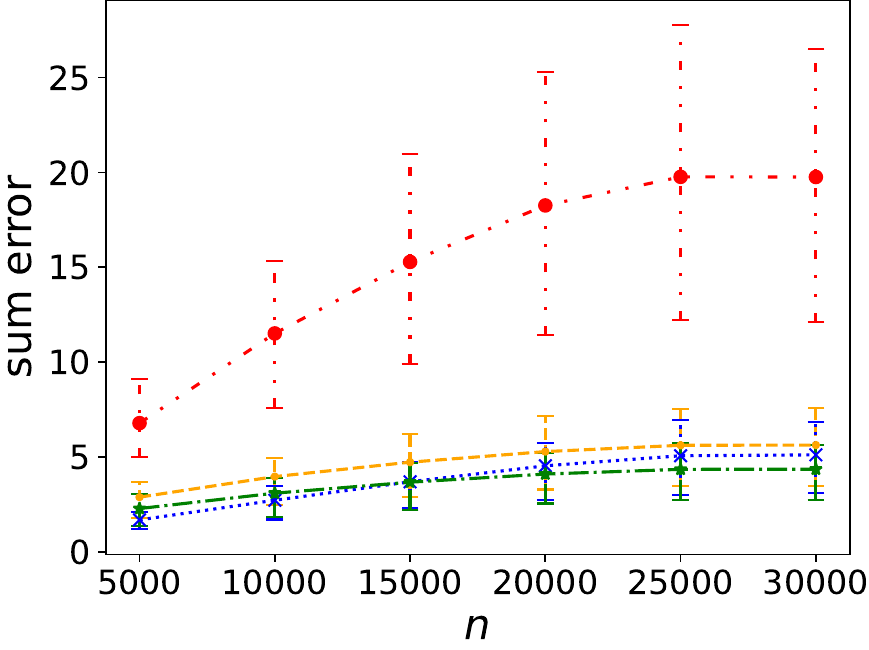}
            \vskip -.08in
            \subcaption{$k=100$.}
         \end{subfigure}
         \hfill
         \begin{subfigure}[t]{0.23\textwidth}
            \centering
            \includegraphics[width=\textwidth]{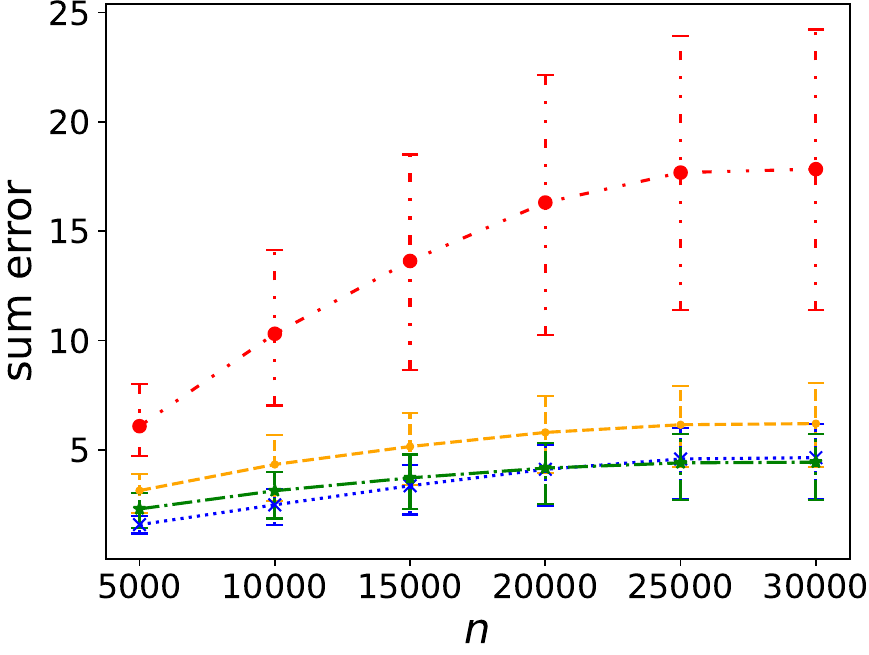}
            \vskip -.08in
            \subcaption{$k=125$.}
         \end{subfigure}
          \vskip -.2in
         \caption{Nearest neighbor: error with respect to tuple size, fixing $\rho=0.01$.}
         \label{fig:sumerr_m_lp}
    \end{figure}
\begin{figure}[H]
     \centering
         \begin{subfigure}[t]{0.23\textwidth}
            \centering
            \includegraphics[width=\textwidth]{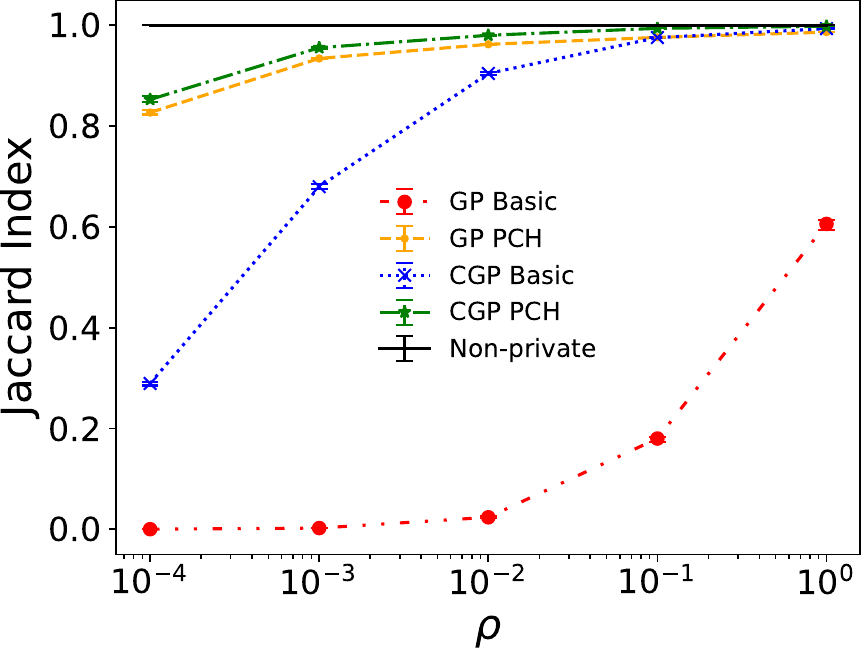}
             \vskip -.08in
            \subcaption{$n=20000$.}
         \end{subfigure}
        \;\;\;\;
         \begin{subfigure}[t]{0.24\textwidth}
            \centering
            \includegraphics[width=\textwidth]{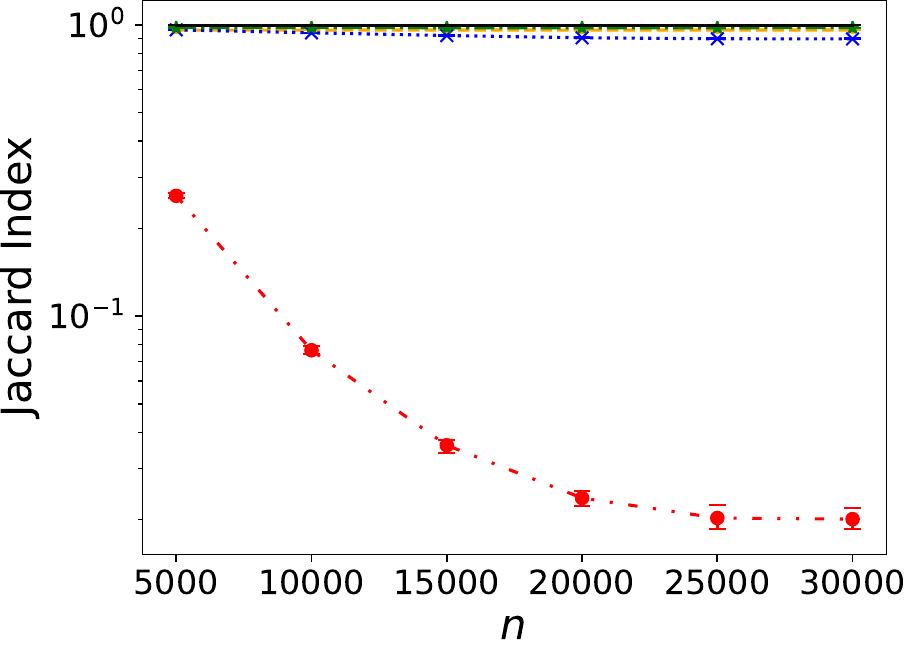}
             \vskip -.08in
            \subcaption{$\rho = 0.01$.}
         \end{subfigure}
          \vskip -.1in
         \caption{Convex hull : error with respect to privacy level (left) and to tuple size (right).}
         \label{fig:convherr_lp}
    \end{figure}

{\section{Metrics for the Central Model}
\label{sec:metrics_central}
For privatization under the central model, where there is a trusted curator who holds $x=(x_1,\dotsb,x_n)$ with $x_i$ corresponding to user $i$, two appropriate metrics are $\dist_1$ and $\dist_2$. In particular, we have
    \begin{equation}
    \nonumber
    \mathrm{dist}_{1}(x,x') := \sum_{i\in [n]} \mathrm{dist}(x_i,x'_i) = 
    \begin{cases}
     \mathrm{dist} (x_j, x'_j) \;, & \text{if}\;x\text{\;and\;}x'\;\text{differ only in entry $j$} \\
    \sum_{i\in [n]} \mathrm{dist}(x_i,x'_i), &\text{otherwise},
    \end{cases}
    \end{equation}
and
    \begin{equation}
    \nonumber
    \mathrm{dist}_{2}(x,x') := \sqrt{\sum_{i\in [n]} \mathrm{dist}(x_i,x'_i)^2} = 
    \begin{cases}
     \mathrm{dist} (x_j, x'_j) \;, & \text{if}\;x\text{\;and\;}x'\;\text{differ only in entry $j$} \\
    \sqrt{\sum_{i\in [n]} \mathrm{dist}(x_i,x'_i)^2}, &\text{otherwise}.
    \end{cases}
    \end{equation}
Thus, for $x\sim x'$ differing in only one entry $j$, both metrics imply $\Pr[M(x)\in S]\le e^{\varepsilon\cdot \dist(x_j, x'_j)} \Pr[M(x')\in S]$.  (The CGP definition is similar.) This captures the privacy guarantee of the central model: The adversary cannot confidently detect the change in any one user's location from the mechanism output and all other users' locations, and the adversary's confidence depends on $\varepsilon\cdot \mathrm{dist} (x_j, x'_j)$.}

\paragraph{The Identity Query.} Suppose we want to privatize the tuple $x=(x_1,\dotsb,x_n)$ where user $i$ contributes $x_i\in \mathbb{R}^2$. 
Under $\dist_2$, we can pose the query as $g:(\mathbb{R}^2)^n\rightarrow \mathbb{R}^{2n}$ defined by 
\[g(x)=[x_{1,1},x_{1,2},x_{2,1},\dotsb,x_{n,1},x_{n,2}]^T.\] 
Then $g$ is $1$-Lipschitz w.r.t. $\mathrm{dist}_2$ (hence, also $\mathrm{dist}_1$). 
Thus, the mechanisms become the $(2n)$-dimensional Laplace mechanism (see Appendix \ref{sec:Rd_gen}) and Gaussian mechanism, respectively. 
As mentioned in Section \ref{sec:Gauss}, the latter is an $\tilde{O}(\sqrt{n})$-factor better in the $\ell_2$ error and much easier to implement. Also, note that under $\dist_2$, there is no need to split the privacy budget since composition is not needed. Thus, the $\ell_2$ error in the $\dist_2$ mechanisms is an $O(\sqrt{n})$-factor better than the $\dist_{\infty}$ mechanisms with composition under the central model. However, the latter gives a stronger privacy guarantee.
    
\end{document}